\documentstyle[12pt,epsf,epsfig]{article}
\begin{document}
\begin{titlepage}
\pagestyle{empty}
\baselineskip=21pt
\rightline{CERN-TH/2001-380}
\rightline{ACT-06/02, CPT-TAMU-16/02}
\rightline{\tt hep-th/0207188}
\vskip 0.25in

\begin{center}
 {\large \bf Universal Calabi-Yau Algebra: }\\
%\vspace{.05in}
{\large \bf   Towards an Unification of Complex Geometry}\\
\end{center}

\vspace{.03in}

\begin{center}
{{\bf F. Anselmo}$^{1}$, {\bf J. Ellis}$^{2}$, {\bf D.V. Nanopoulos}$^{3}$
   and {\bf G. Volkov}$^{4}$}
% \vspace{.25in}

\vspace{.02in}

 {\it $^{1}$ INFN-Bologna, Bologna, Italy\\}
% \vspace{.05in}
 {\it  $^{2}$ Theory Division, CERN, CH-1211 Geneva 23, Switzerland \\}
% \vspace{.05in}
 {\it  $^{3}$ Dept. of Physics,
 Texas A \& M University, College Station, TX~77843-4242, USA,  \\
 HARC, The Mitchell Campus, Woodlands, TX~77381, USA, and \\
 Academy of Athens, 28~Panepistimiou Avenue,
 Athens 10679, Greece\\}
%\vspace{.05in}
{\it  $^{4}$Theory Division, CERN, CH-1211 Geneva, Switzerland,  \\
  LAPP TH, Annecy-Le-Vieux, France, and\\
  St Petersburg Nuclear Physics Institute, Gatchina, 188300 St Petersburg, 
Russia\\}

%\vspace{.05in}

%\newpage
{\bf Abstract}
\end{center}

We present a universal normal algebra suitable for constructing 
and classifying Calabi-Yau spaces in arbitrary dimensions. This algebraic
approach includes natural extensions of reflexive weight vectors to higher
dimensions, related to Batyrev's reflexive polyhedra, and their $n$-ary
combinations. It also includes a `dual' construction based on the
Diophantine decomposition of invariant monomials, which provides explicit
recurrence formulae for the numbers of Calabi-Yau spaces in arbitrary
dimensions with Weierstrass, $K3$, etc., fibrations. Our approach also
yields simple algebraic relations between chains of Calabi-Yau spaces in
different dimensions, and concrete visualizations of their singularities
related to Cartan-Lie algebras. This Universal Calabi-Yau Algebra is a 
powerful tool for decyphering the Calabi-Yau genome in all dimensions.

\vfill
\leftline{CERN-TH/2001-380}
\leftline{July 2002}

\end{titlepage}
\baselineskip=18pt

\section{Introduction: an Algebraic Way to Unify Calabi-Yau Geometry}

Geometrical ideas play ever-increasing r\^oles in the quest to unify all
the fundamental interactions. They were introduced by Einstein in the
formulation of general relativity, and extended to higher dimensions by
Kaluza and Klein in order to include electromagnetism. This is described
by the geometrical principle of gauge invariance, which is also used in
the formulation of the strong and weak interactions. The only known
framework for combining gravity with these other interactions is String
Theory, which has introduced a host of new geometrical ideas into physics.

One of the key problems in String
Theory~\cite{MFrev,rev2,rev3,othermodels} is its reduction from ten to
four dimensions, which is also naturally approached from a geometrical
point of view, though algebraic approaches are also possible. One of the
most powerful ways of compactifying the six surplus dimensions is on a
Calabi-Yau manifold, namely a complex compact manifold with K\"ahler
structure and an $SU(3)$ holonomy
group~\cite{CY,orbifold,Dix,Ler1,can5,Batyrev, K3,Greene1,Sch,
Sha1,Sha2,Dim,Bar,Mir,Lam}. Extensions of the concept of Calabi-Yau
manifolds to other numbers of complex dimensions are also interesting for
various purposes. For example, $K3$ spaces, which may be regarded as
Calabi-Yau manifolds in four real (two complex) dimensions, have been
known and studied for a long time. They may be related to gauge groups
using methods of algebraic geometry, via Coxeter-Dynkin diagrams
~\cite{DuVal, Schlafli, Batyrev,Kodaira,TATE,block1,block3,Can1, SkaKr,
Avram,Skarke, Dais}. Other useful examples are Calabi-Yau spaces in eight real
(four complex) dimensions, which are used in the compactification of $F$
theory from twelve to four dimensions.

One way of constructing at least some Calabi-Yau manifolds is as solutions
of homogeneous polynomial equations in weighted projective spaces,
i.e., polynomial equations in $n$ variables whose monomial exponents
${\vec {\mu}}_{\alpha}$ satisfy the 
relation $\vec{k}  \cdot {\vec {\mu}}_{\alpha} = d$, 
where the $\vec{k}=(k_1,...,k_{n+1})$ are $n+1$ positive
weights that determine the set of  monomials $\vec{m}_{\alpha}$, 
and $d$ is given by the sum of the components of $k_i$, 
i.e.,
$d=k_1+...+k_{n+1}$. The $\vec{\mu}_{\alpha}=(\mu_1,...,\mu_{n+1})_{\alpha}$ 
are the set of ${n+1}$ exponents that characterize the possible monomials 
${\bf m}_{\alpha}$, where $\alpha $ runs over the set of 
integer monomials allowed by the Calabi-Yau equation. All Calabi-Yau 
spaces are not derivable
from weighted projective spaces, but one can obtain many other (possibly
all) Calabi-Yau manifolds as the intersections of two or more such spaces.

A useful technique for constructing Calabi-Yau spaces in any number of
dimensions is to visualize the various possible monomials ${\bf
m}_{\alpha}=(x_1^{\mu_1}x_2^{\mu_2}...x_n^{\mu_{n+1}})_{\alpha} $ as the
$(\mu_1,...,\mu_{n+1})_{\alpha}$ points in the $Z_{n+1}$ integer lattice
of an $n$-dimensional polyhedron. Using this technique,
Batyrev~\cite{Batyrev} demonstrated how to associate by explicit
construction a mirror polyhedron to each Calabi-Yau space. This approach
also established in a very elegant way~\cite{Can1} the corresponding
mirror duality among Calabi-Yau spaces~\cite{duality, block2}.

Some topological properties of Calabi-Yau manifolds in six dimensions,
namely their Euler numbers, are related to the numbers of chiral matter
generations in string theory compactified on them. The Euler and Hodge
numbers are directly calculable in the Batyrev approach, without solving
the Calabi-Yau equations.

We have recently proposed an algebraic approach~\cite{AENV1,AENV2} to the
construction of Calabi-Yau manifolds, in which the geometrical structure
of the Batyrev polyhedra, and therefore the associated gauge groups, are
derived from an algebraic structure of the weight vectors $\vec{k}$. This
opens the way to a generalization of Candelas' results to higher
dimensions. It may also provide a deeper understanding of the origin of
gauge symmetry, which appears naturally in singular limits of Calabi-Yau
spaces.  Moreover, our algebraic approach provides a unified framework for
`decyphering the Calabi-Yau genome', that may ultimately prove suitable
for understanding which of many Calabi-Yau manifolds has been chosen by
Nature to compactify string theory, if any.

In our previous articles~\cite{AENV1,AENV2} we explained how to proceed
constructively by first constructing the weight vectors $\vec{k}$ in a
given dimension $n$ out of sets of between two and $n$ vectors $\vec{k}$
in lower dimensions. Our technique was first to extend the
lower-dimensional vectors by adding zero components in any position in
order to obtain $n$-dimensional `extended' vectors. These were then
combined in binary, ternary, etc., operations to obtain allowed
$n$-dimensional vectors and hence Calabi-Yau spaces, which are associated
into chains characterized by their eldest and youngest vectors. The eldest
vectors substantially determine the structure of all vectors in the chain,
and knowledge of these suffices to obtain much of the important physical
information, reminiscent of the situation with eldest weight vectors and
representations of Cartan-Lie algebras. The resulting Universal Calabi-Yau
Algebra (UCYA) structure of reflexive weight vectors in different
dimensions depends on two integer parameters: the {\it arity} $r$ of the
combination operation $\omega_r$, and the dimension $n$.

As an example of the extension procedure in the case of $K3$ manifolds, we
classified~\cite{AENV1} the 95 different possible weight vectors $\vec{k}$
in 22 binary chains generated by pairs of extended vectors, which included
90 of the total, and 4 ternary chains generated by triplets of extended
vectors, which yielded 91 weight vectors of which 4 were not included in
the binary chains. The one remaining $K3$ weight vector was found in a
quaternary chain \cite{AENV1}.  This algebraic construction provides a
convenient way of generating all the $K3$ weight vectors, and arranging
them in chains of related vectors whose overlaps yield further indirect
relationships.

Moreover, as we discuss in more detail in this paper, our construction
builds higher-dimensional Calabi-Yau spaces systematically out of
lower-dimensional ones, enabling us to enumerate explicitly their
fibrations. As examples, we showed previously~\cite{AENV1,AENV2} how our
construction reveals elliptic and $K3$ fibrations of $CY_3$ manifolds.  
Our approach may also be used to obtain the projective weight vector
structure of a mirror manifold, starting from those of a given
Calabi-Yau manifold.

The main purpose of this article is to develop further the understanding
of this universal algebraic approach to constructing Calabi-Yau manifolds
in various numbers of dimensions with all possible structures, introducing
a new technique based on the Diophantine decomposition of invariant
monomials (IMs).

Our construction of a Universal Calabi-Yau algebra (UCYA) is based on the
two integer parameters, {\it arity} and the {\it dimension} of the
reflexive weight vectors (RWVs), that are connected one-to-one with
Batyrev's reflexive polyhedra. We discussed previously how these could be
classified using the natural extensions of lower-dimensional vectors and
their combination via binary, ternary, etc., operations. The main
innovation in this paper is the introduction of a complementary algebraic
approach to the construction of Calabi-Yau spaces, based on the
construction of suitable monomials ${\vec \mu}$ obeying the `duality'
condition: $\vec{k} \cdot \vec{\mu}_{\alpha} = d$.  
This new `dual' approach is based on suitable
decompositions of invariant monomials (IMs) of given dimensionality,
yielding eldest vectors that could only be obtained by higher-order
$n$-ary operations in the previous approach. This construction supplements
the previous geometrical method related to Batyrev polyhedra, and enables
one to calculate the numbers of eldest vectors, and hence chains, in
arbitrary dimensions. We verify explicitly that the eldest vectors found
in the two different ways agree in several instances for both $CY_3$ and
$CY_4$ spaces, providing increased confidence in our results. The study of
the Calabi-Yau equations and the associated hypersurfaces via the
remarkable composite properties of IMs provides an alternative algebraic
route to reflexive polyhedron techniques. We recall that the
arity-dimension parameter structure is directly connected to the
singularity properties of Calabi-Yau hypersurfaces, and thereby to the
types of Cartan-Lie algebras. Using these remarkable properties one can
hope to decypher the Calabi-Yau genome in any dimension.

We emphasize the complementarity between our approach and that
of~\cite{SkaKr}, which is more geometrical, being based on the
classification of the all reflexive Batyrev polyhedra: see
also~\cite{Avram}.  Our method is more algebraic, being based on the
construction of the weight vectors ${\vec k}$ and/or the corresponding
monomials ${\vec m}$. Central r\^oles are played in our approach by the
composite structures in lower dimensions $\leq (d-1)$ of $CY$ $d$-folds,
and the algebraically dual ways of expansions using weight vectors $\vec 
k$
and invariant monomials (IMs). By analogy with the Galois normal extension
of fields, we term the first way of expanding weight vectors a {\it normal
} extension, and the dual decomposition in terms of {IMs} we call the {\it
Diophantine} expansion. These two expansion techniques are consistently
combined in our algebraic approach, whose composition rules exhibit
explicitly the internal structure of the Calabi-Yau algebra. Our method is
closely connected to the well-known Cartan method for constructing Lie
algebras, and reveal various structural relationships between the sets of
Calabi-Yau spaces of different dimensions. Furthermore, this information
is relatively easy to obtain without large computer facilities. We
interpret our approach as revealing a `Universal Calabi-Yau Algebra'
~\cite{Burris} for the following reasons: `Universal' because it may, in
principle, be used to generate all Calabi-Yau manifolds of any dimension
with all possible substructures, and `Algebra' because it is based on a
sequence of binary and higher $n$-ary operations on weight vectors and
monomials.

We first summarize in Section 2 essential aspects of the UCYA based on
normal extensions of weight vectors, as introduced in our previous
articles~\cite{AENV1,AENV2}. Then, in section 3 we use this method to 
review some results derived previously for $K3$ and $CY_3$ spaces, and to
derive some new results for $CY_4$ spaces. We then develop in Section 4
our new `dual' method based on the decomposition of monomials, and use it
to confirm our previous results on $CY_3$ and $CY_4$ spaces. We also show
that the method of invariant monomials can give recurrence formulae valid
in all dimensions for Calabi-Yau spaces with specific fibrations. We
illustrate this by examples of Calabi-Yau spaces of arity $r=d$,
with some specific types of  elliptic fibres like 
$\{7\}_{\Delta}$,
$\{9\}_{\Delta}$, $\{10\}_{\Delta}$. Finally, Section 5 summarizes the
conclusions and prospects of this algebraic approach.

\section{The Main Elements of  Universal Calabi-Yau Algebra}

\subsection{Basic Framework}

In the search for a universal classification of Calabi-Yau spaces,
one should consider the complex
projective algebraic spaces $CP^n$ with homogeneous coordinates or their
quasihomogeneous generalizations, including toric varieties 
\cite{Fulton, Cox},
since the only complex compact submanifold (with analytic structure)
embedded in $C^n$ is a point~\cite{Sch}. 
The starting point for our algebraic
approach to the classification of Calabi-Yau spaces has therefore been the
construction of `reflexive' weight vectors ${\vec k}$, whose components
specify the complex quasihomogeneous projective spaces
${CP^n(k_1,k_2,...,k_{n+1})}$. These have {$(n+1)$} quasihomogeneous
coordinates ${x_1,...,x_{n+1}}$, which are subject to the following
identification:
\begin{equation}
(x_1, \ldots ,x_{n+1})\,\sim\, (\lambda ^{k_1} \cdot x_1, \ldots ,
\lambda ^{k_{n+1}} \cdot x_{n+1}).
\label{quasihom}
\end{equation}
In the case of $CP^n$ projective spaces there exists a very powerful
conjecture, called Chow's theorem, that each analytic compact (closed)
submanifold in $CP^n$ can be specified by a set of polynomial equations.
The set of zeroes of quasihomogeneous polynomial equations, hereafter
referred to as Calabi-Yau equations, define a projective algebraic variety
in such a weighted projective space.

A $d$-dimensional Calabi-Yau space $X_d$ can be given by the locus of
zeroes of a transversal quasihomogeneous polynomial ${\wp}$ of degree $deg
({\wp})=[d]: [d] = \sum _{j=1}^{n+1} k_j$ in a complex projective space
$CP^n(\vec {k}) \equiv CP^n (k_1,...,
k_{n+1})$
~\cite{K3,Greene1}:
\begin{eqnarray}
X \equiv X^{(n-1)}({k}) \equiv
\{\vec{x}=(x_1,...,x_{n+1})\in CP^n({k})|{\wp}(\vec{x}) = 0\}.
\label{poly1}
\end{eqnarray}
The general quasihomogeneous polynomial of degree $[d]$ is a linear combination
\begin{equation}
{\wp} = \sum_{\vec{\mu}_{\alpha}} c_{\vec{\mu}_{\alpha}}
x^{\vec{\mu}_{\alpha}}
\label{poly2}
\end{equation}
of  monomials  $x^{{\vec{\mu}}_{\alpha}} = {x_1^{{\mu}_{1\alpha}}
x_2^{{\mu}_{2\alpha}}...x_{r+1}^{{\mu}_{(r+1)\alpha}}}$ 
with the condition:
\begin{equation}
{\vec {\mu}}_{\alpha} \cdot \vec {k}\,=\,[d].
\label{poly3}
\end{equation}
This algebraic projective variety is irreducible if and only if
its polynomial is irreducible. A hypersurface will be smooth
for almost all choices of polynomials. To obtain Calabi-Yau $d$-folds one
should choose reflexive weight vectors (RWVs), related to Batyrev's
reflexive polyhedra or to the set of {IMs}. 
Other examples of compact complex manifolds can  be
obtained as the complete
intersections (CICY) of such quasihomogeneous polynomial constraints:
\begin{eqnarray}
X^{(n-r)}_{CICY}
=\{\vec{x}=(x_1,\ldots x_{n+1})\in CP^n \,|\, {\wp}_1(\vec x)
=\ldots ={\wp}_r(\vec x)=0 \},
\end{eqnarray}
where each polynomial ${\wp}_i$ is determined by some weight vector 
$\vec{k}_i$, $i=1,\ldots, r$.

We also recall the existence of {\it mirror symmetry}, relating
each Calabi-Yau manifold to a dual partner, which was first observed
pragmatically in the literature \cite{Dix, Ler1,can5,Roan,block2}
 to which
we return later.

\subsection{The Holomorphic-Quotient Approach to Toric Geometry}

In the toric geometry approach, algebraic varieties are described by a
dual pair of lattices ${ \Lambda}$ and ${ \Lambda^*}$, each isomorphic to
$Z^n$, and a fan $\Sigma$~\cite{Fulton} defined on ${ \Lambda^*_R}$, the
real extension of the lattice ${ \Lambda^*}$. In the new
holomorphic-quotient approach of Batyrev~\cite{Batyrev} and
Cox~\cite{Cox}, a single homogeneous coordinate is assigned to the system
specified by the variety $\mho_{\Sigma}$, in a way similar to the usual
construction of $CP^n$. This holomorphic-quotient construction immediately
gives us the usual description in terms of projective spaces, and moreover
describes naturally the elliptic, $K3$ and other fibrations of Calabi-Yau
spaces that interest us.

The integer points of ${\Delta}^{*} \cap { \Lambda^*} $ define one-dimensional
cones $$({\vec v}_1,...,{\vec v}_{\it N}) = {{\Sigma}^1}_{{\Delta}^{*}} $$
of the fan $\Sigma_{{\Delta}^{*}}$, to each of which one can assign a
coordinate $x_k: k = 1, ..., N$. The one-dimensional cones span the vector
space $\Lambda^*_R$ and satisfy $({\it N} -n)$ linear relations with
non-negative integer coefficients:

\begin{equation}
\sum_l k_j^l {\vec v_l} = 0, \,\,\, k_j^l \geq 0.
\end{equation}
These linear relations can be used to determine relations of equivalence
on the space $C^{\it N} \backslash Z_{\Sigma^{*}}$.
A variety ${\mho}_{\Sigma_{{\Delta}^{*} } }$ is the space 
$C^N \backslash Z_{{\Sigma}_{{\Delta}^{*}}}$ modulo the action of a group 
which is product 
of a finite Abelian group and the torus $(C^{*})^{(\it N-n)}$ :
\begin{equation}
(x_1,...,x_{\it N} )  \sim ({\lambda} ^{k^1_j} x_1,...., 
{\lambda }^{k^N_j} x_N),\,\,\, j = 1,...,{\it N}-n.
\end{equation}
The set $Z_{\Sigma_{{\Delta}^{*}}}$ is defined by the fan in the 
following way:

\begin{equation}
Z_{\Sigma_{{\Delta}^{*} }} \, =\, \bigcup_ I((x_1,...,x_{\it N}) | x_i =0 ,
\forall i \in  I)
\end{equation}

where the union is taken over all index sets $I = ( i_1,...,i_k)$ such
that $({\vec v}_{i_1},..., {\vec v}_{i_k})$ do not belong to the same
maximal cone in ${\Sigma}^{*}$, or several ${x_i}$ can vanish
simultaneously only if the corresponding one-dimensional cones ${\vec
v_i}$ are from the same cone.  The elements of $ {\Sigma}_1^{*}$ are in
one-to-one correspondence with divisors

\begin{equation}
D_{v_i}\,=\, \mho_{\Sigma^{1i}_{{\Delta}^{*}}}.
\end{equation}

The divisors ${D_{v_i}}$ are subvarieties given simply by ${ x_i=0}$.  
The divisors ${ D_{v_i}}$ form a free Abelian group $Div
(\mho_{\Sigma_{{\Delta}^{*} }})$. In general, a divisor $ D \in Div
(\mho_{\Sigma_{{\Delta}^{*}}})$ is a linear combination of some
irreducible hypersurfaces with integer coefficients:

\begin{equation}
{ D=\sum a_i \cdot D_{v_i}}.
\end{equation}
The points of $ \Delta \cap \Lambda$ are in one-to-one correspondence with the
monomials in the homogeneous coordinates $x_i$. A general polynomial is
given by

\begin{equation}
\wp\,=\, \sum_{{\vec \mu} \in {\Delta \cap \Lambda}} c_{\vec \mu} 
\prod_{l=1}^{\Lambda^*}  x_l^{\langle {\vec v}_l, {\vec \mu} \rangle +1}.
\end{equation}   
The equation ${ \wp=0}$ is well defined, and $ \wp$ is holomorphic if the
following condition

\begin{equation}
\langle {\vec v}_l, {\vec \mu} \rangle \, \geq \,-1 \,\,\, for \,\,all\,\,l
\end{equation}
is fulfilled. The $c_{\vec \mu}$ parametrize a family of ${ M_{\Delta}}$ of
Calabi-Yau surfaces defined by the zero locus of $\wp$.

We see that toric varieties can be defined by the quotient $C^k \backslash
Z_{\Sigma}$, not only by a group $(C \backslash {0})^{k-n}$. One should
divide $C^k \backslash Z_{\Sigma}$ also by a finite Abelian group
$G(v_1,...,v_k)$, that is determined from the relations between the
$D_{v_i}$ divisors. In this case, the toric varieties can often have
orbifold singularities, $C^k \backslash G$.
%For example, the toric variety defined by 
%equation {(13)} near points { y=z=0} and { x=z=0} looks locally
%like ${ C^2\backslash Z_2}$ and  ${ C^2\backslash Z_3}$,
%correspondingly.      
Orbifolding of a manifold $M$ by a group ${G}$ gives a new orbit space

\begin{equation}
 M \,\Rightarrow \, M/G
\end{equation} 
which is characterized by the identification ${ x\, \sim \, y}$
of all points ${ x, y \in M}$, such that 

\begin{equation}
x\,=\,g(y),\,\,\,\,g\in G.
\end{equation}

The {isotropy subgroup} ${ G_P }$ is determined by the following
condition:

\begin{eqnarray}
G_{x_P}\,=\,\{ g\in G\,:\,g \cdot x_P\,=\,x_P,\,\,\,x_P\in M\,\}\,\in \,G.
\end{eqnarray}
These points are called fixed points, and yield singularities of the
hypersurface ${ M/G}$.
If ${ G}$ acts { freely}, i.e., if ${ g \cdot x_P=x_P}$
for some ${ g\in G}$ implies ${g=1}$, then the projection
\begin{eqnarray}
p\,:\, M\,\Rightarrow \, M/G
\end{eqnarray} 
is called a covering projection.

Similarly, due to the transformations 
$x_i \rightarrow \lambda ^{k_i} x_i,\, \lambda \in C^*$,
whose orbits define points of $CP^n$ , the weighted projective space 
has singular strata
$$F_I = P^n (\overrightarrow{k})\bigcap 
[x_i=0,\forall i \in (1,...,n+1) \setminus I] $$
if the subset of $([k_i],\,\, i \in I)$, has a non-trivial 
common factor $N_I$.
The possible singular sets on $X$ are either points or 
curves. In the cases of singular points, these singularities are locally 
of the type $C^2/{G}$,
whilst a singular curve has locally a $C^3/{G}$
singularity, where $G$ is a discrete group. 
Both types of singularities and their { resolution}
can be described by the methods of toric geometry,
using the `blow-up' procedure~\cite{K3, Greene1,Dais}.
For example, resolving the ${ C^2/Z_n}$ singularity
gives for rational, i.e., genus zero,
(-2)-curves an intersection matrix
that coincides with the ${ A_{n-1}}$ Cartan matrix.
For a general form of the ${C^2/G}$ singularity, one can show~\cite{K3}
using the condition ${K=0}$ that ${G}$ is a discrete subgroup
of ${ SU(2)}$. Any discrete subgroup of ${SU(2)}$ can be projected into
a subgroup of ${SO(3)}$, and thus can be related to the finite 
symmetry classification of three-dimensional space. Thus, resolving 
the orbifold singularities yields a beautiful interrelation
between the classification of finite group rotations in 
three-space and the {ADE} classification of Cartan-Lie algebras. 

%Changing the values of the moduli parameters 
%$\vec {\mu}$ in a general Calabi-Yau polynomial 
%$\wp$ under the regularity
%condition, one can get a family of C-Y varieties. However, this 
%regularity leaves the possibility to meet the singularities in the 
%ambient space $P^n$. 
%After  resolving these singularities one can come to a 
%smooth C-Y manifold.

\subsection{Calabi-Yau Spaces as Toric Fibrations}

A huge set of the reflexive polyhedra corresponding to Calabi-Yau
manifolds can be classified by their fibration structures. In this way it
is possible, as we show explicitly later, to connect the structures of all
the projective vectors in one specific dimension with the projective
vectors of other dimensions and, as a result, to construct a new algebra
acting on the set of all reflexive weight vectors (RWVs), giving the full
set of ${CY_d}$ hypersurfaces in all dimensions: ${ 1 \leq \,d < \,\inf}$.

To understand this better, we provide more information about two
operations, {\it intersection} and {\it projection}, that give 
information about the possible fibration structures of $CY_d$ spaces 
defined via reflexive polyhedra~\cite{Can1,Skarke}:
\begin{itemize}
\item{There may exist a projection operation 
$\pi:\Lambda \rightarrow \Lambda _{n-k}$, where  $\Lambda_{n-k}$ is an 
$(n-k)$-dimensional sublattice and $\pi (\Delta)$ is also a reflexive 
polyhedron;}
\item{There may exist an intersection operation $\sigma$ through 
the origin
of a reflexive polyhedron, such that $\sigma(\Delta)$ is a again a 
$(n-l)$-dimensional reflexive polyhedron;}
\item{These operations exhibit the following duality properties:}
\end{itemize}
\begin{eqnarray}
 \pi(\Delta)\,&  \Leftrightarrow &\,  \sigma(\Delta^*)  \nonumber\\
  \sigma(\Delta)\,&  \Leftrightarrow &\,\pi(\Delta^*).
\end{eqnarray}

For a reflexive polyhedron $\Delta$ with fan $\Sigma^{*}$ over a
triangulation of the facets of ${\Delta}^*$, the Calabi-Yau hypersurface
in the variety ${\mho}_{\Sigma^{*}}$ is given by the zero locus of the
polynomial $\wp$.
%\begin{equation}
%\wp \,=\, \sum_{\mu \in {\Delta \cap M}}\, c_{\vec {\mu}}\,\cdot
%\prod_{i=1}^{N}  x_i^{\langle \vec{v}_i \cdot \vec{\mu} \rangle +1}. 
%\end{equation}
One can consider the variety ${\mho}_{\Sigma^{*}}$ as a fibration over the
base ${\mho}_{{\Sigma}_{base}}$ with generic fibre
${\mho}_{{\Sigma}_{fibre}}$.  This fibration structure can be written in
terms of homogeneous coordinates. The fibre is determined as an algebraic
subvariety by the polyhedron ${\Delta}^*_{fibre} \subset {\Delta}^*_{CY}$,
and the base can be seen as projection of the fibration along the fibre.
The set of one-dimensional cones in ${\Sigma}_{base}$ (the primitive
generator of a cone is zero or ${\tilde {\vec v}}_i$) is the set of images
of one-dimensional cones in ${\Sigma}_{CY}$ (the primitive generator is
${\vec v}_j$) that do not lie in $N_{fibre}$. The image ${\Sigma}_{base}$
of ${\Sigma}_{CY}$ under $$\Pi : N_{CY} \rightarrow N_{base}$$ gives us
the following relation:

\begin{eqnarray}
\Pi {\vec v}_i\, =\, r_i^j \cdot {\tilde {\vec v}}_j
\end{eqnarray}
if $\Pi {\vec v}_i$ is in the set of one-dimensional cones determined by 
${\tilde{\vec v}}_j$ $r^j_i \in N$, otherwise $r_i^j=0$.

Similarly, the base space is the weighted space with the toroidal
structure:

\begin{eqnarray}
({\tilde x}_1,..., {\tilde x}_{\tilde N})\,\, \, \sim
({\lambda}^{{\tilde k}^1_j} \cdot {\tilde x}_1,...,
{\lambda}^{{\tilde k}^{\tilde N}_j} \cdot {\tilde x}_{\tilde N}),
\,\,\, j=1,...,\tilde N - \tilde n,
\end{eqnarray} 
where the ${\tilde k}^i_j$ are integers such that
$\sum_j {{\tilde k}_i^j} {\tilde {\vec v}}_j = 0$.
The projection map from the variety ${\mho}_{\Sigma}$ to the base can be 
written as
\begin{eqnarray}
{\tilde x}_i\,=\, \prod_j x_j^{r_j^i},
\end{eqnarray}
that corresponds to a redefinition of the torus transformation for
${\tilde x}_i$: 
\begin{eqnarray}
\Pi:
{\tilde x}_i \rightarrow {\lambda}^{{k^j_l}\cdot {r^i_j}} \cdot 
{\tilde x}_i ,\,\,\,\, \sum{ k_l^j} \cdot {r_j^i} \cdot 
{\tilde {\vec v}}_i \,=\,0.
\end{eqnarray}
A simple well-known example with an elliptic fibre and base ${ P^1}$
is given by the following Weierstrass equation for the fibre.
%\begin{equation}
%{ y^2\,=\, x^3\,+\,f(z_1,z_2)\cdot x \cdot z^4\,+\,g(z_1,z_2)\cdot z^6,}
%\end{equation}
%where the coefficients ${ f(z_1,z_2),g(z_1,z_2)}$ are functions on the 
%base. 
To illustrate how this may appear, we consider the case of a dual 
pair of polyhedra ${\Delta}(P^3(1,1,4,6)[12])$ and its dual ${\Delta^*}$.
The mirror polyhedron contains, as intersection $H$ through the
interior point, the elliptic fibre $P^2(1,2,3)$. For all integer points
of ${\Delta}^*$ (excluding the interior point) one can write 
the corresponding complex variables:
\begin{eqnarray}
{\vec v}_1\,&=&\,(0,\underline{-2,-3}) \rightarrow \, x_1 \nonumber \\
{\vec v}_2\,&=&\,(0,\underline{-1,-2}) \rightarrow \, x_2 \nonumber \\
{\vec v}_3\,&=&\,(0,\underline{-1,-1}) \rightarrow \, x_3 \nonumber \\
{\vec v}_4\,&=&\,(0,\underline{0,-1})  \rightarrow \, x_4 \nonumber \\
{\vec v}_0\,&=&\,(0,\underline{0,0})                      \nonumber \\
{\vec v}_6\,&=&\,(0,\underline{1,0})   \rightarrow \, x_6 \nonumber \\
{\vec v}_7\,&=&\,(0,\underline{0,1})   \rightarrow \, x_7
\end {eqnarray}
and
\begin{eqnarray}
{\vec v}_8\,&=&\,(-1,\underline{-4,-6}) \rightarrow \, x_8 \nonumber \\
{\vec v}_9\,&=&\,(1,\underline{0,0}) \rightarrow \, x_9.
\end{eqnarray}
There are some linear relations between integer points inside the fibre:
\begin{eqnarray}
{\vec v}_1\,+\,2\cdot {\vec v}_6\,+\,3 \cdot {\vec v}_7\,&=&\,0,\nonumber\\
{\vec v}_2\,+\,       {\vec v}_6\,+\,2 \cdot {\vec v}_7\,&=&\,0,\nonumber\\
{\vec v}_3\,+\,       {\vec v}_6\,+\, {\vec v}_7       \,&=&\,0,\nonumber\\
{\vec v}_4\,+\,                      {\vec v}_7        \,&=&\,0
\end{eqnarray}
and we have the relation between all the points in ${\Delta}^*$:
\begin{eqnarray}
{\vec v}_8\,+\,{\vec v}_9\,+\,4\cdot {\vec v}_6\,+\,6 \cdot {\vec 
v}_7\,&=&\,0.
\end{eqnarray} 
We see later that, according
our algebraic approach, this example corresponds to the case of the eldest 
RWV $\vec{k}_8=(1,1,1,1,1,1,8,13)[27]$, which defines a $CY_6$ space that
has an elliptic K3 fibre. The 7-dimensional polyhedron corresponding to
this $CY_6$ has an intersection consisting of the 3-dimensional
polyhedron ${\Delta}(P^3(1,1,4,6))$. This example indicates that
the toric description `closes upon itself' and has a natural
extension to higher dimensions. 
This provides additional motivation for an 
algebraic description of Calabi-Yau geometry in 
all dimensions. The toric holomorphic
approach can be embedded naturally in the UCYA: see~\cite{AENV1,AENV2} 
and Sections 3 and 4.

The polyhedron ${\Delta}(P^3(1,1,4,6))$ is  contains 39 points, which can be 
subdivided as follows. The even points of the fibre $P^2(1,2,3)$ are 
determined by the intersection of the plane 
$m_1\,+\,2 \cdot  m_2\, +\,3 \cdot  m_3\,=\,0$
with the positive integer lattice. 
This plane separates the remaining 32 points into 16 `left'
and 16 `right' points. These `left' and `right' points each have 
structures isomorphic to the 
Coxeter-Dynkin diagrams for $E_8$, and their singularities can be used to 
produce these groups, as we now illustrate.

\begin{figure}[th!]
   \begin{center}
   \mbox{
   \epsfig{figure=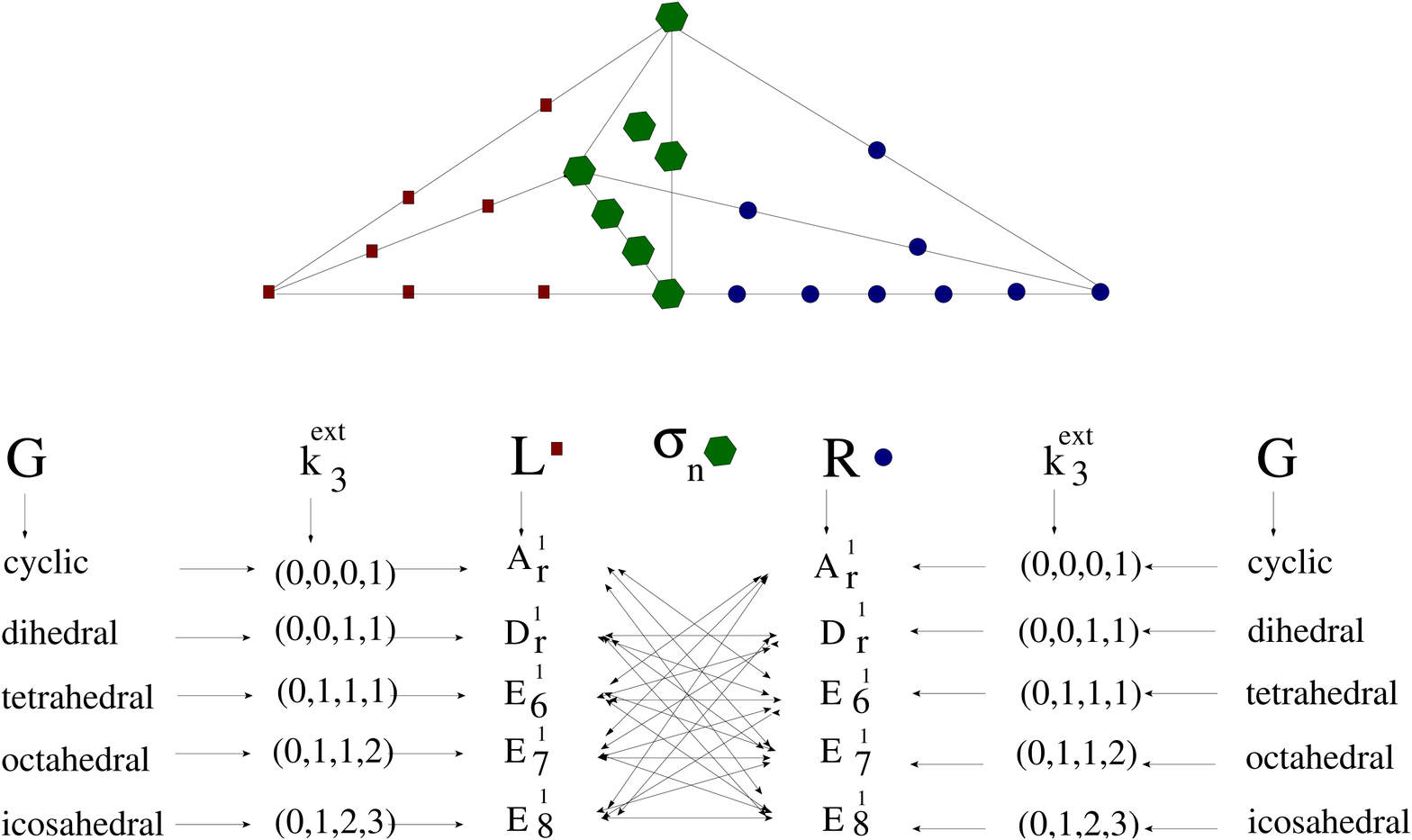,height=16cm,width=18cm}}
   \end{center}
   \caption{\it  
The $K3$ polyhedron determined by the reflexive weight vector
${ \vec{k}_4=(1,1,3,4)[9]}$, which illustrates the appearance of
Coxeter-Dynkin diagrams. The intersection $\sigma$ is determined by {7}
point monomials that correspond to the elliptic fibre $\{7\}_{\Delta}$, and
divides the polyhedron into {7}(+ {3}) points on the left and {9}(+{7}) 
on the right.
These reproduce the Coxeter-Dynkin diagrams for affine $E_6$ and $E_8$,
respectively. Underneath, we also show schematically the general nature of
the highest-weight vectors obtained by arity-2 construction in the UCYA,
displaying the one-to-one link between the 5-dimensional weight vectors  
and the $ADE$ series of Cartan-Lie algebra in $K3$ hypersurfaces. The
r\^oles of the discrete symmetry groups were discussed in~\cite{AENV1}}.
\label{cdjad}
\end{figure}

The $7$ points of the plane $H({\Delta})=m_1+2m_2+3m_3$ are the following:

\begin{eqnarray}
t_1\,&=&\, ({ 5},\underline{-1,-1})\rightarrow ({ x_8^6 x_9^6})\cdot
(\underline {x_1^6 x_2^4 x_3^3x_4^2}),\nonumber\\
t_2\,&=&\, ({ 3},\underline{0,-1})\rightarrow ({ x_8^4 x_9^4})\cdot
(\underline {x_1^4 x_2^3 x_3^2 x_4^2 x_6}),\nonumber\\
t_3\,&=&\, ({ 2},\underline{-1,0})\rightarrow ({ x_8^3 x_9^3})\cdot
(\underline {x_1^3 x_2^2 x_3^2 x_4 x_7}),\nonumber\\
t_4\,&=&\, ({ 1},\underline{1,-1})\rightarrow ({ x_8^2 x_9^2})\cdot
(\underline {x_1^2 x_2^2 x_3 x_4^2 x_6^2}),\nonumber\\
t_5\,&=&\, ({ 0},\underline{0,0})\rightarrow ({ x_8 x_9})\cdot
(\underline {x_1 x_2 x_3x_4x_6x_7}),\nonumber\\
t_6\,&=&\, ({ -1},\underline{2,-1})\rightarrow 
(\underline {x_2 x_4^2 x_6^3}),\nonumber\\
t_7\,&=&\, ({ -1},\underline{-1,1})\rightarrow 
(\underline {x_3 x_7^2}).\nonumber\\
\end{eqnarray}
One can write the Weierstrass equation for the $E_{8_L}$ group
based on the polyhedron ${\Delta}(P^3(1,1,4,6))$ in the form:

\begin{eqnarray}
&&{\underline {x}}_6^3\, +\, \,{\underline {x}}_6^2\cdot ( a_2^{(1)} 
{ x_8 x_9^3} \,+\, a_2^{(2)} { x_9^4})\,+\nonumber\\
&& {\underline {x}}_1^4 \cdot {\underline {x}}_6 
\cdot( a_4^{(1)}{ x_8^3x_9^5}\,+
\,a_4^{(2)} { x_8^2x_9^6}\,+\,
a_4^{(3)} { x_8 x_9^7}\,+\,a_4^{(4)}{ x_9^8})\,+\nonumber\\
&&{\underline {x}}_1^6\cdot ( a_6^{(1)} { x_8^5x_9^7}\,
+\,a_6^{(2)} { x_8^4x_9^8}\,
+\, a_6^{(3)}{ x_8^3x_9^9}\,+\, a_6^{(4)} { x_8^2x_9^{10}}\,+
a_6^{(5)}{ x_8x_9^{11}}\,+\,a_6^{(6)}{ x_9^{12}})\nonumber\\
&&=\, {\underline{x}}_7^2\,+\,a_1\cdot {\underline{x}}_6{\underline{x}}_7
\cdot { x_9^2}\,+  \,
{\underline{x}}_7\cdot (a_3^{(1)} { x_8^2x_9^4}\,
+\,a_3^{(2)}{ x_8x_9^{5}}).
\end{eqnarray}
The second Weierstrass equation for the $E_{8_R}$ group
can be obtained from this equation by interchanging base variables:
$x_8 \leftrightarrow x_9$. The coefficients $a_i$ correspond to the notations 
of the paper by Bershadsky {\it et al.}~\cite{Kodaira}.
The Weierstrass triangle equation 
can be presented in the following general form, where we write
${\underline {x}}_6 \equiv x$, ${\underline{x}}_7 \equiv y$:

\begin{eqnarray}\label{W2}
y^2 \,+\, a_1 \cdot x \cdot y\,+\,a_3 \cdot y \,=\,
x^3 \,+ \, a_2 \cdot x^2\, +\, a_4 \cdot x \,+\, a_6,
\end{eqnarray}
and the $a_i$ are polynomial functions on the base.
In a more simplified form, the Weierstrass equation can be written as:
\begin{equation}\label{W1}
y^2\, =\, x^3\,+\, x \cdot f \,+\, g,
\end{equation}
with discriminant 
\begin{equation}
\Delta \, =\, 4 f^3 \, +\, 27 g^2.
\end{equation}
In the zero locus of the discriminant, some  divisors $D_i$ define
the degeneration of the torus-fibre.
If one can choose the polynomials $f$ and $g$ to be homogeneous
of orders {8} and {12}, respectively, the fibration will 
degenerate over {24} points of the base. For this form of Weierstrass 
equation, there exists an $ADE$ classification of degenerations of 
elliptic fibres, as given by Kodaira~\cite{Kodaira}. In this approach, the 
type of degeneration
of the fibre is determined by the orders $a, b, c$ of the zeroes of the 
functions $f$, $g$ and ${\delta}$. For the case under consideration, one 
has two
singularities at $x_8=0$ and $x_9=0$ with $(a=4,b=5,c=10)$, 
both corresponding to $E_8$. The general algoritm for the $ADE$
classification of elliptic singularities for the general Weierstrass 
equation was considered by Tate~\cite{TATE,Kodaira}. Tate's algorithm
allows one to define in general the type of Lie-algebra singularity. The 
example of the $K3$ space determined by the RWV ${\vec k}_4 = (1, 1, 3, 
4)[9]$ is displayed in Fig.~\ref{cdjad}.

As an example of the use of these observations, we consider $F$-theory,
which can be considered as a decompactification of the type-$IIA$ string.
Our understanding of the duality between the heterotic string and
type-$IIA$ string in $D=6$ dimensions can be used in the duality between
heterotic string on $T^2$ and $F$-theory on an elliptically fibred $K3$
hypersurface. The gauge group is directly defined by the above $ADE$
classification of the quotient singularities of hypersurfaces. The Cartan
matrix of the Lie group in this case coincides, up to a sign, with the
intersection matrix of the blown-down divisors. There are two different
mechanisms for enhancing the gauge groups on the $F$-theory side and on
the heterotic string theory side. On the $F$-theory side, the
singularities of the Calabi-Yau hypersurface give rise to the gauge
groups, but on the heterotic side the singularities can enhance the gauge
group if `small' instantons of the gauge bundle lie on these
singularities. These questions have been studied as a function of the
number of instantons placed on a singularity of type $G$, where $G$ is a
simply laced group.

Furthermore, the elliptic $CY_n$ ($n=3,4$) with $K3$ fibres can be also
used to study $F$-theory dual compactifications of the $E_8 \times E_8$
or $SO(32)$- string theory. To study this in toric geometry, one may
consider a $K3$-polyhedron fibre as a subpolyhedron of the $CY_n$
polyhedron, and the Dynkin diagrams of the gauge groups of the type-$IIA$
string ($F$-theory) compactifications on the corresponding threefold
(fourfold) can be seen in the polyhedron of this $K3$ hypersurface. Of
course, one could also consider the case of an elliptic $CY_4$-fold with
$CY_3$ fibre, where the latter is a Calabi-Yay hypersurface with $K3$
fibre.

We recall~\cite{AENV1} that the lattice structure of the $K3$ projective
vectors obtained by a binary construction exhibits a very interesting
correspondence between the Dynkin diagrams for Cartan-Lie groups in the
$A, D$ series and $E_{6,7,8}$ and particular reflexive weight vectors (see
also Figure~\ref{cdjad}):

\begin{eqnarray}
\vec {k}_1=(1)      & \leftrightarrow& A_r ;\nonumber\\
\vec {k}_2=(1,1)    & \leftrightarrow& D_r ;\nonumber\\
\vec {k}_3=(1,1,1)  & \leftrightarrow& E_6 ;\nonumber\\
\vec {k}_3=(1,1,2)  & \leftrightarrow& E_7 ;\nonumber\\
\vec {k}_3=(1,2,3)  & \leftrightarrow& E_8 .
\label{Dynkin}
\end{eqnarray}
This appearance in Calabi-Yau geometry of the $A, D$ and $E$ series of
Cartan-Lie algebras is connected~\cite{AENV1} with specific one- (two-)
(three-)dimensional structures in an auxiliary complex space.

%\newpage

\subsection{The Arity-Dimension Structure of Universal 
Calabi-Yau Algebra}

Our objective is to construct an universal algebra~\cite{Burris} acting on
the set of reflexive weight vectors in all dimensions, $A_n \equiv
\{$RWV$(n) \}$, and the corresponding set of invariant monomials,
$\{IMs(n) \}$, which is `dual' to $A_n$ in the sense of (\ref{poly3}). We
note that the number of {IM}s is much less the full set of monomials 
${\vec m}_{\alpha}: 1 \leq \alpha \leq \alpha_{max}$ which determine the
Calabi-Yau equation. Through the {IMs} one can determine the highest
vectors of the chains and also the full list of weight vectors in the
corresponding chain. To see this, we start from the unit IM in some
dimension $n$ and then, via a Diophantine expansion, can go on to
determine the conic {IMs}, the cubic {IMs}, the quartic {IMs}, etc..
Similarly, one can continue this process of studying the set of {IMs} via
the Diophantine expansions of conic {IMs}, of cubic {IMs}, etc..
 
\begin{figure}[th!]
   \begin{center}
   \mbox{
   \epsfig{figure=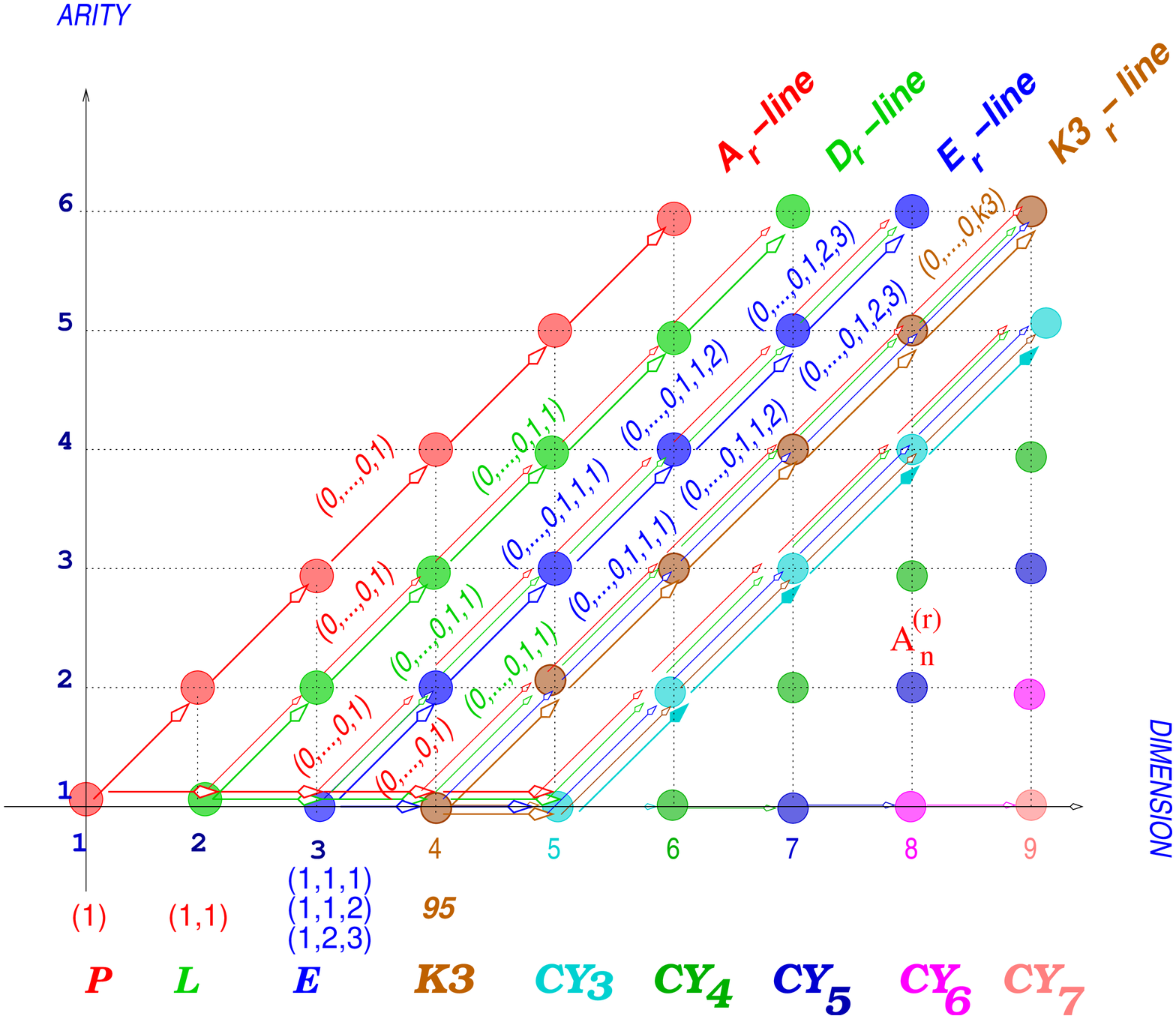,height=16cm,width=18cm}}
   \end{center}
   \caption{\it 
The arity-dimension plane, illustrating the normal expansion of {RWV}s 
by adding zero components to lower-dimensional vectors. For each dimension 
$n$ and arity $r$, it is possible to reconstruct a set of extended 
vectors $\{\AE\}_n^{(r)}$. Along the $A_r$ line, one adds zeroes to the 
trivial vector (1), along the $D_r$ line also to the vector (1,1), etc.. 
Along the $K3_r$ line, one may also add zeroes to any of the 95 $K3$ 
vectors~\cite{AENV1,AENV2}.} 
\label{bas1med} 
\end{figure}

The RWVs and IMs provide independent routes for constructing explicitly
Calabi-Yau spaces of arbitrary dimension (including CICYs). The resulting
UCYA structure of RWVs in different dimensions depends on two integer
parameters, including the `arity' $r$ defined below, as well as the
dimension $n$. An overview in the $(n,r)$ plane is shown in
Fig.~\ref{bas1med}, where the entries $A^{(r)}_n$ label the types
of possible eldest vectors, corresponding to `chains' of related
Calabi-Yau spaces.

%\newpage

As we have just discussed, the algebraic-geometry
realization~\cite{K3,Greene1, Can1} of Coxeter-Dynkin diagrams provides a
general characterization of the possible structures in singular limits of
Calabi-Yau hypersurfaces. Thus, a deeper understanding of the origins of
gauge invariance provides an additional motivation for studying string
vacua via our unification of the complex geometry of $d=1$ elliptic
curves, complex tori, $K3$ manifolds, $CY_3$, $CY_4$, etc. This point is
illustrated in Fig.~\ref{bas1med}, where the points on the the first
three sloping lines, labelled $A_r$ (red), $D_r$ (green) and $E$ (blue),
correspond to those $d$-folds that are characterized by the `maximal'
quotient $A, D, E$ singularities, respectively~\footnote{To be more
precise, the $D$ line includes also $A$-type singularities, and the $E$
line includes also $D$-type and $A$-type singularities.}. As we discuss
later in more detail, this characterization of the types of singularities
is directly connected to the degrees of the associated monomials - linear,
conics, cubics, quartics, etc., that appear along the corresponding
sloping lines.

%\subsection{Operations on Reflexive Weight Vectors}

We now consider in detail important operations in the $(r, n)$ plane.
We seek a map from the sets, $A_i$, of reflexive weight vectors (RWVs) of
lower dimensions $i=1, 2, 3, ..., p < n$ to the set of RWVs of dimension
$n$, $A_n$. For this purpose, we define on the space $A_p = \cup_{i=1}^{p}
A_i$ of RWVs a set $\omega_r: (p=n-r+1)$ of operations combining $r$
elements, where $r = 0,1,2,3,...,n$:
\begin{eqnarray}
\omega_r:\underbrace{A_p \star  A_p \star ...\star A_p}_{r}
\mapsto A_n.\
\end{eqnarray}
We term $r$ the {\it arity} (or rank) of $\omega_r$~\cite{Burris}. An
operation of arity $r=0$ on $A$, called a {\it nullary} (or constant)
operation, selects one element of $A$. An operation of arity $r=1$ on 
$A$
is just a mapping of $A$ into $A$, termed a {\it unary} operation, and
higher-order operations with arity $r=2(3)$ are binary (ternary)
operations, etc.. The precise sense of the symbol ``$\star$" will be
provided later, together with a description of the normal
property of the algebra. It is possible to generate all $n$-dimensional
weight vectors if one already knows all the RWVs of lower dimensions
$1,2,...,n-1$, using the unary, binary, ternary, ... , $r=n$-ary
composition operations $\omega_r$.

The first step in this programme is to define a sequence of unary
operations consisting of extensions of the RWVs $\vec{k}_i \in A_i: i <
n$ to $n$-dimensional RWVs $\in A_n$: $A_i \mapsto \{A_i^{ex}\}_{(n)} $,
obtained simply by adding one or more zero components to this vector in
any location (see Figure \ref{bas1med}):

\begin{eqnarray}
\{A_{n-1}:\vec {k}_{(n-1)}&=&(........)\}   \rightarrow
\{\{A_{n-1}^{ex}\}_n:\vec{k}_n^{(ex)} =(...,0,...)      + all
\,\, permutations \}  \nonumber\\
\{A_{n-2}:\vec {k}_{(n-2)}&=&(........)  \} \rightarrow
\{\{A_{n-2}^{ex}\}_n:\vec{k}_n^{(ex)} =(...,0,...,0,...)+ all
\,\, permutations \}  \nonumber\\
\{A_{n-3}:\vec {k}_{(n-3)} &=&(........)\}  \rightarrow
\{\{A_{n-3}^{ex}\}_n:\vec{k}_n^{(ex)} =(...,0,...,0,...,0,...)
                                     + all \,\,
permutations \}\nonumber\\
........ &=& ...........................................................
\end{eqnarray}
In this first step, the extensions of all reflexive vectors of orders $1,
2, ..., i, ..., (n-1)$ define sets of extended weight-vectors $\AE_n^{i}$
corresponding to operations $\omega_r$ of arity $r=n+1-i=2,3,...n$:

\begin{eqnarray}
 (A_1,A_2,A_3,...,A_{n-1}) &\mapsto&
\cup_{i=1}^{i=n-1} \{A_i^{ex}\}_{(n)}
\equiv {\AE}^{(n-1)}_{(n)}, \nonumber\\
(A_1,A_2,A_3,...,A_{n-2})  &\mapsto&
\cup_{i=1}^{i=n-2} \{A_i^{ex}\}_{(n)}
\equiv {\AE}^{(n-2)}_{(n)}, \nonumber\\
....................&=&.................\nonumber\\
(A_1,A_2,A_3)    \,\,\,\,\, \, \,\,   &\mapsto&
\cup_{i=1}^{i=3}\,\,\,\,\,\{A_i^{ex}\}_{(n)}
\equiv{\AE}^{(3)}_{(n)} \nonumber\\
(A_1,A_2) \qquad \,\,\,\,\,\, &\mapsto&
\cup_{i=1}^{i=2}\,\,\,\, \{A_i^{ex}\}_{(n)}
\equiv{\AE}^{(2)}_{(n)}  \nonumber\\
\,\,\, \qquad(A_1) \qquad \qquad \,\,\, &\mapsto&  
\qquad \,\,\,\,\, \{A_1^{ex}\}_{(n)}
\equiv{\AE}^{(1)}_{(n)}, \nonumber\\
\end{eqnarray}
with the following embeddings
\begin{eqnarray}
{\AE}^{(1)}_{(n)}\subset {\AE}^{(2)}_{(n)}\subset{\AE}^{(3)}_{(n)}\subset
\ldots\subset{\AE}^{(n-2)}_{(n)}\subset {\AE}^{(n-1)}_{(n)}
\end{eqnarray}
defined naturally.

The second step consists of a set of $m$-ary operations
(where the arity $m = 2, 3, ..., r_{max}=n$) to get
the complete list of the RWVs of dimension $n$:
%\newpage
\begin{eqnarray}
arity&=&2:\qquad\,\,\biggl ({\AE}^{(n-1)}_{(n)} \biggl )^2 \qquad  =
{\AE}^{(n-1)}_{(n)} \star {\AE}^{(n-1)}_{(n)} \qquad \qquad \,\,
\Longrightarrow \qquad ( A_{n}^{2})          \nonumber\\
arity&=&3:\qquad\,\,\biggl ({\AE}^{(n-2)}_{{n}} \biggl )^3 \qquad \,=
{\AE}^{(n-2)}_{{n}} \star{\AE}^{(n-2)}_{{n}}
\star{\AE}^{(n-2)}_{{n}} \,\,
\Longrightarrow \qquad ( A_{n}^{3})          \nonumber\\
arity &=&.................................................
............................................ \nonumber\\
arity&=&(n-2):\,\biggl ({\AE}^{(3)}_{(n)} \biggl )^{(n-2)} \,\,\,\, =
{\AE}^{(3)}_{(n)} \star  .... \star{\AE}^{(3)}_{(n)} \qquad \,\,
\qquad \Longrightarrow \qquad ( A_{n}^{n-2})          \nonumber\\
\nonumber\\
arity&=&(n-1):\,\biggl ({\AE}^{(2)}_{(n)} \biggl )^{(n-1)} \,\,\, =
{\AE}^{(2)}_{(n)} \star  .... \star{\AE}^{(2)}_{(n)} \qquad \,\,
\qquad \Longrightarrow \qquad ( A_{n}^{n-1})          \nonumber\\
\nonumber\\
arity&=&n:\,\qquad \,\,\biggl ({\AE}^{(1)}_{(n)} \biggl )^n \qquad \,=
{\AE}^{(1)}_{(n)} \star  .... \star{\AE}^{(1)}_{(n)} \qquad \,\,
\qquad \Longrightarrow \qquad( A_{n}^{n})          \nonumber\\
\nonumber\\
\end{eqnarray}
where  $ A_{n}^{r}$ is the set of RWVs with arity $r$, and 
$A_{n}^{1}\equiv A_{n}$.

The specification of the algebraic formalism for the expansion technique
is then completed by defining the symbol $\star$ appearing in the above
$r$-ary composition operations on the sets $\AE_n^{(n-r+1)}$:

\begin{eqnarray}
\omega_r:\underbrace{{\AE}^{(n-r+1)}_{n} \star
 {\AE}^{(n-r+1)}_{n} \star ...
\star {\AE}^{(n-r+1)}_{n}}_{r}
\Longrightarrow A_n^{r}.
\end{eqnarray}
The symbol $\star$ defines the `intersection'
of $r$ different extended weight vectors, in the sense that they share a
common set of invariant monomials $\vec{\mu}_{\alpha}$
satisfying simultaneously the $r$ conditions:
\begin{equation}
\vec{\mu}_{\alpha} \cdot {{\vec k}_n}^{i(ex)}=d_i,
\label{defintersection}
\end{equation}
where $i=1,2,...,r$.

We term the $\omega_r$ operation on the set $\AE_n^{(n-r+1)}$ {\it normal}
if the intersections ${\vec k}^{(ex)}_{n}\in {\AE}^{(n-r+1)}_{(n)}$ of
some $r$ extended weight vectors, where $2\leq r \leq n$, give a {\it
normal object} in the sense that:

\begin{eqnarray}
{\rm if} \qquad &&\bigcap_{i=1}^{i=r} \biggl( {\vec 
k}^{(ex)}_{n}\biggl)^{(i)}
\qquad  = \qquad \{\Box^{ref}\}_{n-r} \qquad \qquad
\biggl({\vec k}^{(ex)}_{n}\biggl )^{(i)}\in {\AE}^{(n-r+1)}_{(n)}, \nonumber\\
{\rm then} \qquad &&\bigcup_{i=1}^{i=r} \biggl( {\vec 
k}^{(ex)}_{n}\biggl)^{(i)}
\qquad = \qquad  \vec{k}_{n}, \qquad \qquad \qquad \,\, \,\,
\{\vec{k}_{n}\}_R\,\,\,\, \in  A^r_n.\nonumber\\
\end{eqnarray}
The term {\it normal} is appropriate for the following reasons. One is
that the extensions of all the RWVs in lower dimensions $1 \leq i \leq
(n-1)$ to the next dimension $n$ give the complete list of RWVs of
dimension $n$, just as the usual normal Galois extension of a field $K$
includes all the roots of polynomial equations: if a polynomial over the
field $K$ has one root in the normal extension $P$, all the roots must be
in the field $P$.  Another reason why the term {\it normal} is appropriate
is connected with our conjecture that this extension is also complete
under mirror duality, i.e., all the mirror $CY_d$ spaces with $d = n-2$
can also expressed in terms of RWVs with lower dimensions $1\leq i \leq n$
or their extended vectors. This conjecture will be illustrated later in
the case of $K3$ mirror partners.
  
We use below two examples of such a {\it normal object} $
\{\Box^{ref}\}_{n-r}$, one of which can be identified geometrically as a
reflexive Batyrev polyhedron, and the other algebraically as a set of
invariant monomials (IMs). An analysis in which the symbol $
\{\Box^{ref}\}_{n-r}$ is identified as a Batyrev polyhedron~\cite{AENV1}
is illustrated in the Fig.~\ref{basmon1}. It served as our first way of
identifying the 4242 chains of $CY_3$ spaces with arity 2, and we use it
here to derive some new results for $CY_4$ spaces. A second way of
deriving these results, where the symbol $ \{\Box^{ref}\}_{n-r}$ is
identified with sets of IMs, is
discussed in Section 3. There we use the IM approach to rederive previous
results for $K3$ and $CY_3$ spaces, to obtain further results for $CY_4$,
and to derive some results for arbitrary dimension $n$.

In the IM case, the symbol $\bigcup..$ means a simple linear algebraic sum
of the extended RWVs, ${\vec k}^{(ex)}_{n}\in \{\AE\}^{(n-r+1)}_{(n)}$,
which give the following set:

\begin{eqnarray}
\bigcup_{i=1}^{i=r} \biggl( {\vec k}^{(ex)}_{n}\biggl)^{(i)}=
m_1\biggl( {\vec k}^{(ex)}_{n}\biggl)^{(1)}+\ldots
+m_r\biggl( {\vec k}^{(ex)}_{n}\biggl)^{(r)} =
\biggl \{ \vec{k}_n \biggl \}_R\subset A^r_n,
\end{eqnarray}
where $m_1\geq 1,\ldots, m_r\geq 1$ are positive integers
that determine a whole chain, $ \{ \vec{k}_n\}_R$, of RWVs with 
some common properties. In general this normal $\omega_r$ operation gives  
the structure of the chain and directly determines the eldest vector, 
corresponding to the minimal values, $m_1=m_2=...=m_r=1.$ To determine the
complete listing of the chain, we must specify all these coefficients $m_i$.
As discussed later, this may be done using mirror duality or a Diophantine 
expansion.

An operation $\omega_r$ that yields such a normal object is a special
$r$-ary operation that yields specific algebraic relations between the
RWVs in different dimensions, defining a `dual' relation between the
$\bigcup$ and $\bigcap$ operations. In order to specify correctly such an
$\omega_r$ operation, one should study more carefully the sets
$\{\AE\}_n^{(n-r+1)}$.

{\it First}, one should use in $\omega_r$ only $r$ independent extended
RWVs. Independence here means that one should take only those $r$ extended
vectors whose intersections give an object of dimension $n-r$.  This may
conveniently be done by constructing from the $r$ extended RWVs the $r
\times n$ rectangular matrix $A_{r \times n}$ and to check that the
determinant of the $r \times r$ matrix $A_{r \times n} \cdot A_{r \times
n}^T$ does not vanish.

{\it Secondly}, there are requirements concerning the self-consistency of
the algebra with respect to the algebraic operations of different arities.
This is familar in mathematics from the theories of rings and fields, in
which two binary operations exist, and there are additional conditions on
the simultaneous actions of these two operations, such as the law of
distributivity. We term the following analogous property of this algebra,
which plays an important role for arities $r > 2$ also as
`distributivity':  If a RWV can be obtained by two or more operations with
different arities, e.g., with arities 2 and 3, this vector will be
considered a `good' weight vector only if it is determined correctly by
all arity operations.

We must impose this requirement because even weight vectors which can be
got from `good' arity $3, 4, \dots, r$ operations may actually not be
reflexive, because they cannot be obtained legitimately by constructions
of lower arity $2, 3, \dots , r-1$. We call them {\it false} vectors. 
So, if a weight vector is `good' from the point of view of
arity {2}, it cannot be `false' from the point of view of the higher
arities, $\omega_r \geq 3$. Also, if a weight vector is `good' from the
viewpoint of arity {3}, and has no arity {2} structure, it cannot be
`false' with respect to the arities $\omega_r \geq 4$ etc. In other words,
there is an `ordering' of the arity operations, according to which lower
arities `control' higher arities. We exhibit later some examples of
`false' vectors in the $CY_3$ case.

{\it Thirdly}, we must clarify the notions of reducible and irreducible
chains and eldest vectors. Sometimes, one finds by operations with the
same arity $r$ two or more chains/eldest vectors with the same
intersection structure and, moreover, one chain can be expressed in terms
of another chain, i.e.:

\begin{eqnarray} 
a k_1^{ex} + b k_2^{ex}+ c k_3^{ex}= \tilde
a \tilde k_1^{ex}+ \tilde b\tilde k_2^{ex}+\tilde c \tilde k_3^{ex},
\end{eqnarray} 
where all the coefficients $a,b,c$ and $\tilde a$, $ \tilde b $, $\tilde
c$ are positive integers. There exists a `minimal' basis in terms of which
all reducible chains can be constructed as linear combinations with
positive integer coefficients. This issue is transparent in the IM method,
which is powerful enough to provide more solutions than we need to
determine the chains. For example, there is an irreducible chain of $CY_4$
spaces of arity $\omega_r=4$, corresponding to the third $E_r$ line, that
can be constructed from the following extended vectors,

\begin{eqnarray}
\vec{k}^{1(ex)}&=&(0,0,1,0,1,0)=e_1,   \nonumber\\
\vec{k}^{2(ex)}&=&(1,0,0,2,3,0)=e_2,   \nonumber\\
\vec{k}^{3(ex)}&=&(0,1,0,2,3,0)=e_3,   \nonumber\\
\vec{k}^{4(ex)}&=&(0,0,1,0,0,1)=e_4,   \nonumber\\
\end{eqnarray}
Other sets of weight vectors generate the same chain, one example being

\begin{eqnarray}
\vec{k'}^{1(ex)}&=&(1,0,0,2,3,0)=e_1'=e_2,        \nonumber\\
\vec{k'}^{2(ex)}&=&(0,1,0,2,3,0)=e_2'=e_3,        \nonumber\\
\vec{k'}^{3(ex)}&=&(0,0,3,0,1,2)=e_3'=e_1+2e_4,   \nonumber\\
\vec{k'}^{4(ex)}&=&(0,0,3,0,2,1)=e_4'=2e_1+e_4,   \nonumber\\
\end{eqnarray}
One can easily check that the intersections are the same for these two
chains, and that $\sum m_i' \vec{k'}^{i(ex)} \subset \sum m_j
\vec{k}^{j(ex)}$, where $i, j = 1, ..., 4$.

The choice $m_1=1, m_2=1$ is always possible for the binary case, and
determines what we term the {\it eldest} RWV in the chain. With the above
clarification, in general there is also a unique eldest RWV for good
operations with arity $r > 2$. There also exists a set of coefficients
$m_i$ which determines the {\it youngest} RWV~\cite{AENV1} in the chain.  
The set of possible values for the coefficients $m_i$ determining a chain
is easiest to find for the double chains with arity $r=2$. The maximal
values for $m_1$ and $m_2$ are determined by the dimensions of the
extended vectors $\biggl ( {\vec k}^{ex}_n\biggl)^{(2)}$ and $\biggl
({\vec k}^{ex}_n\biggl)^{(1)}$, respectively. There are analogous rules
for triple and higher-order chains~\cite{AENV1}. The RWVs inside a chain
can also be enumerated using the technique of decomposing the IM that
determines the chain~\cite{AENV2}.

Thus, in the framework of the UCYA, it is enough to know the
arity-dimension structure for the eldest RWVs, as seen in Fig.~\ref{basmon1},
just as the eldest weights determine representations of Cartan-Lie
algebras. The tremendous numbers of RWVs and/or reflexive polyhedra may
then be found in any specific cases of interest.

%\subsection{Intersections and Projections}

We have seen above how the composition operations $\omega_r$ with $2 \leq
r \leq n$ give a universal, normal and self-consistent map of the set of
RWVs of lower dimensions $1,2,\dots ,(n-r+1)$ into to the set of
$n$-dimensional RWVs:
\begin{eqnarray}
\omega_r: \qquad \vec{k}_{1},\,\vec{k}_{2},\, \ldots,\,
\vec{k}_{(n-r+1)} \qquad
\Longrightarrow  \qquad \vec{k}_n,
\qquad \qquad 2 \leq r \leq n.
\end{eqnarray}
Thus, a given $RWV(n): \vec{k}_n \in A_n$ can be obtained in $\leq (n-1)$
different ways, depending on the arity structure. More exactly,
constructions with arity $r=2$ correspond to the structure of
$(n-1)$-dimensional sections of the set of RWVs or relflexive polyhedra,
and constructions with arity $r =n$ correspond to the trivial point
sections of the RWVs or reflexive polyhedra. The union of all such subsets
should give the full list of all $n$-dimensional RWVs with all possible
internal structures, as seen in Fig.~\ref{bas1med}:

\begin{eqnarray}
\bigcup_{r=2}^{r=n} ( A_{n}^{r})=(A_{n}^{(1)})\equiv (A_{n}).
\end{eqnarray}
The extension mechanism provides the $\omega_r$ composition operations of
this algebra for all possible arities $ 2 \leq r \leq n$. This
extension is normal, because we consider $d$-dimensional objects
which have definite universal structure properties common for all
dimensions. The condition for such a normal extension applies if there are
composition operations that allow one to reconstruct objects with every
possible internal structure in any dimension $d$, if one has already knows
their structures in lower dimensions $(d-1), (d-2), \dots, 1$. Indeed,
information in only the lowest dimension is sufficient. Conversely, it is
possible to determine algebraically the properties of objects in lower
dimensions, starting from objects in higher dimensions.

The restriction that the geometrical structure of such objects in any
given dimension $d$ remembers the structures of objects in lower
dimensions, $(d-1), (d-2), \dots, 1$ is closely connected with the
intersection-projection mirror duality: $ \sigma \leftrightarrow \pi$.
Concretely, it means that for any $p: d > p > 0$, there should exist at
least one $(d-p)$ intersection or $(d-p)$ projection in which one can see
a $(d-p)$-dimensional object with similar properties. This property is
reflected in the following `normal' expansion series and associated
'factor' set:

\begin{eqnarray}
\mho_0 \lhd \mho_1 \lhd \mho_2 \lhd  \ldots \mho_n\lhd \ldots
\end{eqnarray}
\begin{eqnarray}
\hat \sigma \,\{\mho_1\}=\mho_0
\qquad &{\it or}&  \qquad \hat \pi \,\{\mho_1\}=\mho_0
\nonumber\\
\hat \sigma \,\{\mho_2\}=\mho_1
\qquad \,&{\it or}&\, \qquad \hat \pi \,\{\mho_2\}=\mho_1
\nonumber\\
\hat \sigma \,\{\mho_3\}=\mho_2
\qquad \,&{\it or}&\, \qquad \hat \pi \,\{\mho_3\}=\mho_2
\nonumber\\
     \ldots &&\ldots \nonumber\\
\hat \sigma \,\{\mho_d\}=\mho_{d-1}
\qquad \,&{\it or}&\, \qquad \hat \pi \,\{\mho_d\}=\mho_{d-1}
\nonumber\\
     \ldots &&\ldots \nonumber\\
\end{eqnarray}
where we denote by $\hat \sigma \,\{...\}$ the $(d-1)$-dimensional 
intersection and by $\hat \pi \,\{...\}$ the $(d-1)$-dimensional 
projection of the $d$-dimensional hypersurface from the set $\mho_d$, 
respectively.

This series is reminiscent of the sequence of normal field extensions in
Galois theory, reinforcing our interpretation of the UCYA as a `normal'
algebra. However, the fact that one should use two dual-conjugate
operations, intersection $\hat \sigma$ and projection $\hat \pi$, to
complete the procedure of extension is an essential new feature beyond
Galois theory. For the construction of the Universal Calabi-Yau 
Algebra via the $r$-ary composition operations $\omega_r$ of
the RWVs, it is not enough to use only the intersection operators, but one
should also use the intersection-projection duality.

Combined with the arity structure, this duality gives very naturally the
fibre structures of $CY_d$ hypersurfaces. To see these, one needs to use a
link between the two polyhedra in a mirror pair. In the holomorphic
quotient formalism~\cite {Cox}, one can easily see that the intersection
of the mirror partner determines a corresponding fibre structure of
$CY_d$. For example, one can easily see the fibre structures in $K3$
spaces, which are determined by the interplay of the integer
point-monomials in the intersection, ${ \sigma( \Delta)}$, and in the
projection, ${ \pi({\Delta}^*)}$. These fibre structures are defined by
the following dual relations between point-monomials in the intersection
of the polyhedron and the projection in mirror polyhedron, respectively: 
$\sigma(\Delta)=4  \rightarrow \pi(\Delta^*)=10$, 
$\sigma(\Delta)=5  \rightarrow \pi(\Delta^*)=9$, 
$\sigma(\Delta)=6  \rightarrow \pi(\Delta^*)=8$, 
$\sigma(\Delta)=7  \rightarrow \pi(\Delta^*)=7$,
$\sigma(\Delta)=8  \rightarrow \pi(\Delta^*)=6$, 
$\sigma(\Delta)=9  \rightarrow \pi(\Delta^*)=5$, 
$\sigma(\Delta)=10 \rightarrow \pi(\Delta^*)=4$  
producing planar fibres, and
$\sigma(\Delta)=3  \rightarrow \pi(\Delta^*)=3$ producing a line-segment 
fibre~\cite{AENV1}. 

Some of the $K3, CY_3,...$ chains satisfy the following condition:

\begin{eqnarray}
\sigma(\Delta ^n)   \,&=&\, \pi (\Delta^n)   \,=\,{\Delta}^{n-1} ,\nonumber\\
\pi({\Delta^n}^*)\,&=&\,  \sigma ({\Delta^n}^*)\,=\,{{\Delta}^{n-1}}^*,
\end{eqnarray}
which means that the number of integer points in the intersection of the
polyhedron (mirror polyhedron) forming the reflexive polyhedron of lower
dimension is equal to the number of projective lines crossing these
integer points of the polyhedron (mirror). The projections of these lines
on a plane in the polyhedron and a plane in its mirror polyhedron
reproduce, of course, the reflexive polyhedra of lower dimension. Only for
self-dual polyhedra can one have
\begin{equation}
 \sigma(\Delta ^n)  \,=\, \pi (\Delta^n)   \,=\,
\pi({\Delta^n}^*)\,=\,  \sigma ({\Delta^n}^*)\,=\,
{\Delta}^{n-1}\,=\,{{\Delta}^{n-1}}^*,
\end{equation}
namely the most symmetrical form of these relations.

\subsection{Mirror symmetry and CICY in the Universal Calabi-Yau Algebra}

We have already mentioned briefly the significance of mirror duality in
the Universal Calabi-Yau Algebra (UCYA). In this subsection we stress the
link between mirror symmetry and CICY spaces in the UCYA.

The normal universal composition operations defined earlier give an
algebraic description of the full underlying set of reflexive weight
vectors $\bigcup_r A^r_n$, where each set $A^r_n$ contains the full list
of reflexive weight vectors in each dimension $n$ and with all internal
arity structures $2 \leq r \leq n$. The hypersurfaces corresponding to the
set $A_n$, with fixed $n = d + 2$ can be called the {\it single} or {\it
zero-level} $\{CY_d\}^{0}$ (where in general the level $l = r - 1$),
which are each described by a single polynomial equation:

\begin{equation}
\{CY_d\}^{(0)} \qquad =  \qquad A_{d+2}
\end{equation}
The set $\{CY_d\}^{0}$ with fixed $d$ gives only 
a small part of the full set of $CY_d$ hypersurfaces:

\begin{eqnarray}
\{CY_d\}^{0} \subset \sum_l \{CY_d\}^{l}
\end{eqnarray}
where 
\begin{eqnarray}
\{CY_d\}^{1} &=& A^2_{d+3} , \nonumber\\            
\{CY_d\}^{2} &=& A^3_{d+4} , \nonumber\\            
\{CY_d\}^{3} &=& A^4_{d+5} , \nonumber\\            
............. &=&.............................
\end{eqnarray}
where positive integer number $l = r - 1$ characterize the level (arity)
of the set. We illustrate these relations in the arity-dimension $(n,
r)$ plot Fig.~\ref{basmon1}, where the full set of $CY_3$ is seen to be
obtained from the collection of all integer points on the line $n = r + d
+ 1: d = 3$, i.e., $\{CY_3\} = A_5 + A^2_6 + A^3_7 + A^4_8 + ...$. The
integers on this plot, $A^r_n$, describe not only the number of the
arity-$r$ chains of $CY_{d = n - 2}$ spaces, but due to the UCYA also give
the $CICY$s of fixed dimension on the different levels.

The hypersurfaces $\{CY_d\}^{l}$ on the following levels $l = 1 , 2 , 3,
...$ are described by two, three, four, ... polynomial equations in
$CP^{d+2}$, $CP^{d+3}$, $CP^{d+4}$, ... spaces, respectively, i.e., for
the equations of a Calabi-Yau hypersurface $\{CY_d\}^l$ on the level $l =
r - 1$, one should consider the $r$ polynomial equations in $CP^{d + r }$
complex projective space:

\begin{eqnarray}
 A^2_{d+3}\rightarrow \{CY_d\}^{(1)}\,&=&
\{\vec{x}\in CP^{d+2}\,|\, {\wp}_{i_1}(\vec x)= {\wp}_{i_2}(\vec x)=0 \} 
\nonumber\\
A^3_{d+4}\rightarrow \{CY_d\}^{(2)}\,&=&
\{\vec{x}\in CP^{d+3}\,|\, {\wp}_{i_1}(\vec x)= {\wp}_{i_2}(\vec x)=
{\wp}_{i_3}(\vec x)=0 \} 
\nonumber\\
A^3_{d+5}\rightarrow \{CY_d\}^{(3)}\,&=&
\{\vec{x}\in CP^{d+4}\,|\, {\wp}_{i_1}(\vec x)= {\wp}_{i_2}(\vec x)=
{\wp}_{i_3}(\vec x)= {\wp}_{i_4}(\vec x)=0 \} 
\nonumber\\
\ldots \qquad &=& \qquad  \ldots,
\nonumber\\
\end{eqnarray} 
where $\vec x \equiv (x_1, \ldots, x_{d+r+1}) \in CP^{d+r}$. 
The action of the polynomial equations  is 
equivalent to the simultaneous action
of the normal sets of extended RWVs, i.e.,

\begin{eqnarray}
 \{{\wp}_{i_1}(\vec x)= {\wp}_{i_2}(\vec x)=0\}
&\Leftrightarrow &
\biggl ({\vec k^{(ex)(i_1)}}_{d+3}
\cap {\vec k^{(ex)(i_2)}}_{d+3}
 \biggl)\nonumber\\
\{ {\wp}_{i_1}(\vec x)= {\wp}_{i_2}(\vec x)=
{\wp}_{i_3}(\vec x)=0 \} &\Leftrightarrow &
\biggl ({\vec k^{(ex)(i_1)}}_{d+4}
\cap {\vec k^{(ex)(i_2)}}_{d+4} \cap 
{\vec k^{(ex)(i_3)}}_{d+4}\biggl)
\nonumber\\
\{{\wp}_{i_1}(\vec x)= \ldots
= {\wp}_{i_4}(\vec x)=0 \}  & \Leftrightarrow &  
\biggl ({\vec k^{(ex)(i_1)}}_{d+5}\cap 
\ldots \cap {\vec k^{(ex)(i_4)}}_{d+5}\biggl)
\nonumber\\
\ldots \qquad &\Leftrightarrow& \qquad  \ldots
\nonumber\\
\end{eqnarray} 
Thus, within the UCYA the description of the CICY is very simple.
These manifolds correspond to the `normal' submanifolds 
of the manifolds of the higher dimensions, and can very easily be
described by the intersections of extended RWVs. This can be seen 
from the following identity:

\begin{eqnarray}
\biggl ({\vec k^{(ex)(i_1)}}_{d+r+1} \cap 
\ldots  \cap {\vec k^{(ex)(i_s)}}_{d+r+1}\cap
\ldots {\vec k^{(ex)(i_r)}}_{d+r+1} \biggl) \equiv
\biggl ({\vec k^{(ex)(i_1)}}_{d+r+1}\cap 
\ldots \cap {\vec k}_{d+r+1 }\cap \ldots {\vec k^{(ex)(i_r)}}_{d+r+1} \biggl)
\nonumber\\ 
\end{eqnarray} 
Thus, the set of hypersurfaces with fixed $d$ corresponding to all the
non-zero level numbers $l = r - 1 = 1, 2, ...$ contains the full list
of CICYs in a fixed number $d$ of dimensions:

\begin{equation}
\{CICY\}_d=\sum_{l\geq 1} \{CY\}_d^{l} = 
\sum_{r \geq 2, n} \{A^r_{d+r+1}\}, \qquad where \qquad n=d+r+1.
\end{equation}

Mirror symmetry can be expressed in the UCYA as a projection-intersection
symmetry between hypersurfaces and their mirror partners, respectively.  
For example, a $d$-dimensional hypersurface $X_d$ should have in the
$(d-r)$-dimensional intersection (projection) a subspace of similar
structure with lower dimension: $X_{d - r}$ and, consequently, the mirror
hypersurface $X^*_d$ should have in the corresponding projection
(intersection) the mirror subspace $X_{d - r}^*$.
  
The main conjecture of the UCYA is that it is closed under the mirror
symmetry transformation, after taking into account all these $r$-ary
composition operations. The set of lowest-level $\{CY_d\}^{(0)}$ spaces is
not closed under mirror duality: only the full list of $\{CY_d\}^{(0)}$
and $CICY_d^{l}$ spaces, corresponding to the integer points on the line
$n = d + r + 1$ of the $(r-n)$ plot is closed under this duality.

\section{The Mechanisms of Normal Expansion}

To understand the origin of the normal extension,
i.e., to find in the set $\AE_n^{(n-r+1)}$ `good' $r$-extended
RWVs that define a normal intersection,
one must consider generally the set of $r$ RWVs 
and analyze their intersections. To this end, we now formulate more 
precisely the correspondence defining normal $\omega_r$ composition
operations on RWVs, considering first
the case with arity $r = 2$.
Let two fixed RWVs of dimension $n$, 
${\vec k_{n}}^a=(k^a_1|k^a_2,...,k^a_{n})[d_a] \in A_{n}$
and 
${\vec k_{n}}^b=(k^b_1|k^b_2,...,k^b_{n})[d_b]\in A_{n}$ 
have the following sets of monomials, 
\begin{eqnarray}
{\vec \mu}^i  &=&(\mu_1|\mu_2,...,\mu_{n})^i , 
\nonumber\\  
{\vec \nu}^j  &=&(\nu_1|\nu_2,...,\nu_{n})^j , 
\end{eqnarray}
such that ${\vec \mu}^i \cdot {\vec k_{n}}^a = d_a$ and 
${\vec \nu}^j \cdot {\vec k_{n}}^b = d_b$. 

We have separated by the symbol $\{|\}$ the first 
components from the other components of the RWVs ${\vec k_n}^a$ 
and ${\vec k_n}^b$, 
and similarly in the corresponding monomials.
For each of the vectors ${\vec k_{n}}^a$ and ${\vec k_{n}}^b$,
we consider subsets of the monomials,
\begin{eqnarray}
{\vec \mu}^{i'}  =(\mu_1|\mu_2,...,\mu_{n})^{i'} \subset S_{n}^a, 
\nonumber\\
{\vec \nu}^{j'}  =(\nu_1|\nu_2,...,\nu_{n})^{j'} \subset S_{n}^b, 
\end{eqnarray}
with the following two properties: 
\begin{eqnarray}
\,\,\,\mu_2=\nu_2,\,\,\,\,\,\,\mu_3=\nu_3,\,\,\, 
\ldots,\,\,\mu_{n} =\nu_{n}
\end{eqnarray}
or 
\begin{eqnarray}
S_{n}^a|_{x_1=0}=S_{n}^b|_{x_1=0}=S_{n-1};
\end{eqnarray}
and, moreover, the set of vectors 
\begin{eqnarray}
S_{n-1}:({\vec \mu}_{red})^i=(\mu_2, \ldots , \mu_{n})^i
\end{eqnarray}
corresponds in the integer lattice  to a `reduced' reflexive polyhedron
of dimension $(n-1)$.

%The definition of the reduced $(n-1)$-dimenional reflexive polyhedron is
%given later. 
For now, we note that the weight vectors ${\vec k_{n}}^a$ 
and ${\vec k_{n}}^b$ also determine in the integer lattice 
the reduced $(n-1)$-dimenional
reflexive polyhedron. Thus, when we consider the extensions of these two
RWVs:

\begin{eqnarray}
{\vec k_{n+1}^{(ex)a}} &=&(0,k_1|k_2,\ldots, k_{n})^a, \nonumber\\
{\vec k_{n+1}^{(ex)b}} &=&(k_1,0|k_2,\ldots, k_{n})^b,
\end{eqnarray} 
and, correspondingly, their common monomials which can be obtained
from the extensions of the set $S_{n-1}$, i.e.,
\begin{eqnarray}
(\vec \mu^{ex})^i=S_{n+1}^{ex}  
=(\nu_1,\mu_1|\mu_2=\nu_2,\ldots,\mu_{n}=\nu_{n})^i  
\end{eqnarray} 
which satisfy the following two conditions:
\begin{eqnarray}
{\vec k_{n+1}^{(ex)a}} \cdot (\vec \mu^{ex})^i &=&[d_a] \nonumber\\
{\vec k_{n+1}^{(ex)b}} \cdot (\vec \mu^{ex})^i &=&[d_b],
\end{eqnarray} 
again the set of extended monomials $S_{n+1}^{ex}$ produces in the integer
$n+1$-dimensional lattice the reduced $(n-1)$-dimensional reflexive
polyhedron.

One can now see that the weight vector
$\vec{k}_{n+1}[d]= {\vec{k}_{n+1}^{(ex)a}} +{\vec{k}_{n+1}^{(ex)b}}$
satisfies the  condition:
\begin{eqnarray}
\vec{k}_{n+1} \cdot (\vec \mu^{ex})^i =[d]=[d_a]  + [d_b],
\end{eqnarray} 
what means that the $n$-dimensional polyhedron corresponding to this new 
weight vector contains itself the reduced $(n - 1)$-dimensional
reflexive polyhedron. We illustrate how this works for arity {2},
the expansion giving the condition that the central point-monomial  
is inside new polyhedron.
For this one can consider in the reduced reflexive polyhedron 
an expanded point 
$P_0= (\nu_1^{0},\mu_1^{0}|\mu_2,...,\mu_{n}) $ and check that 
its right and left neighbour-points,
$P_{+,(-)}=(\nu_1^{0} +(-){\xi^+}  ,\mu_1^{0} -(+)\xi^- |\mu_2,...,\mu_{n}) $,
satisfy the condition $\vec k_n \cdot P_{+,(-)}=d$.
One can easily check this, for example,  when the first expanded components
of the $n$-dimensional refllexive weight vectors,
${\vec k_n}^a$ and ${\vec k_n}^b$, are both units, i.e.,
$k_1^a=k_2^b=1$. It is easy to see that each $n$-dimensional reflexive 
weight vector has 
at least $n$ normal extensions into $(n + 1)$-dimensional 
RWVs~\footnote{Also, one can check that if some extension of the eldest 
vector is good, then all vectors in the chain will also have analgous good
extensions.}.

The generalization to other cases: RWVs in different dimensions, `good'
triples, etc. ... can be done in a similar way. Thus, we can reformulate
the conjecture of a normal arity extension, for any arity, in the language
of RWVs. As an immediate consequence, we can get chains in higher
dimensions. For example, each $RFW(n-1)$ can give us $(n-1)$ chains of
dimension $n$, and it is very easy to understand the origin of
$n$-dimensional chains in terms of $(n-1)$-dimensional chains. This gives
us a very convenient rule for getting arity-2 $n$-dimensional chains, by
first taking into account the eldest vectors in $(n-1), (n-2), ...$
dimensions.

\begin{figure}[th!]
   \begin{center}
   \mbox{
   \epsfig{figure=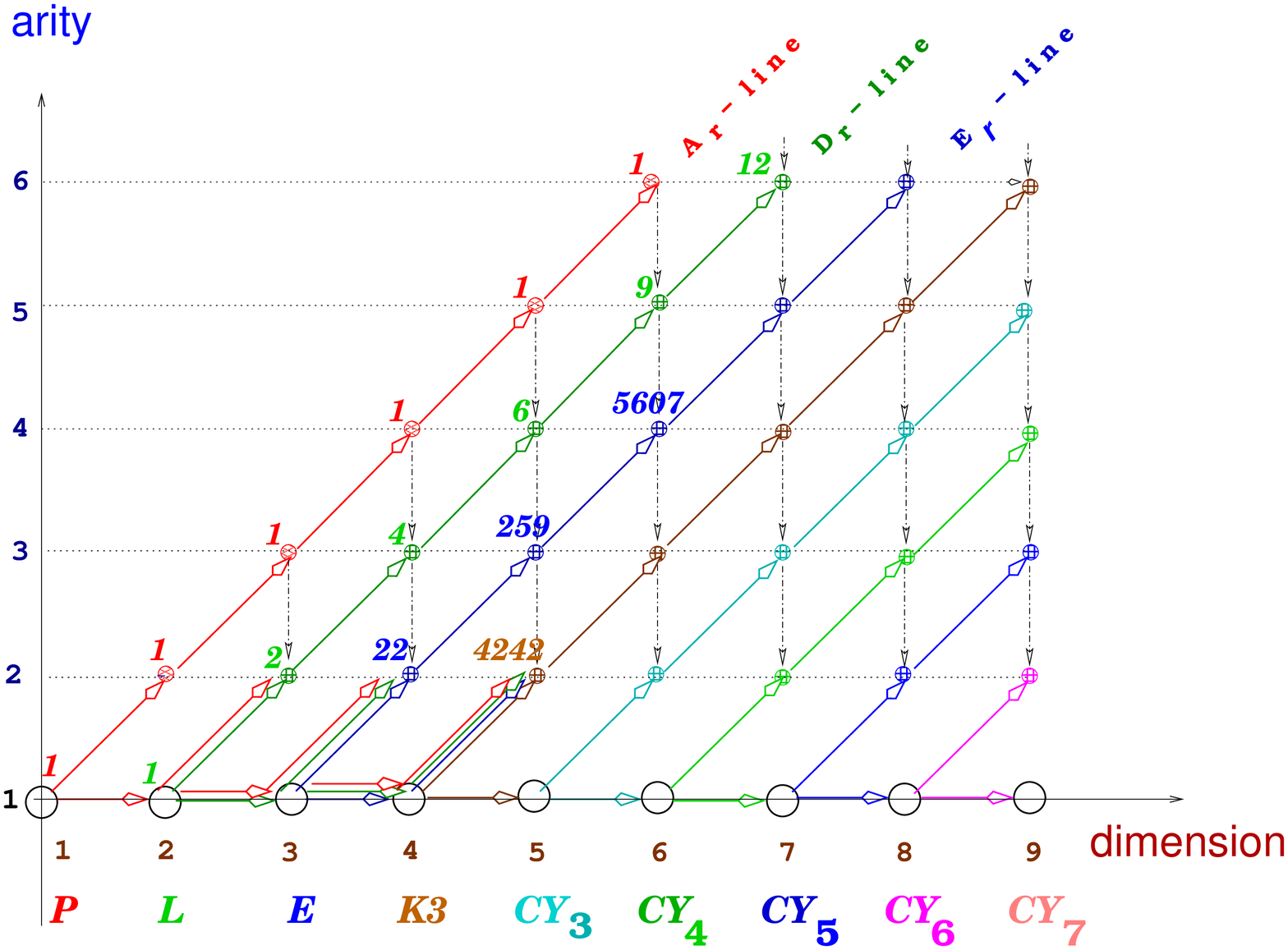,height=16cm,width=18cm}}
   \end{center}
   \caption{\it  
The arity-dimension plot, showing the numbers of eldest 
vectors/chains obtained by normal extensions of RWVs, including previous 
results for $CY_3$ and lower-dimensional spaces, and new results for 
$CY_4$ and $CY_5$ spaces.} 
\label{basmon1}
\end{figure}

We are now ready to demonstrate more explicitly how our construction acts
in the arity-dimension plots, Fig.~\ref{bas1med} and Fig.~\ref{basmon1}.
In so doing, we also show how this construction makes explicit the
structural systematics of the singularities of $CY_d$-folds. Our method in
this section is based on normal expansions of RWVs, which are related to
Batyrev's reflexive polyhedra, but we shall also use some results obtained
by the Diophantine decomposition of IMs.

%\newpage

\subsection{Analysis of $A_r$ and $D_r$ lines }

Starting from dimension ${n=1}$, where we have
$A_1^{1}=\vec{k}_1^{eld}=(1)$, the extension to $n = 2$ is immediate, 
and it is trivial to see that there is just chain of arity 2:
\begin{eqnarray}
\omega_2:{\AE}^{(2)}_{2} \star
 {\AE}^{(2)}_{2}  \mapsto A_2^{2},
\end{eqnarray}
where the set ${\AE}^{(2)}_{2}$ contains only two extended vectors:
$\vec{k}_1^{1(ex)}=(0,1)$ and $\vec{k}_1^{2(ex)}=(1,0)$.
There is one `good' intersection:
\begin{eqnarray}
\vec{k}_1^{1(ex)}\cap \vec{k}_1^{2(ex)}=(1)',
\end{eqnarray}
where the symbol $1'$ signifies that this intersection has only point,
corresponding to the unit monomial $E_1=(1,1)$, for which the algebra gives
just one eldest RWV:
\begin{eqnarray}
\vec{k}_1^{1(ex)}\cup \vec{k}_1^{2(ex)}=\vec{k}_2=(1,1).
\end{eqnarray}
Thus $A_2^{1}=A_2^{2}=\vec{k}_2=(1,1)$. This vector corresponds to a trivial
reflexive polyhedron that just consists of a line segment, and to the
three monomials:
\begin{eqnarray}
C_2=(0,2), \qquad E_2=(1,1),\qquad C_2=(2,0).
\end{eqnarray}
Following the diagonal red $A_r$ line to higher dimensions $n \geq 3$, in
each case we also get only one chain of arity $\omega_r=n$, with a single
eldest vector: $\vec{k}_n^{eld}=(1,...,1)$. Thus, every chain with the
maximal arity $n$ is determined by just the one unit RWV $\vec{k}_1=(1)$,
making the following set of extensions to higher-dimensional vectors:
\begin{eqnarray}
\vec{k}_n^{1(ex)}&=&(1,0,0,\ldots,0,0,0), \nonumber\\
\vec{k}_n^{2(ex)}&=&(0,1,0,\ldots,0,0,0), \nonumber\\
\ldots &=& \ldots \ldots \ldots \nonumber\\
\vec{k}_n^{(n-1)(ex)}&=&(0,0,0\ldots,0,1,0), \nonumber\\
\vec{k}_n^{n(ex)} &=&(0,0,0\ldots,0,0,1), \nonumber\\
\end{eqnarray}
This set of extended vectors has only one `good' intersection:
\begin{eqnarray}
\cup_{i=1}^{i=n}\vec{k}_n^{i(ex)}= 1,
\end{eqnarray}
corresponding to the existence of just one chain with the
eldest vector $\vec{k}_n^{eld}=(1,...,1)$.
The number of points in the intersection with the integer
$Z_n$ lattice defining the corresponding polyhedron is given by the
binomial coefficient $C_{2n-1}^n$.

In the $K3$ case, for example, the only chain of arity 4
has as eldest vector $k_4^{eld}=(1,1,1,1)$, i.e.:
\begin{eqnarray}
\omega_4:\underbrace{{\AE}^{(1)}_{4}  \star \ldots
\star {\AE}^{(1)}_{4}}_{4}
\mapsto A_4^{4},
\end{eqnarray}
where $ {\AE}^{(1)}_{4}$ contains 4 extended vectors.
To obtain the complete listing of $K3$ spaces in this
chain, one can use mirror symmetry to find the maximal values of the
integer parameters ${m}$ describing the chain, or use the maximal 
expansion of the unit monomial $E_4=(1,1,1,1)$ in terms of four monomials
$P_1,...,P_4$ with the following condition: $1/4(P_1+...+P_4)=(1,1,1,1)$.

We introduce the following notation for the
Diophantine decomposition of any monomial:
$ {D}_{n}[p_1,p_2,...,p_i/d]$ with
$2\leq \,i\, \leq n$, $n$ is the dimension and
$p_1,...,p_i$ are integers with $d = p_1 + ... + p_i$, as
defined by the following equation:
\begin{equation}
{D}_{n}[p_1,p_2,...,p_i/d_D] \cdot \vec{\mu}_0\,=\,
( p_1 P_1+ p_2 P_2+....+p_i P_i)/d,
\label{DE}
\end{equation}
where $\vec{\mu}_0$ is a primordial monomial of dimension $n$ and
$\vec{\mu}_1,...,\vec{\mu}_i$ are the expanded  monomials.
According to this notation, the above
Diophantine decomposition can be represented in the form
${D}_4[1,1,1,1/4] \cdot E_4=\{P_,P_2,P_3,P_4 \}$ with the usual
Diophantine property: $(P_1+...+P_4)/4=E_4$, in this case, 
$p_1=...=p_4=1$ and $d=4$.
The definition of an invariant monomial (IM) is very closely
connected with this definition of Diophantine expansion.
We call the monomials participating in such a Diophantine expansion 
(\ref{DE}) a {\it set of IMs}. The unit monomial $E_n$ is and {IM} by 
definition, then with the decomposition $D_n[1,1/2]$ we produce a triple 
of {IMs}, $E_n, C_1,C_2$ obeying the condion $(C_1+C_2)=E_n$, etc..

This expansion has just {42} different irreducible solutions, which is much
less than the complete list of $K3$ RWVs. This example re-emphasizes that
one cannot get full information about Calabi-Yau $d$-folds by following
only the first diagonal line on the arity-dimension plot. Fortunately,
this also means that the internal structures of $K3$, $CY_3$,... $d$-folds
are much more complicated and interesting.

The next example is the $CY_3$ case with $n=5$, where there is again
just one chain with eldest vector $k_5^{(eld)}=(1,1,1,1,1)$ that can be
obtained with arity 5. The listing of this chain again can again be
obtained using mirror symmetry or by the maximal expansion of the unit
monomial $E_5=(1,1,1,1,1)$ in terms of five monomials $P_1,...,P_5$. The
corresponding Diophantine equation, $E_5=1/5(P_1+P_2+...+P_5)$, has 7269
solutions, giving the number of RWVs in $A_5^{(5)}$, again much smaller
than the total number of $CY_3$ spaces. {\it A fortiori}, the same is true
along this line for all dimensions.

These simple examples show that, although each $CY_d$-fold must contain a
central point, one cannot obtain all the $d$-folds simply by expanding the
unit monomial. The geometrical reason why the expansion of the unit
monomial $E_n$ as the sum of $n$ monomials cannot give the full list of
$CY_d$-folds lies in the complex composite structure of its geometry. The
geometry of $CY_d$-folds has a composite `Russian doll' structure. In the
intersection by the $(d-1)$-hyperplane through the centre of a $d$-fold,
one can find a $(d-1)$-fold, then in the next intersection by the
$(d-2)$-hyperplane one can find a $(d-2)$-fold, etc.. The picture is
completed by mirror symmetry and the intersection-projection symmetry of
$CY_d$-folds. We note that, according to the arity-dimension structure, if
an intersection by a hyperplane does not contain a $d$-fold, then the
corresponding $d$-folds have to be seen by projection on this hyperplane.

\begin{figure}[th!]
   \begin{center}
   \mbox{
   \epsfig{figure=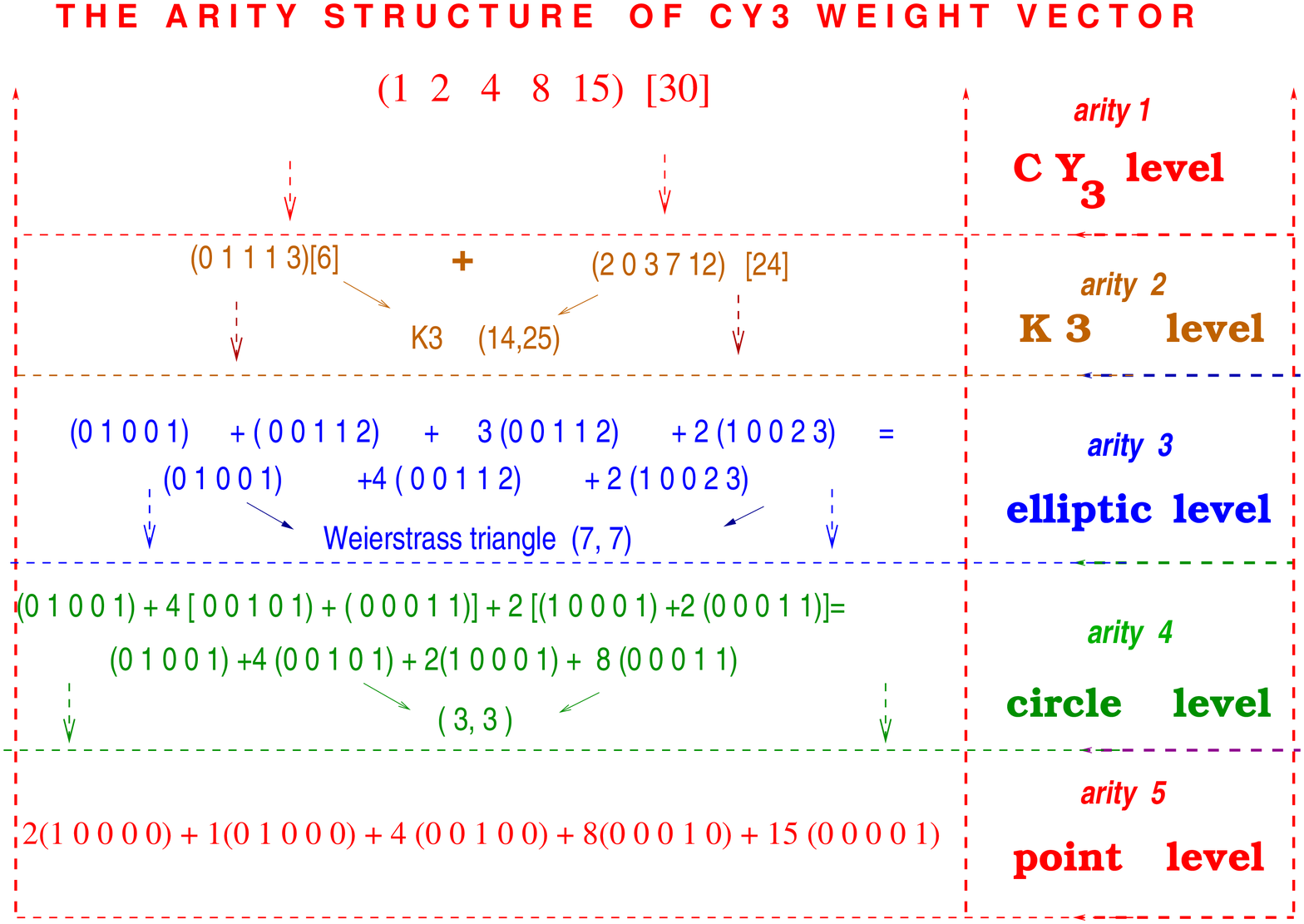,height=12cm,width=16cm}}
   \end{center}
   \caption{\it 
An example of the arity structure of a sample $CY_3$ weight vector, 
obtainable by normal expansion of higher-arity RWVs, showing the 
relation to specific $K3$ and Weierstrass spaces.}
\label{rwv1220}
\end{figure}

\begin{figure}[th!]
   \begin{center}
   \mbox{
   \epsfig{figure=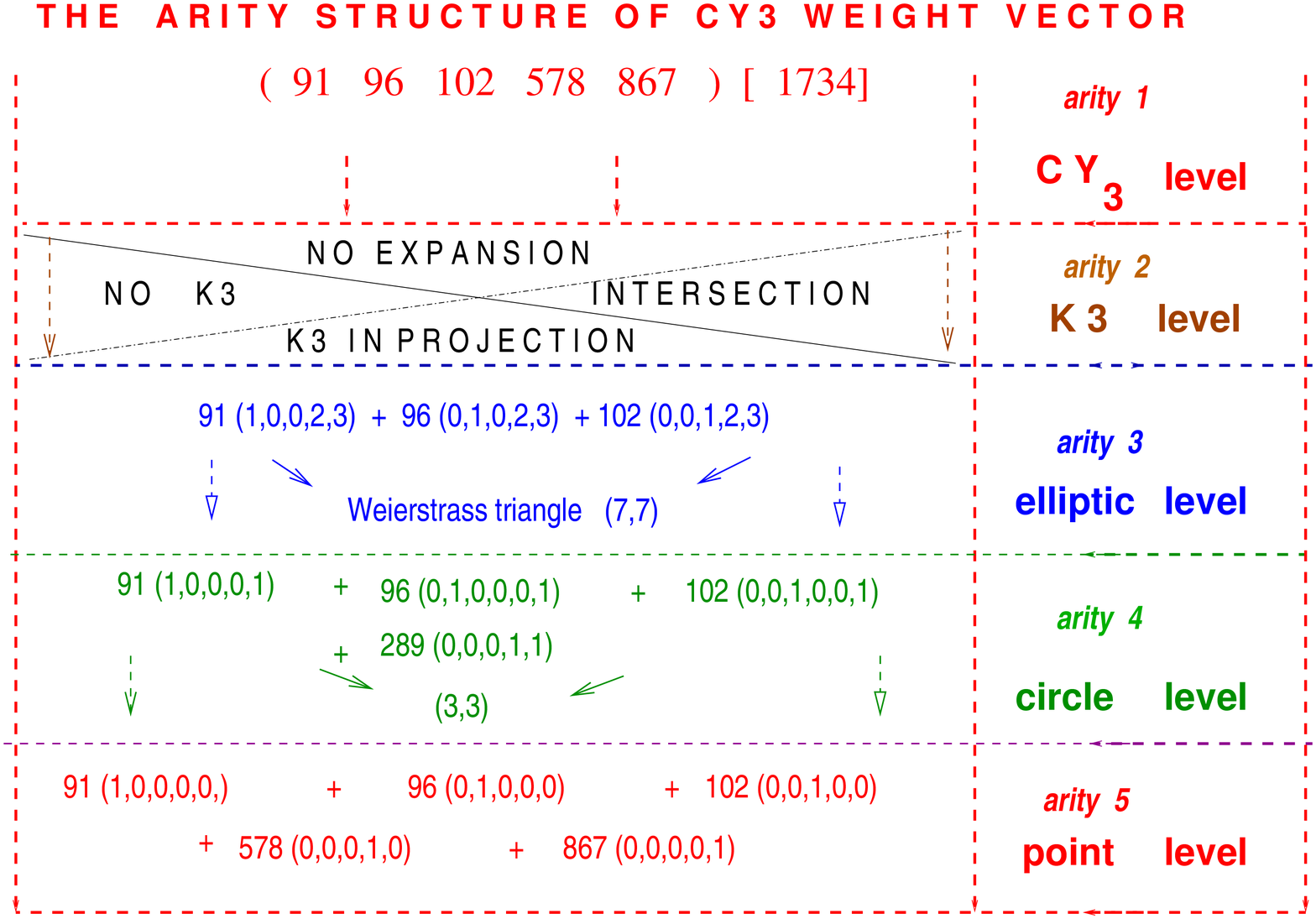,height=12cm,width=16cm}}
   \end{center}
   \caption{\it 
Another example of the arity structure of a sample $CY_3$ weight vector,
obtainable by normal expansion of higher-arity RWVs, which has 
counterparts at arities 3, 4 and 5, but not arity 2.}
\label{rwv1734}
\end{figure}

Only if we know all the $\omega_r$ operations with which a given vector
can be obtained can we know all the possible sets of expansions of this
reflexive vector, corresponding to the different chains in which it
appears in different dimensions, as illustrated in Figs~\ref{basmon1},
\ref{rwv1220}, \ref{rwv1734}. To get new information on the structures of
$CY_d$-folds, we should study $CY_d$-folds obtained by operations
corresponding to the points on the second sloping line. This $A_n^{(n-1)}$
set of $CY_d$-folds has a somewhat more complicated structure and,
according to our approach, is described by the weight vectors (1) and
(1,1), extended to dimension $n$. This already implies that the internal
geometrical structures of these $CY_d$ folds is connected algebraically
with the properties of conic monomials.

The two examples shown in Figs.~\ref{rwv1220} and \ref{rwv1734} display
different expansions of RWVs that illustrate how
the arity structure of the Universal Calabi-Yau Algebra reveals the full
algebraic structure of all RWVs. It gives a complete algebraic expansion
of all RWVs via the sets of extended RWVs constructed already from RWVs of
lower dimension.

There appear constructions with two different arities $r$ already in three
dimensions: apart from the maximal arity 3, also the lower arity 2
construction can give some information about the composite structure of
the RWVs and the corresponding polyhedra. The points on the second (green)
diagonal line in the Figures start with this case. In dimension three, it
is easy to construct the complete list of {RWV}s in the arity-3 chain
whose eldest vector is $A_3^{(1)}=\vec{k}_3^{eld}=(1,1,1)$, using the
expansion technique or mirror symmetry: they are $(1,1,1), (1,1,2),
(1,2,3)$.  Analyzing the decomposition of the RWVs in chains with arity 2,
we find these three vectors in two different chains of arity 2:

\begin{eqnarray}
\omega_2:{\AE}^{(2)}_{3} \star
 {\AE}^{(2)}_{3}  \mapsto A_3^{2},
\end{eqnarray}
with two different eldest RWVs. To generate these two chains, one should
consider two RWVs of dimensions one and two, and extend them in all
possible ways to dimension three, i.e.:
\begin{eqnarray}
\vec{k}_1^{i(ex)}&=& (0,0,1) \qquad + \qquad 3\; permutations \nonumber\\
\vec{k}_2^{i(ex)}&=& (0,1,1) \qquad + \qquad 3\; permutations \nonumber\\
\end{eqnarray}
Then, in the set of six extended vectors, one should consider all $C_{6}^2$
possible cases, so as to find the only `good' triple intersections,
corresponding to the arity two construction. The two chains of arity two
are the following:
\begin{eqnarray}
\vec{k}_1^{1(ex)}= (1,0,0), \,  \, \vec{k}_2^{2(ex)}=(0,1,1)
\end{eqnarray}
and
\begin{eqnarray}
\vec{k}_2^{1(ex)}= (0,1,1), \,  \, \vec{k}_2^{2(ex)}=(1,0,1)
\end{eqnarray}
with the eldest  vectors being, respectively:
\begin{eqnarray}
\vec{k}^{(eld)}= (1,1,1) 
\end{eqnarray}
and
\begin{eqnarray}
\vec{k}^{(eld)}= (1,1,2).
\end{eqnarray}
The two different good
intersections giving us line-segment reflexive polyhedra
can be described by the following sets of linear and conic monomials:
\begin{eqnarray}
C_1=(1,2,0), \qquad C_2=(1,0,2),\qquad E_3=(1,1,1)
\end{eqnarray}
and
\begin{eqnarray}
C_1=(2,2,0), \qquad C_2=(0,0,2),\qquad E_3=(1,1,1)
\end{eqnarray}
with the following universal property: $C_1+C_2=2 \cdot E_3$.

In the $K3$ case in four dimensions, there are already four chains of
arity 3 on the second diagonal $D_r$ line:
\begin{eqnarray}
\omega_3:{\AE}^{(2)}_{4} \star
 {\AE}^{(2)}_{4} \star {\AE}^{(2)}_{4}
\mapsto A_4^{3},
\end{eqnarray}
with four corresponding eldest RWVs. To get these four chains, one can
consider two RWVs of dimension one and two, and extend them in all
possible ways to dimension four, i.e.:
\begin{eqnarray}
\vec{k}_1^{ex}&=& (0,0,0,1) \qquad + \qquad 4\; permutations \nonumber\\
\vec{k}_2^{ex}&=& (0,0,1,1) \qquad + \qquad 6\; permutations \nonumber\\
\end{eqnarray}
Now, within the set of ten extended vectors, one should look for
`good' triple intersections corresponding to the arity {3},
among the $C_{10}^3$ possible cases.
The four chains of arity 3 found in this way are the following:
\begin{eqnarray}
\vec{k}^{i(ex)}&=& (1,0,0,0), \, (0,1,0,0), \, (0,0,1,1)\nonumber\\
\vec{k}^{i(ex)}&=& (0,0,1,0), \, (1,0,0,1), \, (0,1,0,1)\nonumber\\
\vec{k}^{i(ex)}&=& (0,0,1,1), \, (0,1,0,1), \, (1,0,0,1)\nonumber\\
\vec{k}^{i(ex)}&=& (1,0,0,1), \, (0,1,0,1), \, (1,0,1,0)\nonumber\\
\end{eqnarray}
with the following eldest vectors:
\begin{eqnarray}
\vec{k}^{(eld)}&=& (1,1,1,1)\nonumber\\
\vec{k}^{(eld)}&=& (1,1,1,2)\nonumber\\
\vec{k}^{(eld)}&=& (1,1,1,3)\nonumber\\
\vec{k}^{(eld)}&=& (2,1,1,2)\nonumber\\
\end{eqnarray}
These four different intersections give us segment reflexive polyhedra 
that can be described by the following sets of conic monomials:
\begin{eqnarray}
C_1=(1,1,2,0), &&\qquad C_2=(1,1,0,2), \nonumber\\
C_1=(2,2,1,0), &&\qquad C_2=(0,0,1,2), \nonumber\\
C_1=(2,2,2,0), &&\qquad C_2=(0,0,0,2), \nonumber\\
C_1=(2,2,0,0), &&\qquad C_2=(0,0,2,2), \nonumber\\
\end{eqnarray}
with the following universal property: $C_1+C_2=2 \cdot E_4$.
To list these four chains, one can again use mirror symmetry, or
one may consider the decompositions of the following conic monomials:
\begin{eqnarray}
&&C_1=(2,2,2,0) \rightarrow {P_1,P_2,P_3: 1/3(P_1+P_2+P_3)=C_1},
\qquad C_2=(0,0,0,2); \nonumber\\
&&C_1=(2,2,1,0)\rightarrow {P_1,P_2,P_3: 1/3(P_1+P_2+P_3)=C_1},
\qquad C_2=(0,0,1,2); \nonumber\\
&&C_1=(2,1,1,0)\rightarrow {P_1,P_2,P_3: 1/3(P_1+P_2+P_3)=C_1},
 \qquad C_2=(0,1,1,2); \nonumber\\
&&C_1=(2,2,0,0)\rightarrow {P_1,P_2: 1/2(P_1+P_2)=C_1}\, \nonumber\\
&&C_2=(0,0,2,2)\rightarrow {P_3,P_4: 1/2(P_3+P_4)=C_2}. \nonumber\\
\end{eqnarray}
For example, as was already shown in~\cite{AENV2}, there are 34 vectors in
the first chain (III). The eldest and youngest vectors of this chain are
$k_4^{eld}=(1,1,1,3)$ and $k_4^{eld}=(7,8,10,25)$, respectively. The full
number of RWVs in $A_n^{(3)}$ with conic structures is 90. Thus, to get
the full list of $K3$ with new conic structures, one can use a composite
expansion technique, starting from the unit monomial $E_4=(1,1,1,1)$. In
the first step, one can use the expansion of unit monomials to produce the
conics $C_1$ and $C_2$, giving rise to the eldest vectors and chains with
arity 3. Secondly, to get the complete listings of the chains, one should
construct the decompositions of the conics in terms of three monomials, as
found by solving the corresponding Diophantine equations.

Similarly, in the $CY_3$ case there are already six chains of arity 4:
\begin{eqnarray}
\omega_4:\underbrace{{\AE}^{(2)}_{5} \star
 \ldots \star {\AE}^{(2)}_{5}}_{4}
\mapsto A_5^{4},
\end{eqnarray}
They can be obtained by extending the two simple RWVs (1) and (1,1)
to dimension five. The number of distinct extended vectors is now
${15=5+10}$. Then one should look for `good' quadruple
intersections among the $C_{15}^4$ possible combinations.
The resulting six chains of arity {4} are the following:
\begin{eqnarray}
\vec{k}^{i(ex)}&=& (1,0,0,0,0), \, (0,1,0,0,0),\,  (0,1,0,0,0), \,
(0,0,0,1,1)\nonumber\\
\vec{k}^{i(ex)}&=& (0,0,1,0,0), \, (0,0,0,1,0), \, (1,0,0,0,1), \,
(0,1,0,0,1)\nonumber\\
\vec{k}^{i(ex)}&=& (1,0,0,0,0), \, (0,0,0,1,1), \, (0,0,1,0,1), \,
(0,1,0,0,1)\nonumber\\
\vec{k}^{i(ex)}&=& (0,0,0,1,1), \, (0,0,1,0,1), \, (0,1,0,0,1), \,
(1,0,0,0,1)\nonumber\\
\vec{k}^{i(ex)}&=& (0,0,1,0,0), \, (0,1,0,1,0), \, (1,0,0,1,0), \,
(0,1,0,0,1)\nonumber\\
\vec{k}^{i(ex)}&=& (1,0,0,1,0), \, (0,1,0,1,0), \, (0,0,1,1,0), \,
(1,0,0,0,1)\nonumber\\
\end{eqnarray}
with the following six eldest vectors:
\begin{eqnarray}
\vec{k}^{(eld)}&=& (1,1,1,1,1)\nonumber\\
\vec{k}^{(eld)}&=& (1,1,1,1,2)\nonumber\\
\vec{k}^{(eld)}&=& (1,1,1,1,3)\nonumber\\
\vec{k}^{(eld)}&=& (1,1,1,1,4)\nonumber\\
\vec{k}^{(eld)}&=& (1,2,1,2,1)\nonumber\\
\vec{k}^{(eld)}&=& (2,1,1,3,1)\nonumber\\
\end{eqnarray}
The intersections giving the segment reflexive polyhedra
can be described by the following sets of conic monomials:
\begin{eqnarray}
C_1=(1,1,1,2,0), &&\qquad C_2=(1,1,1,0,2), \nonumber\\
C_1=(2,2,1,1,0), &&\qquad C_2=(0,0,1,1,2), \nonumber\\
C_1=(1,2,2,2,0), &&\qquad C_2=(1,0,0,0,2), \nonumber\\
C_1=(2,2,2,2,0), &&\qquad C_2=(0,0,0,0,2), \nonumber\\
C_1=(2,2,1,0,0), &&\qquad C_2=(0,0,1,2,2), \nonumber\\
C_1=(2,2,2,0,0), &&\qquad C_2=(0,0,0,2,2), \nonumber\\
\end{eqnarray}
with the following Diophantine property: $C_1+C_2=2 \cdot E_5$.

To find the complete list of 14,017 RWVs in the first chain,
one can consider the following
expansions of the monomial $C_1$ ($C_2$):
\begin{eqnarray}
C_1=(2,2,2,2,0) \rightarrow {P_1,P_2,P_3,P_4: 1/4(P_1+P_2+P_3+P_4)=C_1},
 \qquad C_2=(0,0,0,0,2). \nonumber\\
\end{eqnarray}
The eldest and the youngest vectors of this chain are
$k_5^{eld}=(1,1,1,1,4)[8]$  and $k_5^{yng}=(75,84,86,98,343)[686]$,
respectively.

So far we have mainly been recapitulating results for lower-dimensional 
$CY_d$-folds, explaining them in a manner suitable for extensions to 
higher
dimensions, and now we give some first results for the $CY_4$ case. It is
easy to see that there are already {9} chains of arity 5:
\begin{eqnarray}
\omega_5:\underbrace{{\AE}^{(2)}_{6} \star
 \ldots \star {\AE}^{(2)}_{6}}_{5}
\mapsto A_6^{5},
\end{eqnarray}
These can be obtained by extending the two RWVs (1) and (1,1) to dimension
{6}, in all ${21=6+15}$ different ways. There are $C_{21}^5$
different combinations to search for
`good' quadruple intersections, among which we find the following
{9} chains of arity {5}:
\begin{eqnarray}
\vec{k}^{i(ex)}&=& (1,0,0,0,0,0), \, (0,1,0,0,0,0), \,
(0,0,1,0,0,0), \,  (0,0,0,1,0,0), \, (0,0,0,0,1,1)  \nonumber\\
\vec{k}^{i(ex)}&=& (0,0,1,0,0,0), \, (0,0,0,1,0,0), \,
(0,0,0,0,1,0), \,  (0,1,0,0,0,1), \, (1,0,0,0,0,1)\nonumber\\
\vec{k}^{i(ex)}&=& (0,0,0,1,0,0), \, (0,0,0,0,1,0),  \,
(0,0,1,0,0,1)  \,  (0,1,0,0,0,1), \, (1,0,0,0,0,1)\nonumber\\
\vec{k}^{i(ex)}&=& (0,0,0,0,1,0), \, (0,0,0,1,0,1),  \,
(0,0,1,0,0,1), \,  (0,1,0,0,0,1), \, (1,0,0,0,0,1)\nonumber\\
\vec{k}^{i(ex)}&=& (0,0,0,0,1,1), \, (0,0,0,1,0,1),  \,
(0,0,1,0,0,1), \,  (0,1,0,0,0,1), \, (1,0,0,0,0,1)\nonumber\\
\vec{k}^{i(ex)}&=& (0,0,1,0,0,0), \, (0,0,0,1,0,0),  \,
(1,0,0,0,1,0), \,  (1,0,0,0,0,1), \, (0,1,0,0,0,1)\nonumber\\
\vec{k}^{i(ex)}&=& (0,0,0,1,0,0), \, (1,0,0,0,0,1),  \,
(0,1,0,0,0,1), \,  (0,0,1,0,0,1), \, (1,0,0,0,1,0)\nonumber\\
\vec{k}^{i(ex)}&=& (1,0,0,0,0,1), \, (0,1,0,0,0,1),  \,
(0,0,1,0,0,1), \,  (0,1,0,0,1,0), \, (0,0,1,1,0,0)\nonumber\\
\vec{k}^{i(ex)}&=& (1,0,0,0,0,1), \, (0,1,0,0,0,1),  \,
(0,0,1,0,0,1), \,  (0,0,0,1,0,1), \, (0,0,0,1,1,0)\nonumber\\
\end{eqnarray}
with the following {9} eldest vectors:
\begin{eqnarray}
\vec{k}^{(eld)}&=& (1,1,1,1,1,1)\nonumber\\
\vec{k}^{(eld)}&=& (1,1,1,1,1,2)\nonumber\\
\vec{k}^{(eld)}&=& (1,1,1,1,1,3)\nonumber\\
\vec{k}^{(eld)}&=& (1,1,1,1,1,4)\nonumber\\
\vec{k}^{(eld)}&=& (1,1,1,1,1,5)\nonumber\\
\vec{k}^{(eld)}&=& (2,1,1,1,1,2)\nonumber\\
\vec{k}^{(eld)}&=& (2,1,1,1,1,3)\nonumber\\
\vec{k}^{(eld)}&=& (1,2,2,1,1,3)\nonumber\\
\vec{k}^{(eld)}&=& (1,1,1,2,1,4)\nonumber\\
\end{eqnarray}
These {9} different `good'
intersections yield segment reflexive polyhedra
described by the following sets of conic monomials:
\begin{eqnarray}
C_1=(1,1,1,1,2,0), &&\qquad C_2=(1,1,1,1,0,2), \nonumber\\
C_1=(2,2,1,1,1,0), &&\qquad C_2=(0,0,1,1,1,2), \nonumber\\
C_1=(2,2,2,1,1,0), &&\qquad C_2=(0,0,0,1,1,2), \nonumber\\
C_1=(2,2,2,2,1,0), &&\qquad C_2=(0,0,0,0,1,2), \nonumber\\
C_1=(2,2,2,2,2,0), &&\qquad C_2=(0,0,0,0,0,2), \nonumber\\
C_1=(2,2,1,1,0,0), &&\qquad C_2=(0,0,1,1,2,2), \nonumber\\
C_1=(2,2,2,1,0,0), &&\qquad C_2=(0,0,0,1,2,2), \nonumber\\
C_1=(2,2,2,0,0,0), &&\qquad C_2=(0,0,0,2,2,2), \nonumber\\
C_1=(2,2,2,2,0,0), &&\qquad C_2=(0,0,0,0,2,2), \nonumber\\
\end{eqnarray}
with the following Diophantine property: $C_1 + C_2 = 2 \cdot E_5$.

Continuing along this line to the highest arity in each higher dimension,
we find the following numbers of different eldest vectors and chains: 1,
2, 4, 6, 9, 12, 16, 20, 25, 30, 36, 42, 49, 56, 64, 72, 81, 90, 100, 110,
121, 132, 144, ..., which are given by a simple recurrence relation, as we
show in the next section.

\subsection{Analysis of the $E_r$ line}

We now consider Calabi-Yau $d$-folds with more complicated
structures, corresponding to the sets $A_n^{(n-2)}$ on the arity-dimension
plot. These can be constructed from RWVs of lower dimensions: (1); (1,1);
(1,1,1),(1,1,2),(1,2,3), suitably extended to dimension ${n}$.

\subsubsection{The $K3$ Case}

The $E_r$ line in this case corresponds to arity 2:
\begin{eqnarray}
\omega_2\,:\,{\AE}^{(3)}_{4}
\star {\AE}^{(3)}_{4}
\mapsto A_4^{2},
\end{eqnarray}
where ${\AE}^{(2)}_4$ contains ${50=4+6+6+12+24}$ extended vectors. We
find 22 eldest vectors and chains of arity 2, corresponding to $K3$ spaces
with a composite structure containing a $d=1$ complex curve. To find the
full lists of RWVs in these {22} chains, one may use the argument that the
maximal values of integer parameters $m$ are bounded by the dimensions of
the corresponding extended vectors, as was done in~\cite{AENV1}.

The composite structure of $d$-folds along this
third diagonal line on the arity-dimension plot is determined
by cubic and quartic monomials, of which we now give some examples.
Chain I in Table~\ref{TabXL} is determined by the `good'
intersection of two extended vectors
\begin{equation}
\vec{k}_1^{(ex)}=(1,0,0,0) , \qquad
\vec{k}_3^{(ex)}=(0,1,1,1).
\end{equation}
This intersection contains {10}
monomial points, producing a plane reflexive polyhedron, corresponding
to the RWV $\vec{k}_3=(1,1,1)$~\cite{AENV1}. These {10} monomials include
the following three cubic monomials:
\begin{eqnarray}
P_1=(1,3,0,0) , \qquad P_2=(1,0,3,0) , \qquad P_2=(1,0,0,3), \nonumber\\
\end{eqnarray}
with the Diophantine property:
\begin{equation}
\frac{1}{3} (P_1+P_2+P_3) =E_4=(1,1,1,1).
\end{equation}
Among the total of {22} chains there are {20} chains with the same
property. In turn, among these {20} cases, the intersection for each of
these chains contains itself some triple monomials, $P_1,P_2,P_3$,
of which at least one is cubic, i.e., has  degree {3}.

Only chains VIII, IX and XX have no such triples. However, the
intersections of the chains VIII and IX have another Diophantine property,
as they are determined by the intersection of the following pairs of
extended vectors:
\begin{eqnarray}
\vec{k}_3^{1(ex)}=(0,2|1,1) &\qquad &
\vec{k}_3^{2(ex)}=(1,0|1,2)\nonumber\\
\vec{k}_3^{1(ex)}=(1,0|1,2) &\qquad &
\vec{k}_3^{2(ex)}=(0,2|1,3) \nonumber\\
\end{eqnarray}
For these intersections, which contain {5} monomial points,
one can see in both cases the following pairs of quartic
and conic monomials, respectively:
\begin{eqnarray}
P_1&=&(0,0,4,0),\, P_2=(4,2,0,0), \qquad C_1=(2,1,2,0), \,
C_2=(0,1,0,2), \nonumber\\
P_1&=&(4,3,0,0),\, P_2=(0,1,4,0), \qquad C_1=(2,2,2,0), \, C_1=(2,2,2,0),
\nonumber\\
\end{eqnarray}
with the following common Diophantine property in both cases:
\begin{eqnarray}
\frac{1}{2}(P_1+P_2)=C_1,\qquad \frac{1}{2}( C_1+C_2)=E_4.
\end{eqnarray}
These relations can be rewritten in the form:
\begin{equation}
\frac{1}{4}(P_1+P_2+2\cdot  C_2)=E_4.
\end{equation}
In the full list of {22} chains there are {20} cases with such cubic and 
quartic Diophantine properties. 

However, the chain X is not included in these two sets. It is
determined by the following intersection:
\begin{eqnarray}
\vec{k}_2^{(ex)}=(0,0,|1,1) \qquad
\vec{k}_2^{(ex)}=(1,1,|0,0)\nonumber\\
\end{eqnarray}
which has among its {9} monomial points two conic pairs:
\begin{eqnarray}
C_1=(2,0,2,0),\,&\qquad &  C_2=(0,2,0,2),  \nonumber\\
\tilde C_1=(2,0,0,2), \,&\qquad &\tilde C_2=(0,2,2,0)\nonumber\\
\end{eqnarray}
There are {17} such cases among all the {22} chains. 

These examples show how different chains can be identified by decomposing
IMs in terms of conic, cubic, quartic, etc., monomials. In the next
section we use these Diophantine properties to study the possible
structures of $CY_d$-folds in arbitrary dimensions. As we show there in
more detail, the use of such invariant monomials provides an alternative
to Batyrev's reflexive polyhedron way of constructing $CY_d$ spaces.

The  structure of the projective ${\vec{k}_4}$ vectors in
the 22 chains reveals interesting interrelations between the five
classical regular dual polyhedron pairs in three-dimensional space
(the one-dimensional point, two-dimensional line-segment
and three dimensional tetrahedron, octahedron-cube and
icosahedron-dodecahedron) and Coxeter-Dynkin diagrams ${CD}$ for the
five types of Lie algebras:
$A, D, E_{6,7,8}$, as illustrated in Fig.~\ref{cdjad}. The structures of
the vectors ${\vec {k}_4}$ in the ${CD_{\sigma,\pi}}$ diagrams, which
can be seen in the corresponding polyhedra of projective vectors,
follow completely follow those of the only possible five `extended' 
vectors.
%\begin{eqnarray}
%%\vec {k}_1^{ext}=(0,0,0,1)  & \leftrightarrow& A_r ;\nonumber\\
%\vec {k}_2^{ext}=(0,0,1,1)  & \leftrightarrow& D_r ;\nonumber\\
%\vec {k}_3^{ext}=(0,1,1,1)  & \leftrightarrow& E_6 ;\nonumber\\
%\vec {k}_3^{ext}=(0,1,1,2)  & \leftrightarrow& E_7 ;\nonumber\\
%\vec {k}_3^{ext}=(0,1,2,3)  & \leftrightarrow& E_8 .
%\end{eqnarray}
We show in Table~\ref{TabXL} the ${A,D,E}$ structures and the
$CD_{\sigma}$ diagrams of all the eldest double ${K3}$ projective
vectors from the {22} chains~\cite{AENV1}.
In the first column, the Roman numbers enumerate the {22} 
chains, and we indicate by $^*$ the cases 
where the planar polyhedron in the
intersection coincides with the polyhedron in the projection, i.e.,
$\sigma=\pi$.

{\scriptsize
\begin{table}[!ht]
\centering
\caption{\it The list of the eldest vectors
for all 22 $K3$ chains:  $m = n = 1$, enumerating the corresponding
Coxeter-Dynkin $CD_{\sigma}$ diagrams.}
\label{TabXL}.
\begin{tabular}{|c|c|c|c||c|}
\hline\hline\hline
${N} $&${{\vec k}_i(eldest)} $&$ Structure$&$
max \,\Delta( \sigma)$&${CD_\sigma}
$\\
\hline\hline
$I$&$(1,1,|1,1)[4]
$&$ (0,1,|1,1)_{e6}+(1,0,|0,0)_{a}
$&$35={\bf 10_{e6}}+(\sigma={\{10\}}_{\triangle})+{\bf 15_a}
$&${E_{6}^{(1)}\leftrightarrow  A_{12}^{(1)}}
$ \\
$II^* $&$(1,1,|2,2)[6]
$&$(0,1,|1,1)_{e6}+(1,0,|1,1)_{e6}
$&$ 30={\bf 10_{e6}}+(\sigma={\{10\}}_{\triangle})
+ {\bf 10_{e6}}
$&${E_6^{(1)} \leftrightarrow E_6^{(1)}}
$ \\
$III $&$(3,1,|2,3)[9] $&$(0,1,|1,1)_{e6}+(3,0,|1,2)_{e8}
$&$23={\bf 4_{e6}}+(\sigma= {\{4\}}_{\triangle})+ {\bf 15_{e8}}
$&${{G_2}^{(1)}\leftrightarrow E_8^{(1)}}
$ \\
\hline\hline
$IV  $&$(1,1,|1,2)[5] $&$(0,1,|1,2)_{e7}+(1,0|0,0)_{a}
$&$34={\bf 13_{e7}}+(\sigma= {\{9\}}_{\triangle})+ {\bf 12_{a}}
$&${E_7^{(1)}\leftrightarrow A_{9}^{(1)}}
$ \\
$V $&$(1,1,|1,3)[6] $&$(0,1,|1,2)_{e7}+(1,0,|0,1)_{d}
$&$39={\bf 13_{e7}}+(\sigma={\{9\}}_{\triangle})+{\bf 17_{d}}
$&${E_7^{(1)}\leftrightarrow D^{(1)}_{10}}
$ \\
$VI^* $&$(1,1,|2,4)[8]  $&$(0,1,|1,2)_{e7}+(1,0,|1,2)_{e7}
$&$ 35={\bf 13_{e7}}+(\sigma= {\{9\}}_{\triangle})+{\bf 13_{e7}}
$&${E_7^{(1)}\leftrightarrow  E_7^{(1)}}
$ \\
$VII $&$(1,1,|1,1)[4] $&$(0,2,|1,1)_{e7}+(2,0,|1,1)_{e7}
$&$ 35={\bf 13_{e7}}+(\sigma= {\{9\}}_{\triangle})+{\bf 13_{e7}}
$&${E_7^{(1)}\leftrightarrow E_7^{(1)}}
$ \\
$VIII  $&$ (1,2,|2,3)[8] $&$(1,0,|1,2)_{e7}+(0,2,|1,1)_{e7}
$&$ 24={\bf 12_{e7}} +(\sigma= {\{5\}}_{\triangle})+{\bf 7_{e7}}
$&${E_7^{(1)}\leftrightarrow F_4^{(1)}}
$\\

$IX  $&$ (1,2,|1,5)[10] $&$(1,0,|1,2)_{e7}+(0,2,|1,3)_{e8}
$&$ 28={\bf 7_{e7}}+ (\sigma= {\{5\}}_{\triangle}) +{\bf 16_{e8}}
$&${F_4^{(1)}\leftrightarrow   E_8^{(1)}}
$\\
\hline\hline
$X   $&$(1,1,||1,1)[4]  $&$(0,0,||1,1)_{d}+(1,1,||0,0)_{d}
$&$ 35={\bf 13_d}+(\sigma= {\{9\}}_{\Box})+{\bf 13_{d}}
$&${D_8^{(1)}\leftrightarrow D_8^{(1)}}
$\\
$XI    $&$(1,1,|1,2)[5] $&$(1,0,|0,1)_d + (0,1,|1,1)_{e6}
$&$ 34={\bf 15_d}+(\sigma= {\{9\}}_{\Box})+{\bf 10_{e6}}
$&${D_8^{(1)}\leftrightarrow E_6^{(1)} }
$\\
$XII^*   $&$(1,1,|1,1)[4] $&$(0,3,|1,2)_{e8}+(3,0,|2,1)_{e8}
$&$ 13={\bf 4_{e8}}+(\sigma= {\{5\}}_{\Box})+{\bf 4_{e8}}
$&${G_2^{(1)}\leftrightarrow G_2^{(1)} }
$\\
$XIII^*  $&$(1,1,|2,3)[7] $&$(0,1,|1,1)_{e6}+(1,0,|1,2)_{e7}
$&$ 31={\bf 10_{e6}}+(\sigma= {\{8\}}_{\Box})+{\bf 13_{e7}}
$&${E_6^{(1)}\leftrightarrow E_7^{(1)}}
$\\
$XIV^*  $&$(1,1,|1,2)[5]  $&$(0,2,|1,1)_{e6}+(2,0,|1,3)_{e8}
$&$ 18={\bf 7_{e6}}+(\sigma= {\{6\}}_{\Box})+{\bf 5_{e8}}
$&${F_4^{(1)}\leftrightarrow G_2^{(1)} }
$\\
$XXII^* $&$ (1,1,|2,2)[6]  $&$(0,1,|1,2)_{e7}+(1,0,|1,0)_{d}
$&$ 30={\bf 13_{e7}}+(\sigma={\{7\}}_{\Box})+{\bf 10_{d}}
$&${E_7^{(1)}\leftrightarrow  {D_7} }
$\\
\hline\hline
$XV   $&$ (1,1,|2,3)[7] $&$(0,1,|2,3)_{e8}+ (1,0,|0,0)_{a}
$&$ 31={\bf 16_{e8}}+(\sigma= {\{7\}}_{\triangle})+{\bf 8_a}
$&${E_8^{(1)}\leftrightarrow A_6^{(1)} }
$\\
$XVI  $&$ (1,1,|2,4)[8] $&$(0,1,|2,3)_{e8}+(1,0,|0,1)_{d}
$&$ 35={\bf 16_{e8}}+(\sigma= {\{7\}}_{\triangle})+{\bf 12_{d}}
$&${E_8^{(1)}\leftrightarrow D_8 }
$\\
$XVII^*     $&$(1,1,|3,4)[9]  $&$ (0,1,|2,3)_{e8}+(1,0,|1,1)_{e6}
$&$ 33={\bf 16_{e8}}+(\sigma= {\{7\}}_{\triangle})+{\bf 10_{e6}}
$&${E_8^{(1)}\leftrightarrow  E_6^{(1)} }
$\\
$XVIII^*   $&$(1,1,|3,5)[10]  $&$ (0,1,|2,3)_{e8}+(1,0,|1,2)_{e7}
$&$ 36={\bf 16_{e8}}+(\sigma= {\{7\}}_{\triangle})+{\bf 13_{e7}}
$&${E_8^{(1)}\leftrightarrow  E_7^{(1)} }
$\\
$XIX^*   $&$ (1,1,|4,6)[12]  $&$ (0,1,|2,3)_{e8}+(1,0,|2,3)_{e8}
$&$ 39={\bf 16_{e8}}+(\sigma= {\{7\}}_{\triangle})+{\bf 16_{e8}}
$&${E_8^{(1)}\leftrightarrow  E_8^{(1)} }
$\\
$XX^*      $&$ (1,1,|1,3)[6]    $&$(0,2,|1,3)_{e8}+(2,0,|1,3)_{e8}
$&$ 21={\bf 7_{e8}}+(\sigma= {\{7\}}_{\triangle})+{\bf 7_{e8}}
$&${{F_4^{(1)}}\leftrightarrow  {F_4^{(1)}} }
$\\
$XXI^*     $&$ (3,3,|2,4)[12] $&$(0,3,|1,2)_{e8}+(3,0,1,2)_{e8}
$&$ 15={\bf 4_{e8}}+(\sigma= {\{7\}}_{\triangle})+{\bf 4_{e8}}
$&${{G_2^{(1)}} \leftrightarrow  {G_2^{(1)}} }
$\\
\hline\hline\hline
\end{tabular}
\end{table}
}

\subsubsection{The $CY_3$ Case}

One can make similar considerations in the next dimension $n=5$, i.e.,
the case of $CY_3$ spaces with arity 3 structure. As indicated in
Fig.~\ref{bas1med}, one must first extend the five RWVs: $(1); (1,1);
(1,1,1),(1,1,2),(1,2,3)$ to dimension 5, and then find all the normal
intersections corresponding to arity 3: \begin{eqnarray}
\omega_3\,:\,{\AE}^{(3)}_{5} \star {\AE}^{(3)}_{5} \star {\AE}^{(3)}_{5}
\mapsto A_5^{3}, \end{eqnarray} where ${\AE}^{(3)}_5$ contains
${115=5+10+10+30+60}$ extended vectors. A search of the normal triple
intersections gave 259 candidates~\cite{AENV2}, of which ${11}$
intersections have false eldest vectors that cannot be obtained
consistently by an arity-2 construction. Among these {259} chains, {161}
are irreducible, and 7 of these have false eldest vectors with $m=n=l=1$.

We do not discuss these here in further detail, except to mention the
biggest irreducible chain with an arity-3 structure, which consists of the
RWVs $(m,n,k,2(m+n+k),3(m+n+k))[6(m+n+k)]$, all of which have elliptic
Weierstrass fibre bundles. The total number of $CY_3$ hypersurfaces in
this chain is 20,796, and its eldest and youngest RWVs are (1,1,1,6,9)[18]
and (91,96,102,578,867)[1734], respectively. Its listing can be obtained
by mirror symmetry or via a triple expansion of the monomial
$P_0=(6,6,6,0,0)$ of the Weierstrass triangle, which is the normal
intersection of the following three extended vectors from ${\AE}^{(3)}_5$:
$\vec {k}_5^{ex} = (0,0,1,2,3),(0,1,0,2,3), (1,0,0,2,3)$. More precisely,
the listing of this chain can be obtained from the following Diophantine
expansion ~\cite{AENV2} :

\begin{eqnarray}
D_5[1,1,1/3] \,\,:\,\,P_0 \rightarrow \{P_1,P_2,P_3: 
\frac{1}{3}(P_1+P_2+P_3)=P_0\}.
\end{eqnarray}
In this definition, the subscript 5 labels the dimension of the
monomial $P_0$, and the combination
[111/3] labels the type of Diophantine expansion: in this case,
the monomial $P_0$ is decomposed into three monomials with the condition 
$(P_1+P_2+P_3)/3=P_0$. In Table~\ref{Tabelquin1}, one can sometimes see
that the same vector can give different Coxeter-Dynkin diagrams, for 
example, {14} and {14'}.

{\scriptsize
\begin{table}[!ht]
\centering
\caption{ \it The eldest vectors for all $S_3$-symmetric chains of $CY_3$ 
hypersurfaces.}
\label{Tabelquin1}
%\scriptsize
\vspace{.05in}
\begin{tabular}{|c|c||c|c||c|c||c|c||}
\hline
${ N} $&${ {\vec k}_i(eldest)} $&$ h^{2,1}  $&$ h^{1,1}
$&$V  $&$ V^*
$&$ Cox-Dyn
$&$ P_0 section
$\\
\hline\hline
${ 1} $&$(1,1,1|1,1)        $&$ 101      $&$ 1
$&$5  $&$ 5
$&$(E_6)(A)^2
$&$(3,1,1|0,0)
$\\
\hline
${ 2} $&$(1,1,1|1,2)        $&$ 103      $&$ 1
$&$5  $&$ 5
$&$(E_7)(A)^2
$&$(4,1,1|0,0)
$\\
\hline
${ 2'} $&$(1,1,1|1,2)        $&$ 103      $&$ 1
$&$5  $&$ 5
$&$(E_6)(D)(A)
$&$(3,2,1|0,0)
$\\
\hline
${ 3} $&$(1,1,1|1,3)        $&$ 122      $&$ 2
$&$8 $&$ 6
$&$(E_6)(D)^2
$&$(3,2,2|0,0)
$\\
\hline
${ 4} $&$(1,1,1|1,4)        $&$ 149      $&$ 1
$&$5  $&$ 5
$&$(E_7)(D)^2
$&$(4,2,2|0,0)
$\\
\hline \hline
${ 5} $&$(1,1,1|2,2)        $&$ 95       $&$ 2
$&$9  $&$ 6
$&$(E_6)^2(A)
$&$(3,3,1|0,0)
$\\
\hline
${ 6} $&$(1,1,1|2,3)        $&$ 106      $&$ 2
$&$8  $&$ 6
$&$(E_6)^2(D)
$&$(3,3,2|0,0)
$\\
\hline
${ 6'} $&$(1,1,1|2,3)        $&$ 106      $&$ 2
$&$8  $&$ 6
$&$(E_7)(E_6)(A)
$&$(4,3,1|0,0)
$\\
\hline
${ 6''} $&$(1,1,1|2,3)        $&$ 106      $&$ 2
$&$8  $&$ 6
$&$(E_8)(A)^2
$&$(6,1,1|0,0)
$\\
\hline
${ 7} $&$(1,1,1|2,4)        $&$ 123      $&$ 3
$&$9  $&$ 6
$&$(E_7)(E_6)(D)
$&$(4,3,2|0,0)
$\\
\hline
${ 7'} $&$(1,1,1|2,4)        $&$ 123      $&$ 3
$&$9  $&$ 6
$&$(E_8)(D)(A)
$&$(6,2,1|0,0)
$\\
\hline
${ 8} $&$(1,1,1|2,5)        $&$ 145      $&$ 1
$&$5  $&$ 5
$&$(E_7)^2(D)
$&$(4,4,2|0,0)
$\\
\hline
${ 8'} $&$(1,1,1|2,5)        $&$ 145      $&$ 1
$&$5  $&$ 5
$&$(E_8)(D)^2
$&$(6,2,2|0,0)
$\\
\hline
${ 9} $&$(1,1,1|3,3)        $&$ 112      $&$ 4
$&$5  $&$ 5
$&$(E_6)^3
$&$(3,3,3|0,0)
$\\
\hline
${10} $&$(1,1,1|3,4)        $&$ 126      $&$ 4
$&$10 $&$ 7
$&$(E_7)(E_6)^2
$&$(4,3,3|0,0)
$\\
\hline
${10'} $&$(1,1,1|3,4)        $&$ 126      $&$ 4
$&$10 $&$ 7
$&$(E_8)(E_6)(A)
$&$(6,3,1|0,0)
$\\
\hline
${11} $&$(1,1,1|3,5)        $&$ 144      $&$ 4
$&$10 $&$ 7
$&$(E_7)^2(E_6)
$&$(4,4,3|0,0)
$\\
\hline
${11'} $&$(1,1,1|3,5)        $&$ 144      $&$ 4
$&$10 $&$ 7
$&$(E_8)(E_7)(A)
$&$(6,4,1|0,0)
$\\
\hline
${12} $&$(1,1,1|,3,6)        $&$ 165      $&$ 3
$&$5  $&$ 5
$&$(E_7)^3
$&$(4,4,4|0,0)
$\\
\hline
${12'} $&$(1,1,1|,3,6)        $&$ 165      $&$ 3
$&$5  $&$ 5
$&$(E_8)(E_7)(D)
$&$(6,4,2|0,0)
$\\
\hline
${13} $&$(1,1,1|4,5)        $&$ 154      $&$ 4
$&$7  $&$ 6
$&$(E_8)(E_6)^2
$&$(6,3,3|0,0)
$\\
\hline
${14} $&$(1,1,1|4,6)        $&$ 173      $&$ 5
$&$9  $&$ 6
$&$(E_8)^2(A)
$&$(6,6,1|0,0)
$\\
\hline
${14'} $&$(1,1,1|4,6)        $&$ 173      $&$ 5
$&$9  $&$ 6
$&$(E_8)(E_7)(E_6)
$&$(6,4,3|0,0)
$\\
\hline
${15} $&$(1,1,1|4,7)        $&$ 195      $&$ 3
$&$7  $&$ 6
$&$(E_8)(E_7)^2
$&$(6,4,4|0,0)
$\\
\hline
${15'} $&$(1,1,1|4,7)        $&$ 195      $&$ 3
$&$7  $&$ 6
$&$(E_8)^2(D)
$&$(6,6,2|0,0)
$\\
\hline
${16} $&$(1,1,1|5,7)        $&$ 208      $&$ 4
$&$7  $&$ 6
$&$(E_8)^2(E_6)
$&$(6,6,3|0,0)
$\\
\hline
${17} $&$(1,1,1|5,8)        $&$ 231      $&$ 3
$&$7  $&$ 6
$&$(E_8)^2(E_7)
$&$(6,6,4|0,0)
$\\
\hline
${18} $&$(1,1,1|6,9)        $&$ 272      $&$ 2
$&$5  $&$ 5
$&$(E_8)^3
$&$(6,6,6|0,0)
$\\
\hline \hline
\end{tabular}
\end{table}
}
%\normalsize

\subsubsection{New Results for $CY_4$ Spaces}

We now apply the technique explained in the earlier subsections to find
the arity-4 chains for the 6-dimensional $CY_4$ case:
\begin{eqnarray}
\omega_4\,:\,{\AE}^{(3)}_{6}
\star {\AE}^{(3)}_{6} \star {\AE}^{(3)}_{6} \star {\AE}^{(3)}_{6}
\mapsto A_6^{3},
\end{eqnarray}
where the set ${\AE}^{3}_6$ contains
vectors constructed from the following five {RWV}s:
$\vec{k}_1=(1)$, $\vec{k}_2=(1,1)$,
$\vec{k}_3=(1,1,1),(1,1,2),(1,2,3)$ by extending them to dimension {6}.
It is trivial to see that there are 221 such extended vectors:
\begin{eqnarray}
\vec{k}_1^{(ex)}&=&(0,..,0,1) \qquad +\qquad 6\;{permutations} ; 
\nonumber\\
\vec{k}_2^{(ex)}&=&(0,...,1,1)\qquad +\qquad 15\;{permutations} ; 
\nonumber\\
\vec{k}_3^{(ex)}&=&(0,..,1,1,1)\qquad +\qquad 15\;{permutations} , 
\nonumber\\
\vec{k}_3^{(ex)}&=&(0,..,1,1,2)\qquad +\qquad 60\;{permutations} , 
\nonumber\\
\vec{k}_3^{(ex)}&=&(0,..,1,2,3)\qquad +\qquad120\;{permutations} 
,\nonumber\\
\end{eqnarray}
Among these {221} extended vectors one should look for the
quadruple `good' intersections, which involves considering
a total of $C_{221}^{4}$ combinatorial possibilities.

Some of the quadruples turn out to be equivalent to other quadruples,
after some permutation of the vectors and/or their components.  Checking
this is by far the most time-consuming task for our computer program.  To
give some idea, there are 720,768 quadruples that do not all have zeroes
in the same component (which is definitely not a good case) and where the
number of intersection monomials is between 4 and 10 (the intersection of
four 6-dimensional vectors is a polygon, and all good cases on the plane
are known to have between 4 and 10 monomials). When checking if two
quadruples are equivalent, one has to check $720,768 \times 720,767$
divided by 2 combinations, and this, in general, for all possible
permutations of the 4 vectors and the 6 components, {\it i.e.}, $4! \times
6!$ times. The final result is that, in dimension 6, there are 5,607
different 4-vector chains obtainable by extending the 5 `good' vectors
existing in dimensions 1, 2 and 3. Many of these 5,607 chains of arity 4
could be reducible, so the next step is to identify the irreducible
chains. This is done later using the method of invariant monomials (IMs).

In this case, the number of irreducible chains turns out to be 2,111, and
the number of different eldest RWVs that generate all these 2,111 chains is
only 397. This reveals that these {397} eldest vectors have a very rich
composite substructure. We show in Table~\ref{Tabelsix1} the list of
{RWV}s with maximal symmetry at least as large as $S_4$. This Table also
shows the structures of the Coxeter-Dynkin diagrams included in the
reflexive polyhedra. It is easy to convince oneself that the number of
Coxeter-Dynkin factors grows with the arity. In particular, for arity 4, 
the number of Coxeter-Dynkin diagrams is also equal to {4}, as seen in 
Table~\ref{Tabelsix1}. 

{\scriptsize
\begin{table}[!ht]
\centering
\caption{\it  The eldest vectors for all the $S_4$-symmetric 
arity-4 chains of $CY_3$ hypersurfaces.}
\label{Tabelsix1}
%\scriptsize
\vspace{.05in}
\begin{tabular}{||c|c|c|c||c|c|c|c||}
\hline
${ N} $&${ {\vec k}_i(eld)}   $&$Coxeter-Dynkin
$&$P_0 \; section
$&${ N} $&${ {\vec k}_i(eld)} $&$Coxeter-Dynkin
$&$section\, P_0              $\\
\hline\hline
${ 1} $&$(1,1,1,1,|1,1)        $&$(E_6)(A)^3
$&$(3 1 1 1 0 0)
$&${ 16}$&$(1,1,1,1,|4,4)     $&$(E_6)^4
$&$(3 3 3 3 0 0)              $\\
\hline
${2} $&$(1,1,1,1,|1,2)         $&$(E_6)(D)(A)^2
$&$(3 2 1 1 0 0)
$&${17}$&$(1,1,1,1,|4,5)      $&$(E_8)(E_6)^2(A)
$&$(6 3 3 1 0 0)              $\\
\hline
${2'} $&$(1,1,1,1,|1,2)        $&$(E_7)(A)^3
$&$(4 1 1 1 0 0)
$&${18}$&$(1,1,1,1,|4,6)      $&$(E_7)^2(E_6)^2
$&$(4 4 3 3 0 0)              $\\
\hline
${ 3} $&$(1,1,1,1,|1,3)        $&$(E_6)(D)^2(A)
$&$(3 2 2 1 0 0)
$&$   $&$                     $&$
$&$                           $\\
\hline
${3'} $&$(1,1,1,1,|1,3)        $&$(E_7)(D)(A)^2
$&$(4 2 1 1 0 0)
$&${18'}$&$(1,1,1,1,|4,6)     $&$(E_8)(E_7)(E_6)(A)
$&$(6 4 3 1 0 0)              $\\
\hline
${ 4} $&$(1,1,1,1,|1,4)        $&$(E_6)(D)^3
$&$(3 2 2 2 0 0)
$&${18''}$&$(1,1,1,1,|4,6)    $&$(E_8)(E_6)^2(D)
$&$(6 3 3 2 0 0)              $\\
\hline
${4'} $&$(1,1,1,1,|1,4)       $&$(E_7)(D)^2(A)
$&$(4 2 2 1 0 0)
$&${18'''}$&$(1,1,1,1|,4,6)  $&$(E_8)^2(A)^2
$&$(6 6 1 1 0 0)             $\\
\hline
${ 5} $&$(1,1,1,1,|1,5)       $&$(E_7)(D)^3
$&$(4 2 2 2 0 0)
$&${19}$&$(1,1,1,1,|4,7)     $&$(E_7)^3(E_6)
$&$(4 4 4 3 0 0)             $\\
\hline \hline
${ 6} $&$(1,1,1,1,|2,2)      $&$(E_6)^2(A)^2
$&$(3 3 1 1 0 0)
$&${19'}$&$(1,1,1,1,|4,7)      $&$(E_8)(E_7)^2(A)
$&$(6 4 4 1 0 0)             $\\
\hline
${7}  $&$(1,1,1,1,|2,3)      $&$(E_6)^2(D)(A)
$&$(3 3 2 1 0 0)
$&${19''}$&$(1,1,1,1,|4,7)     $&$(E_8)^2(D)(A)
$&$(6 6 2 1 0 0)             $\\
\hline
${7'} $&$(1,1,1,1,|2,3)      $&$(E_7)(E_6)(A)^2
$&$(4 3 1 1 0 0)
$&${ 20}$&$(1,1,1,1,|4,8)      $&$(E_7)^4
$&$(4 4 4 4 0 0)             $\\
\hline
${7''} $&$(1,1,1,1,|2,3)     $&$(E_8)^2(A)^3
$&$(6 1 1 1 0 0)
$&${ 20'}$&$(1,1,1,1,|4,8)     $&$(E_8)(E_7)^2(D)
$&$(6 4 4 2 0 0)             $\\
\hline
${ 8} $&$(1,1,1,1,|2,4)      $&$(E_6)^2(D)^2
$&$(3 3 2 2 0 0)
$&${ 20''}$&$(1,1,1,1,|4,8)    $&$(E_8)^2(D)^2
$&$(6 6 2 2 0 0)             $\\
\hline
${8'} $&$(1,1,1,1,|2,4)      $&$(E_7)^2(A)^2
$&$(4 4 1 1 0 0)
$&${ 21}$&$(1,1,1,1,|5,6)    $&$(E_8)(E_6)^3
$&$(6 3 3 3 0 0)             $\\
\hline
${8''} $&$(1,1,1,1,|2,4)     $&$(E_8)(D)(A)^2
$&$(6 2 1 1 0 0)
$&${ 22}$&$(1,1,1,1,|5,7)    $&$(E_8)(E_7)(E_6)^2
$&$(6 4 3 3 0 0)             $\\
\hline
${ 9} $&$(1,1,1,1,|2,5)      $&$(E_7)^2(D)(A)
$&$(4 4 2 1 0 0)
$&${ 22'}$&$(1,1,1,1,|5,7)   $&$(E_8)(E_7)^2(D)
$&$(6 4 4 2 0 0)             $\\
\hline
${9'} $&$(1,1,1,1,|2,5)      $&$(E_7)(E_6)(D)^2
$&$(4 3 2 2 0 0)
$&${ 22''}$&$(1,1,1,1,|5,7)  $&$(E_8)^2(E_6)(A)
$&$(6 6 3 1 0 0)             $\\
\hline
${9''} $&$(1,1,1,1,|2,5)     $&$(E_8)(D)^2(A)
$&$(6 2 2 1 0 0)
$&${ 23}$&$(1,1,1,1,|5,8)    $&$(E_8)(E_7)^2(E_6)
$&$(6 4 4 3 0 0)             $\\
\hline
${10}$&$(1,1,1,1,|2,6)       $&$(E_7)^2(D)^2
$&$(4 4 2 2 0 0)
$&${ 23'}$&$(1,1,1,1,|5,8)   $&$(E_8)^2(E_7)(A)
$&$(6 6 4 1 0 0)             $\\
\hline
${10'}$&$(1,1,1,1,|2,6)      $&$(E_8)(D)^3
$&$(6 2 2 2 0 0)
$&${ 23''}$&$(1,1,1,1,|5,8)  $&$(E_8)^2 (E_6)(D)
$&$(6 6 3 2 0 0)             $\\
\hline
${ 11}$&$(1,1,1,1,|3,3)      $&$(E_6)^3(A)
$&$(3 3 3 1 0 0)
$&${ 24}$&$(1,1,1,1,|5,9)    $&$(E_8) (E_7)^3
$&$(6 4 4 4 0 0)             $\\
\hline
${12}$&$(1,1,1,1,|3,4)       $&$(E_6)^3(D)
$&$(3 3 3 2 0 0)
$&${ 25}$&$(1,1,1,1,|6,8)    $&$(E_8)^2 (E_6)^2
$&$(6 6 3 3 0 0)             $\\
\hline
${12'}$&$(1,1,1,1,|3,4)      $&$(E_7)(E_6)^2(A)
$&$(4 3 3 1 0 0)
$&${ 26}$&$(1,1,1,1,|6,9)    $&$(E_8)^2 (E_7)(E_6)
$&$(6 6 4 3 0 0)             $\\
\hline
${12''}$&$(1,1,1,1,|3,4)     $&$(E_8)(E_6)(A)^2
$&$(6 3 1 1 0 0)
$&${ 26'}$&$(1,1,1,1,|6,9)   $&$(E_8)^3 (A)
$&$(6 6 6 1 0 0)             $\\
\hline
${13}$&$(1,1,1,1,|3,5)       $&$(E_7)^2(E_6)(A)
$&$(4 4 3 1 0 0)
$&${ 27}$&$(1,1,1,1,|6,10)   $&$(E_8)^3 (D)
$&$(6 6 6 2 0 0)             $\\
\hline
${13'}$&$(1,1,1,1,|3,5)      $&$(E_8)(E_6)(D)(A)
$&$(6 3 2 1 0 0)
$&${27'}$&$(1,1,1,1,|6,10)   $&$(E_8)^2 (E_7)2
$&$(6 6 4 4 0 0)             $\\
\hline
${13''}$&$(1,1,1,1,|3,5)     $&$(E_8)(E_7)(A)^2
$&$(6 4 1 1 0 0)
$&${ 28}$&$(1,1,1,1,|7,10)   $&$(E_8)^3(E_6)
$&$( 6 6 6 3 0 0 )           $\\
\hline
${ 14}$&$(1,1,1,1,|3,6)      $&$(E_7)^3(A)
$&$(4 4 4 1 0 0)
$&${ 29}$&$(1,1,1,1,|7,11)   $&$(E_8)^3(E_7)
$&$( 6 6 6 4 0 0 )           $\\
\hline
${14'}$&$(1,1,1,1,|3,6)      $&$(E_8)(E_7)(D)(A)
$&$(6 4 2 1 0 0)
$&${ 30}$&$(1,1,1,1,|8,12)   $&$(E_8)^4
$&$( 6 6 6 6 0 0 )           $\\
\hline
${ 15}$&$(1,1,1,1,|3,7)      $&$(E_7)^3(D)
$&$(4 4 4 2 0 0)
$&$   $&$                    $&$
$&$                          $\\
\hline
${15'}$&$(1,1,1,1,|3,7)      $&$(E_8)(E_7)(D)^2
$&$(6 4 2 2 0 0)
$&$   $&$                    $&$
$&$                          $\\
\hline  \hline
\end{tabular}
\end{table}
}
%\normalsize

Along the $E_r$ line, the structures of the $CY_d$ surfaces are described
by cubic and quartic monomials, and sometimes pairs of conics, as we
illustrated in the cases of $K3$ and $CY_3$, and show more concretely in
the next Section. There we use Diophantine relations between the cubic and
quartic IMs to obtain complete information about the chains and eldest
vectors  (see Fig. \ref{diof}). The method of IMs is independent of Batyrev's 
reflexive polyhedra, and has an exclusively algebraic character.

\section{The Monomial Decomposition Route to the Universal Calabi-Yau 
Algebra}

\begin{figure}[th!]
   \begin{center}
   \mbox{
   \epsfig{figure=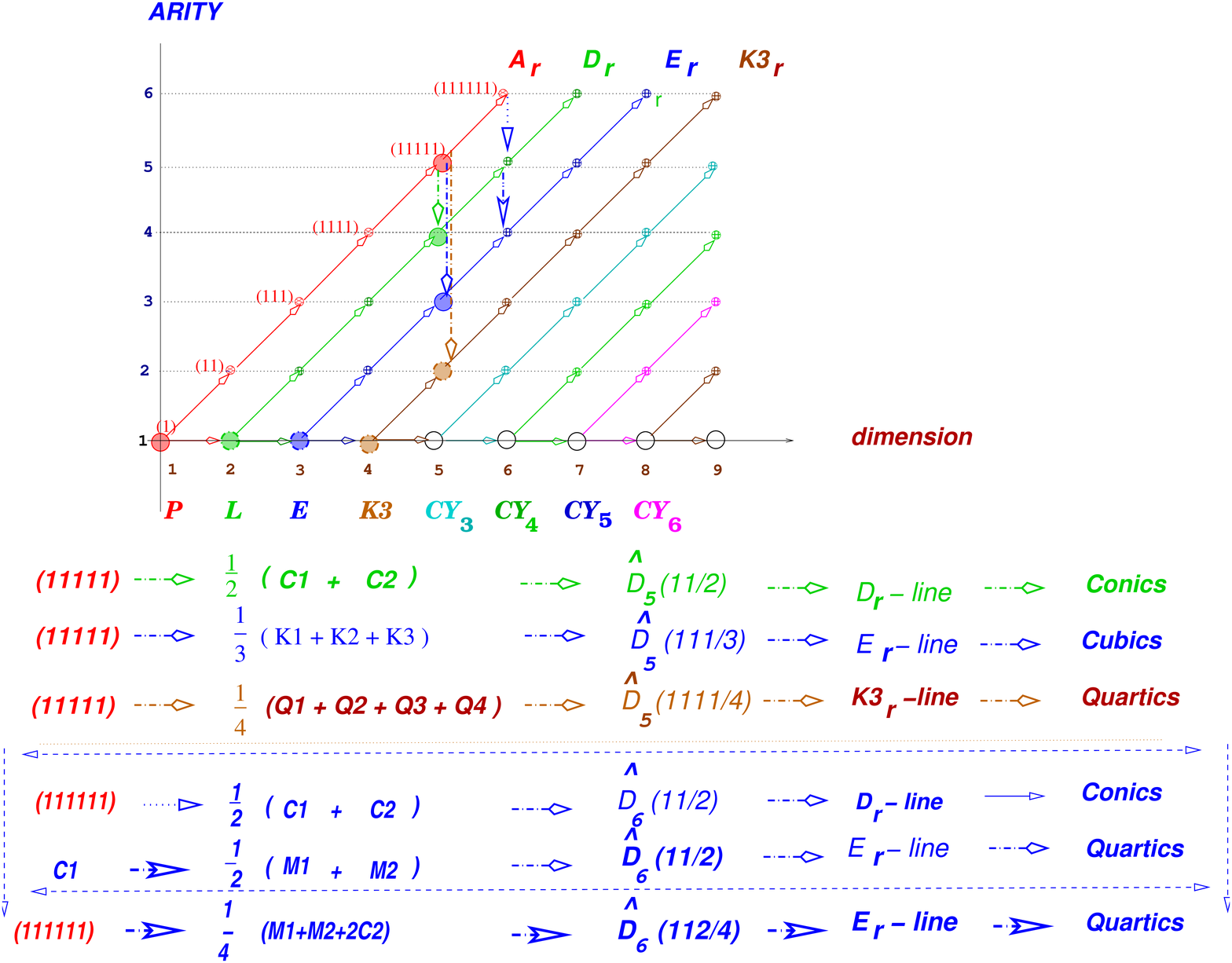,height=16cm,width=18cm}}
   \end{center}
   \caption{\it 
Examples of the Diophantine decompositions of unit monomials, illustrating 
how conics, cubics and quartics may be used to construct spaces with lower 
arities.}
\label{diof}
\end{figure}

We have seen that the Universal Calabi-Yau Algebra (UCYA) systematizes all
the $CY_d$ manifolds according to the arity of their construction as well
as their dimensionality. In this way, the UCYA enables us to explore
systematically relations between different $CY_d$ spaces, as well as their
internal structures, such as the singularities which are related to
possible new gauge-group structures. In the previous section, we
concentrated on the normal expansion technique, related to Batyrev's reflexive
polyhedron method, that inter-relates $CY_d$ spaces in different numbers
of dimensions. We also gave some examples of the extra information that can be
gleaned by decomposing IMs. In this Section, we explore further this
technique based on the  Diophantine expansion of invariant monomials (IMs), 
illustrated in Fig.~\ref{diof}, that relates spaces with
different arities. As we show later in more detail, this algebraic method
uses the Diophantine properties of the equations defining
Calabi-Yau spaces. It enables us to construct and enumerate the chains of
$CY_d$ spaces that we discussed previously. Moreover, we find
recurrence relations for the numbers of chains with different arities in
different dimensions.

%\newpage 

\subsection{Analysis of the $A_r$ and $D_r$ Lines}

We start with some simple examples, taken from the leading diagonal $A_r$
line, corresponding to the maximal arity
$r = n$. As we have already discussed, the IMs along this line have a
very simple structure, since in each dimension $n$ there is just one such
monomial, the unit $E_n=(1,...,1)_n$: see Table~\ref{Tabquad}. To find
the chain of $CY_d$ spaces corresponding to each such IM, we should solve
the equation

\begin{equation}
\vec{k}^{i(ex)} \cdot E_n=d(\vec{k}^{i(ex)})
\end{equation}
which immediately gives us
\begin{equation}
(\vec{k}^{i(ex)})_1+ \ldots +(\vec{k}^{i(ex)})_n=d(\vec{k}^{i(ex)}),
\end{equation}
where $(\vec{k}^{i(ex)})_1$,..., $(\vec{k}^{i(ex)})_1$ are the $n$
components of the extended vector $\vec{k}^{i(ex)}$. In general, such an
equation has an infinite number of solutions, but these are restricted to
a finite number by the fact that in the UCYA the extended vectors along
this diagonal line can be constructed only by extending the unit vector
$\vec{k}_1=(1)$ to dimension $n$. In this simple
case the Diophantine equation can immediately be rewritten in the
following form:
\begin{equation}
(\vec{k}^{i(ex)})_1+ \ldots (\vec{k}^{i(ex)})_n=1,
\end{equation}
which gives exactly $n$ different extended vectors:
$(\vec{k}^{i(ex)})_j=\delta^i_j$, where $ i,j=1,2,...,n$.  These $n$
different solutions $\vec{k}^{i(ex)}, (i=1,...,n)$, actually produce
only one chain of arity $n$, with the eldest vector
\begin{equation}
\vec{k}_n=\cup_{i=1}^{i=n} (0,...0,1_i,0,...)=(1,...,1).
\end{equation}
One can check this result by the previous method of looking for the
intersection of these $n$ extended vectors:
\begin{equation}
\cap_{i=1}^{i=n} (0,...0,1_i,0,...) =(1')_n.
\end{equation}
Thus, one can see that the unit IM in any dimension $n$ uniquely
determines a single chain of arity $n$ with unit reflexive eldest vector. 

It is a consequence of the UCYA that all RWVs can be
decomposed in terms of the extended vectors of this $A_r$ line:
\begin{equation}
\vec{k}_n=\sum_{i=1}^{i=n} a_i \vec{k}_1^{i(ex)},
\end{equation}
with some non-negative coefficients $a_i$. The problem of finding the
coefficients in this decomposition is connected directly with the internal
structure of the $CY_d$ spaces. For instance, if the $CY_d$ space had only
a simple substructure related just to the unit invariant monomial
$E_n$~\footnote{The single central point in one of the Batyrev
polyhedra.}, one might think that the complete set of $CY_d$ spaces could
be found by simple extension of this unit $E_n$ monomial to $n$ monomials
$P_i$, with the property $(1/n)(P_1+...+P_n)=E_n$, in which case the genome
for all $CY_d$ spaces would be solved. However, fortunately $CY_d$ spaces
have a much richer and more intriguing internal substructure, and such a
direct Diophantine expansion cannot give the full number and
substructure of all the $CY_d$ spaces. The Diophantine expansions of the
unit monomials $E_4$ in the $K3$ case and $E_5$ in the $CY_3$ case give
only the {42} and {7,269} {RWV}s in these two chains, respectively. 

Due to the complicated internal structure of $CY_d$ spaces, the Diophantine
expansions can in many cases can be degenerate. In these cases, the $E_n$
cannot simply be expanded in terms of $n$ points $P_i: i=1,...,n$, with
the property $E_n=(1/n)(P_1+...+P_n)$. However, some other weaker conditions
could still apply:
\begin{eqnarray}
E_n \, \mapsto \{P_1,...,P_{(n-r+1)}| 
E_n=\frac{1}{(n-r+1)}(P_1+\ldots +P_{n-r+1})\}
\end{eqnarray}
for $r = (n - 1), (n - 2), \ldots, 3, 2.$ This is another way to see why we 
must go on to
study further the substructure of the $CY_d$ spaces, corresponding to the
arities $r = (n - 1), (n - 2), ..., 3, 2$, i.e., we should go on to study
points on the next lines on the arity-dimension plot. The corresponding
$CY_d$ spaces will have substructures described by higher-degree
monomials, such as conics, cubics, quartics, quintics, etc.. For example, 
the $CY_d$ spaces with substructures corresponding to the $E_r$
line have elliptic fibres, whose substructure is described by cubic and
quartic monomials. The $CY_d$ spaces with richer substructure,
corresponding to the next (fourth) line, have $K3$ fibres, and are
described by the properties of quintic, sextic, septic and octic
monomials. Some of these $CY_d$ spaces also have a hyperelliptic
substructure.

These arguments show that, in order to study the structures of the chains
of fixed arity and dimension appearing along the next lines: $D_r$,
$E_r$, ..., we should first determine the types of IMs corresponding to
these lines. Next, we should find suitable sets of extended vectors by
solving the following equations:
\begin{equation}
 \vec{k}^{i(ex)} \cdot (IM)_a=d(\vec{k}^{i(ex)}).
\end{equation}
In looking for the solutions of these equations, one may use the fact 
that, in the UCYA, each sloping line is described by a particular type of 
extended weight vectors.

We consider the $CY_d$ spaces with substructures corresponding to
the second $D_r$ line with arity $r = (n - 1)$. We have the following
sets of possible different types of conic monomials in any dimension $n$,
as given in the Table~\ref{Tabquad}, which can be generalized by a
recurrence relation to any dimension $n$:
\begin{equation}
N_{conics}=\frac{(n)(n-1)}{2}.
\end{equation}
We propose to determine all the possible conic IMs for this line, starting
from the unit monomial $E_n$ and two conic monomials, $C_{i(n)}$ and
$C_{j(n)}$, taken from Table~\ref{Tabquad}~\footnote{In this Table, one
can use all possible permutations of the expressions for the conic
monomials.}. These monomials should satisfy the following Diophantine
property: $r = (n - 1)$:
\begin{equation}
C_{i(n)}+C_{j(n)}=\frac{1}{2}E_n,
\end{equation}
where the index $n$ notes the dimension being considered.

{\scriptsize
\begin{table}[!ht]
\centering
\caption{\it The invariant monomials (linear and quadratic) for the
(1)+(11)
RWVs extended to Weierstrass, $K3$, $CY_3$ and $CY_4$ spaces, 
corresponding to the $D_r$ line on the arity-dimension plot \ref{basmon1}.
In the case of the $A_r$ line, there can be only linear invariant 
monomials.}
\label{Tabquad}
\vspace{.05in}
\begin{tabular}{||c||c||c||c|c||c|c|c||c|c|c|c||}
\hline
$   P                
$&$ L        
$&$ W          
$&$ K3               $&$ K3
$&$ Qu               $&$ in       $&$tic 
$&$ Se               $&$ x        $&$t         $&$ ic 
$\\ \hline\hline
$   (1)    
$&$ (11)             
$&$ (111)
$&$ (1111)           $&$ 
$&$ (11111)          $&$          $&$
$&$ (111111)         $&$          $&$          $&$  
$\\ \hline\hline
$
$&$ (20)
$&$ (200)      
$&$ (2000)           $&$ 
$&$ (20000)          $&$          $&$
$&$ (200000)         $&$          $&$          $&$  
$\\ \hline
$
$&$
$&$ (210)
$&$ (2100)           $&$ (2110)
$&$ (21000)          $&$ (21100)  $&$ (21110)
$&$ (210000)         $&$ (211000) $&$ (211100) $&$(211110) 
$\\ \hline
$
$&$
$&$ (220)
$&$ (2200)           $&$ (2210) 
$&$ (22000)          $&$ (22100)  $&$ (22110) 
$&$ (220000)         $&$ (221000) $&$ (221100) $&$(221110) 
$\\ 
$
$&$
$&$
$&$                  $&$ (2220)
$&$                  $&$ (22200)  $&$ (22210)
$&$                  $&$ (222000) $&$ (222100) $&$(222110)
$\\
$ 
$&$
$&$
$&$                  $&$
$&$                  $&$          $&$ (22220)
$&$                  $&$          $&$ (222200) $&$(222210)
$\\
$ 
$&$
$&$
$&$                  $&$
$&$                  $&$          $&$
$&$                  $&$          $&$          $&$ (222220)
$\\ \hline \hline
\end{tabular}
\end{table}
}
%\normalsize 

In order to enumerate these chains, we should find all possible
pairs of conic monomials with the required
Diophantine property, solving the following equations: 
\begin{equation}
\vec{k}^{i(ex)} \cdot C_{1(n)}=\vec{k}^{i(ex)} \cdot
E_n=d(\vec{k}^{i(ex)}).  
\end{equation} 
To give sense to these equations
and, consequently, to evaluate the finite number of solutions for the
chains and their eldest vectors corresponding to the arity
$r = (n - 1)$, one should recall that, in the UCYA, the points on this
line are determined by $n$-dimensional extensions of the two eldest
vectors $\vec{k}_1=(1)$ and $\vec{k}_2=(1,1)$.  This means that the
possible values of $d(\vec{k}^{i(ex)})$ in these equations are only 1 and
2. Also, the components of the extended vectors can only be 0, 1 or 2. The
$n$-dimensional recurrence relation for conic monomials allows one to find
all the IMs and to solve this problem for all dimensions $n$, in the sense
of finding all the corresponding eldest vectors. The recurrence formula
for the numbers of chains along this line is: 
\begin{eqnarray}
N_{chains}&=&k\cdot(k+1), \qquad if \;\,\,n=(2k+1) \nonumber\\ N_{chains}&=&
k^2, \qquad if \;\,\,n=(2k). \nonumber\\ 
\end{eqnarray} 
Accordingly, along
this line the numbers of the eldest vectors and chains in the dimensions
$n=2, 3, 4, ...$ are the following: 1, 2, 4, 6, 9, 12, 16, 20, 25, 30, 36,
42, 49, 56, 64, 72, 81, 90, 100, 110, 121, 132, 144, .... 

In order to understand this recurrence formula, one can construct the list
of all different pairs of conics that appear in any dimension, as
illustrated in
Table~\ref{Tabquad}. For example, for $CY_3$ ($n = 5$) one can get $6 = 2
\cdot 3$ chains of arity $4$, for $CY_4$ ($n = 6 = 2k$) there are $9 =
3^2$ chains of arity $5$ and for $CY_5$ ($n = 7 = 2k+1, k = 3$) there are
$12 = 3 \cdot 4$ chains of arity $6$. Thus, one of the main results of the
IM approach for the first two diagonal lines, $A_r$ and $D_r$, is to solve
this aspect of the $CY_d$ `genome' in any dimension.

There is another interesting observation connected to the one-to-one
correspondence between the IMs and chains. As an example, consider the
following pair of $n$-dimensional conics:
\begin{equation}
C_{1(n)}=(2,2,2,\ldots ,2,0),\qquad C_{2(n)}=(0,0,0,\ldots ,0,2).
\end{equation}
The solutions of the equations
\begin{equation}
\vec{k}^{i(ex)} \cdot C_{1(n)}=\vec{k}^{i(ex)} \cdot E_n=d(\vec{k}^{i(ex)}). 
\end{equation}
can be expressed in the following form:
\begin{equation}
(\vec{k}^{i(ex)})_1+\ldots +(\vec{k}^{i(ex)})_{n-1}=1,\qquad 
(\vec{k}^{i(ex)})_{n}=1.
\end{equation}
So we get exactly $(n-1)$ different solutions for the extended vectors,
producing one chain of arity $r = (n - 1)$:
\begin{eqnarray}
\vec{k}^{1(ex)}  &=&(1,0,0,\ldots, 0,0,1)\nonumber\\
\vec{k}^{2(ex)}  &=&(0,1,0,\ldots, 0,0,1)\nonumber\\
... &=&\ldots\nonumber\\
\vec{k}^{n-2(ex)}&=&(0,0,0,\ldots, 1,0,1)\nonumber\\
\vec{k}^{n-1(ex)}&=&(0,0,0,\ldots, 0,1,1)\nonumber\\
\end{eqnarray}
with the eldest vector $\vec{k}_n=(1,1,1,\ldots,1,1,n-1)$.

To find the sets of RWVs inside the chains, one may study the following
Diophantine condition for the conics:
\begin{equation}
C_{1(n)}\, \mapsto \, \{P_1, \ldots ,P_{(n-1)}|
C_{1(n)}=\frac{1}{(n-1)}(P_1+\ldots P_{(n-1)}) \}
\end{equation}
and then solve the following $n$ equations for the eldest vectors:
\begin{eqnarray}
\vec{k}^{i(ex)} \cdot P_j\, &=&\vec{k}^{i(ex)} \cdot C_{2(n)}\,=
d(\vec{k}^{i(ex)}),\nonumber\\
C_{1(n)}\,+\,C_{2(n)}\,&=&\,\frac{1}{2}E_n.
\end{eqnarray}
In the case of $K3$ chains, we can compare the list of the RWVs obtained
from the Diophantine expansion with the complete list, which contains {34}
and {48} possibilities, respectively. We can see that, in this case, there
are {14} RWVs that cannot be obtained directly by the expansion of conics
$C_{1(n)}$ in terms of order-$(n-1)$ monomials.  Therefore, we should
study the possibilities for expanding in terms of order-$(n-2), (n-3),
...$ monomials, going to the next diagonal lines of the arity-dimension
plot. Due to the more complicated structure of $CY_{d = (n - 2)}$ spaces
with other arities $r = (n - 2), ..., 3, 2$, corresponding to the
third $E_r$ line, the fourth line, etc., the Diophantine condition for the
conics is not closed.

\subsection{Analysis of the $E_r$ Line.}

As the next step, we consider the situation along the diagonal $E_r$
line.

\subsubsection{Recurrences of Sets of Invariant Monomials}

Continuing our previous approach, the first
step is to enumerate the cubic and quartic monomials, from which we can
find all the IMs along this $E_r$ line. The appearance of cubic monomials
is connected with the following new Diophantine condition for the
expansion of the unit monomials $E_n$ of the $A_r$ line:
\begin{eqnarray}
E_n \mapsto \{P_1,P_2,P_3|\frac{1}{3}(P_1+P_2+P_3)=E_n\}.
\end{eqnarray}
However, the set of appropriate cubic monomials is somewhat more
restricted. Similarly, the appearence of quartic monomials is connected
with the possible Diophantine expansion of the conic monomials $C{i(n)}$
of the second $D_r$ line:
\begin{eqnarray}
C_{i(n)}\, \mapsto\, \{P_1,P_2|\frac{1}{2}(P_1+P_2)=C_{i(n)}\}.
\end{eqnarray}
Again, there are some further restrictions on the list of possible quartic
monomials. We now give the list of cubic and quartic monomials that are
relevant in the framework of the UCYA. 

%\newpage
{\scriptsize
\begin{table}[!ht]
\centering
\caption{ \it The invariant monomials
 (cubics and quartics) for the five weight vectors
(1)+(11)+(111)+(112)+(123),
extended to Weierstrass, $K3$, $CY_3$ and $CY_4$ spaces, corresponding 
to the $E_r$ line on the arity-dimension plot.}
\label{Tabcub}
\vspace{.05in}
\begin{tabular}{||c||c|c|c||c|c|c|c||c|c|c|c|c||}
\hline\hline
$   W             
$&$ K3            $&$  K3               $&$  N
$&$ Quin          $&$  ti               $&$  c        $&$ N 
$&$ Se            $&$  xt               $&$  i        $&$ c
$&$ N
$\\ \hline\hline    
$   (300)   
$&$ (3000)        $&$                   $&$  1
$&$ (30000)       $&$                   $&$           $&$ 1 
$&$ (300000)      $&$                   $&$           $&$
$&$  1
$\\ \hline
$  (310)
$&$(3100)         $&$ (3110)            $&$ 2
$&$(31000)        $&$ (31100)           $&$(31110)    $&$ 3
$&$(310000)       $&$ (311000)          $&$(311100)   $&$(311110) 
$&$ 4
$\\ \hline      
$  
$&$(3200)         $&$ (3210)            $&$
$&$(32000)        $&$ (32100)           $&$(32110)    $&$
$&$(320000)       $&$ (321000)          $&$(321100)   $&$(321110) 
$&$ 
$\\
$ 
$&$               $&$                   $&$
$&$               $&$ (32200)           $&$(32210)    $&$ 
$&$               $&$ (322000)          $&$(322100)   $&$(322110) 
$&$
$\\    
$             
$&$               $&$                   $&$  2       
$&$               $&$                   $&$           $&$ 5
$&$               $&$                   $&$(322200)   $&$(322210) 
$&$ 9
$\\ \hline
$
$&$ (3300)        $&$ (3310)            $&$
$&$(33000)        $&$ (33100)           $&$(33110)    $&$ 
$&$(330000)       $&$ (331000)          $&$(331100)   $&$(331110) 
$&$
$\\
$
$&$               $&$                   $&$
$&$               $&$(33200)            $&$(33210)    $&$  
$&$               $&$(332000)           $&$(332100)   $&$(332110) 
$&$
$\\ 
$                 
$&$               $&$                   $&$
$&$               $&$                   $&$           $&$
$&$               $&$                   $&$(332200)   $&$(332210) 
$&$
$\\ 
$               
$&$               $&$                   $&$
$&$               $&$ (33300)           $&$(33310)    $&$  
$&$               $&$ (333000)          $&$(333100)   $&$(333110) 
$&$
$\\
$                  
$&$               $&$                   $&$ 
$&$               $&$                   $&$           $&$ 
$&$               $&$                   $&$(333200)   $&$(333210) 
$&$
$\\ 
$                 
$&$               $&$                   $&$ 2
$&$               $&$                   $&$           $&$ 7
$&$               $&$                   $&$(333300)   $&$(333310) 
$&$ 16             
$\\  \hline \hline 
$   (400)          
$&$ (4000)        $&$                   $&$ 1
$&$ (40000)       $&$                   $&$           $&$ 1 
$&$ (400000)      $&$                   $&$           $&$
$&$ 1
$\\ \hline 
$                
$&$ (4100)        $&$ (4110)            $&$  2
$&$ (41000)       $&$ (41100)           $&$(41110)    $&$ 3
$&$ (410000)      $&$ (411000)          $&$(411100)   $&$(411110) 
$&$  4
$\\ \hline
$              
$&$ (4200)     $&$ (4210)               $&$   
$&$ (42000)    $&$ (42100)              $&$ (42110)   $&$ 
$&$ (420000)   $&$ (421000)             $&$ (421100)  $&$(421110) 
$&$
$\\ 
$              
$&$            $&$                      $&$ 
$&$            $&$ (42200)              $&$(42210)    $&$
$&$            $&$ (422000)             $&$(422100)   $&$(422110) 
$&$
$\\  
$             
$&$            $&$                      $&$ 2
$&$            $&$                      $&$           $&$ 5
$&$            $&$                      $&$(422200)   $&$(422210) 
$&$ 9
$\\ \hline
$              
$&$ (4300)     $&$ (4310)               $&$ 
$&$ (43000)    $&$ (43100)              $&$(43110)    $&$ 
$&$ (430000)   $&$ (431000)             $&$(431100)   $&$(431110) 
$&$
$\\ 
$              
$&$            $&$                      $&$ 
$&$            $&$ (43200)              $&$(43210)    $&$ 
$&$            $&$ (432000)             $&$(432100)   $&$(432110) 
$&$
$\\ 
$
$&$            $&$                      $&$ 
$&$            $&$                      $&$           $&$ 
$&$            $&$                      $&$(432200)   $&$(432210) 
$&$
$\\
$               
$&$            $&$                      $&$
$&$            $&$ (43300)              $&$(43310)    $&$ 
$&$            $&$ (433000)             $&$(433100)   $&$(433110) 
$&$
$\\ 
$              
$&$            $&$                      $&$
$&$            $&$                      $&$           $&$
$&$            $&$                      $&$(433200)   $&$(433210) 
$&$
$ \\ 
$              
$&$            $&$                      $&$ 2
$&$            $&$                      $&$           $&$ 7
$&$            $&$                      $&$(433300)   $&$(433310)
$&$ 16
$ \\ \hline
$              
$&$ (4400)     $&$ (4410)               $&$ 
$&$ (44000)    $&$ (44100)              $&$(44110)    $&$ 
$&$ (440000)   $&$ (441000)             $&$(441100)   $&$(441110) 
$&$
$\\ 
$              
$&$            $&$                      $&$ 
$&$            $&$ (44200)              $&$(44210)    $&$ 
$&$            $&$ (442000)             $&$(442100)   $&$(442110) 
$&$
$\\  
$             
$&$            $&$                      $&$
$&$            $&$                      $&$           $&$
$&$            $&$                      $&$(442200)   $&$(422210) 
$&$
$ \\
$               
$&$            $&$                      $&$
$&$            $&$ (44300)              $&$(44310)    $&$ 
$&$            $&$ (443000)             $&$(443100)   $&$(443110) 
$&$
$\\
$               
$&$            $&$                      $&$          
$&$            $&$                      $&$           $&$ 
$&$            $&$                      $&$(443200)   $&$(443210) 
$&$
$ \\
$               
$&$            $&$                      $&$
$&$            $&$                      $&$           $&$
$&$            $&$                      $&$(433300)   $&$(433310) 
$&$
$ \\
$               
$&$            $&$                      $&$
$&$            $&$ (44400)              $&$(44410)    $&$
$&$            $&$ (444000)             $&$(444100)   $&$(444110) 
$&$
$ \\
$               
$&$            $&$                      $&$ 
$&$            $&$                      $&$           $&$
$&$            $&$                      $&$(444200)   $&$(444210) 
$&$
$ \\ 
$              
$&$            $&$                      $&$ 
$&$            $&$                      $&$           $&$
$&$            $&$                      $&$(444300)   $&$(444310) 
$&$
$ \\ 
$              
$&$            $&$                      $&$ 2 
$&$            $&$                      $&$           $&$ 9
$&$            $&$                      $&$(444400)   $&$(444410) 
$&$ 25
$ \\ \hline\hline
\end{tabular}
\end{table}
}
%\normalsize

\begin{figure}[th!]
   \begin{center}
   \mbox{
   \epsfig{figure=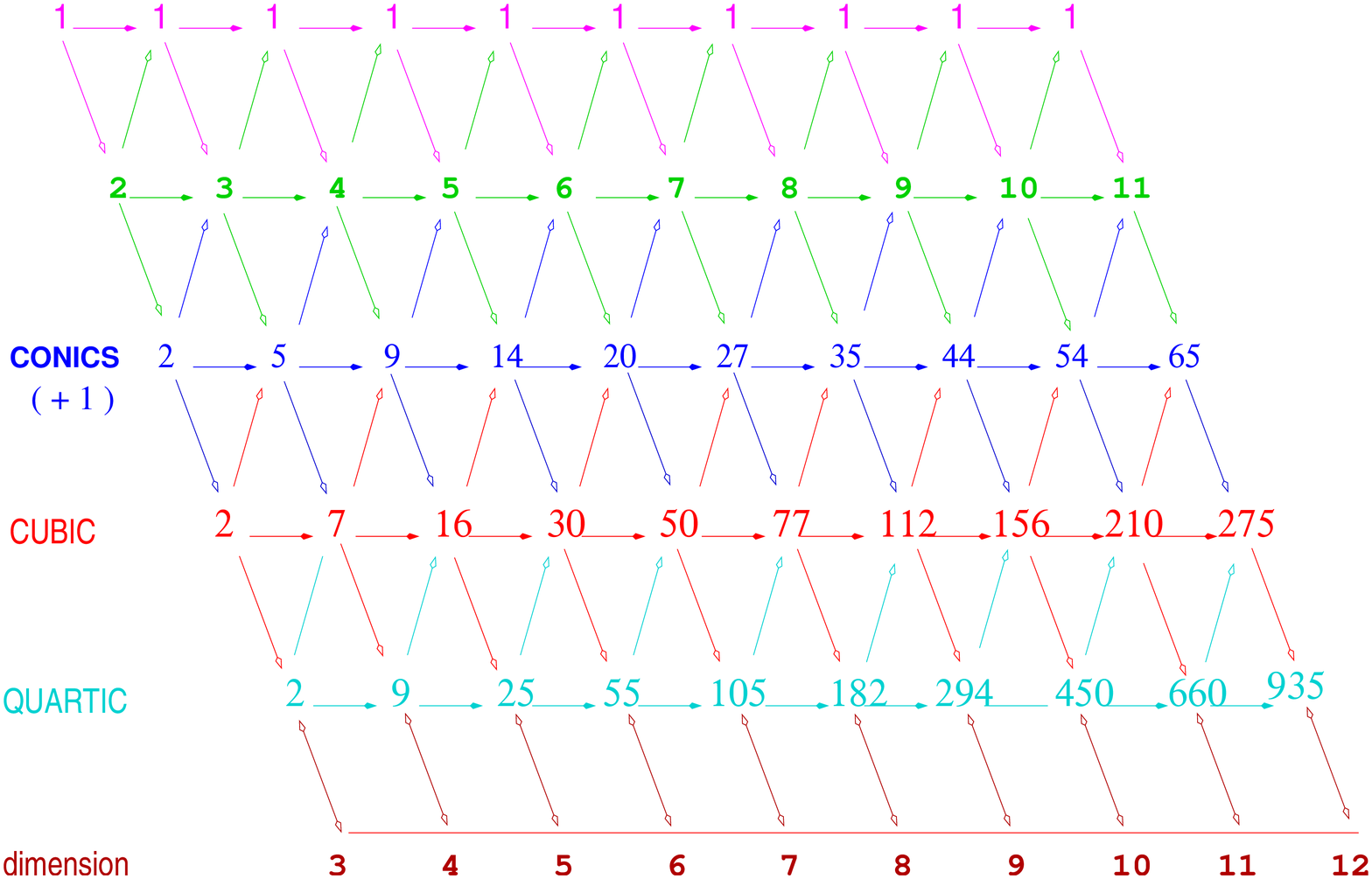,height=16cm,width=18cm}}
   \end{center}
   \caption{\it 
Lattice illustrating recurrence relations for the numbers 
of conic, cubic and quartic monomials.}
\label{recur}
\end{figure}

Examining Tables~\ref{Tabcub} and \ref{Tabsix}, one can convince oneself 
that there 
exist
recurrence formulae for the IMs in any dimension. 
These formulae are obvious for the leading (red and green) lines.
The expressions for the numbers of cubic and quartic monomials are,
respectively:
\begin{eqnarray}
N_{cubics}&=&\frac{1}{6}(n-2) (n-1) (n+3)\nonumber\\
N_{quartics}&=&\frac{1}{24} (n-2)(n-1)(n)(n+5)\nonumber\\
\end{eqnarray}
There are remarkable links between the numbers of conics, cubics and
quartics. For example, to obtain the number of quartics in dimension $n$,
one should sum over all the cubics in dimensions $3, 4, ..., n$, i.e.,
$N_{Quart}^{(n)}={\sum_i=3^{i=n}} N_{Cub}^{(i)}$. Thus, as seen in
Fig.~\ref{recur}, the number 105 of quartic monomials in the septic
Calabi-Yau case can be represented as follows: $2_{dim=3}+7_{dim=4}+
16_{dim=5}+ 30_{dim=6}+50_{dim=7}$.

%\newpage 

Based on these tables, one can convince oneself that there should also
exist $n$-dimensional recurrence formulae for the IMs applicable along 
other diagonal lines in any dimension, as
we have found for the first two lines on the arity-dimension plot in
Fig.~\ref{basmon1}. However, the situation can become complicated,
because, in the construction of the cubic and quartic IMs, one must also
take into account conic and conic + cubic monomials, respectively.
In the case of Calabi-Yau spaces with Weierstrass fibres, it is also
important to know the list of sextic monomials that we give in
Table~\ref{Tabsix}. The recurrence formula for the number of sextic 
monomials can be obtained from this Table:
\begin{eqnarray}
C_{n+2}^{n-3}= 
\frac{(n+2)!}{(n-3)!5!},
\end{eqnarray}
where $n \geq 3$ is the dimension of the weight-vector space.

{\scriptsize
\begin{table}[!ht]
\centering
\caption{ \it The sextic  monomials, $M_6$,
for the five weight vectors $(1) + (11) + (111) + (112) + (123)$,
extended to $K3$, $CY_3$ and $CY_4$ spaces, corresponding to the
$E_r$ line on the arity-dimension plot.}
\label{Tabsix}
%\scriptsize
\vspace{.05in}
\begin{tabular}{||c|c||c|c|c||c|c|c|c||}
\hline
$   K3            $&$  N 
$&$ Quin          $&$  tic              $&$   N 
$&$ Se            $&$  xt               $&$  ic        
$&$ N
$\\ \hline\hline       
$   (6000)        $&$ 1
$&$ (60000)       $&$                   $&$ 1 
$&$ (600000)      $&$                   $&$
$&$ 1
$\\ \hline                 
$   (6100)        $&$ 1 
$&$ (61000)       $&$ (61100)           $&$ 2
$&$ (610000)      $&$ (611000)          $&$(611100)   
$&$  3
$\\ \hline              
$   (6200)     $&$   
$&$ (62000)    $&$ (62100)              $&$ 
$&$ (620000)   $&$ (621000)             $&$ (621100)  
$&$
$\\               
$              $&$ 
$&$            $&$ (62200)              $&$
$&$            $&$ (622000)             $&$(622100)    
$&$
$\\               
$              $&$  1
$&$            $&$                      $&$ 3
$&$            $&$                      $&$(622200)   
$&$ 6 
$\\ \hline              
$   (6300)     $&$ 
$&$ (63000)    $&$ (63100)              $&$ 
$&$ (630000)   $&$ (631000)             $&$(631100) 
$&$
$\\               
$              $&$ 
$&$            $&$ (63200)              $&$ 
$&$            $&$ (632000)             $&$(632100)
$&$
$\\ 
$              $&$
$&$            $&$                      $&$ 
$&$            $&$                      $&$(632200) 
$&$
$\\               
$              $&$
$&$            $&$ (63300)              $&$ 
$&$            $&$ (633000)             $&$(633100) 
$&$
$\\               
$              $&$
$&$            $&$                      $&$
$&$            $&$                      $&$(633200)
$&$
$ \\               
$              $&$ 1
$&$            $&$                      $&$ 4
$&$            $&$                      $&$(633300)
$&$ 10
$ \\ \hline              
$   (6400)     $&$ 
$&$ (64000)    $&$ (64100)              $&$ 
$&$ (640000)   $&$ (641000)             $&$(641100)
$&$
$\\               
$              $&$
$&$            $&$ (64200)              $&$
$&$            $&$ (642000)             $&$(642100)
$&$
$\\               
$              $&$
$&$            $&$                      $&$
$&$            $&$                      $&$(642200)   
$&$
$ \\               
$              $&$
$&$            $&$ (64300)              $&$ 
$&$            $&$ (643000)             $&$(643100)   
$&$
$\\               
$              $&$          
$&$            $&$                      $&$ 
$&$            $&$                      $&$(643200)    
$&$
$ \\               
$              $&$
$&$            $&$                      $&$
$&$            $&$                      $&$(643300)    
$&$
$ \\               
$              $&$
$&$            $&$ (64400)              $&$
$&$            $&$ (644000)             $&$(644100)    
$&$
$ \\               
$              $&$  
$&$            $&$                      $&$
$&$            $&$                      $&$(644200)    
$&$
$ \\               
$              $&$               
$&$            $&$                      $&$    
$&$            $&$                      $&$(644300)    
$&$
$ \\               
$              $&$ 1                    
$&$            $&$                      $&$  5       
$&$            $&$                      $&$(644400)    
$&$ 15
$ \\ \hline
$   (6600)     $&$                       
$&$ (66000)    $&$ (66100)              $&$            
$&$ (660000)   $&$ (661000)             $&$(661100)    
$&$
$\\               
$              $&$                      
$&$            $&$ (66200)              $&$            
$&$            $&$ (662000)             $&$(662100)    
$&$
$\\               
$              $&$                      
$&$            $&$                      $&$           
$&$            $&$                      $&$(662200)    
$&$
$ \\               
$              $&$                      
$&$            $&$ (66300)              $&$           
$&$            $&$ (663000)             $&$(663100)   
$&$
$\\               
$              $&$                               
$&$            $&$                      $&$            
$&$            $&$                      $&$(663200)    
$&$
$ \\               
$              $&$                      
$&$            $&$                      $&$           
$&$            $&$                      $&$(663300)    
$&$
$ \\               
$              $&$                      
$&$            $&$ (66400)              $&$        
$&$            $&$ (664000)             $&$(664100)    
$&$
$ \\               
$              $&$                       
$&$            $&$                      $&$           
$&$            $&$                      $&$(664200)    
$&$
$ \\               
$              $&$                       
$&$            $&$                      $&$           
$&$            $&$                      $&$(664300)    
$&$
$ \\               
$              $&$                        
$&$            $&$                      $&$           
$&$            $&$                      $&$(664400)    
$&$ 
$ \\ 

$              $&$                        
$&$            $&$  (66600)             $&$           
$&$            $&$  (666000)            $&$(666100)    
$&$ 
$ \\ 
$              $&$                      
$&$            $&$                      $&$           
$&$            $&$                      $&$(666200)    
$&$ 
$ \\ 
$              $&$                      
$&$            $&$                      $&$           
$&$            $&$                      $&$(666300)    
$&$
$ \\
$              $&$                      
$&$            $&$                      $&$            
$&$            $&$                      $&$(666400)    
$&$ 
$ \\
$              $&$ 1                     
$&$            $&$                      $&$ 6           
$&$            $&$                      $&$(666600)    
$&$ 21
$ \\ \hline\hline
\end{tabular}
\end{table}
}
%\normalsize 

\subsubsection{Applications to $K3$ Spaces and their Fibrations}

We now consider the $K3$ case in more detail, studying the third $E_r$
line. From previous work, we already know that there are $22$ chains of
arity 2, a result we now rederive using the IM method. To obtain all the
chains and the corresponding eldest weight vectors, one needs to consider
three types of Diophantine expansions, as described above. The first
mechanism to consider is the expansion of the unit monomial in terms of
three monomials, whose maximal degree is 3:
\begin{eqnarray} 
E_4 \, \longrightarrow \{P_1,P_2,P_3:\frac{1}{3} \, (P_1+P_2+P_3) = E_4\}.
\end{eqnarray}
We see easily from Tables~\ref{Tabelda}, \ref{Tabeldb} and \ref{Tabeldc}
that there exist {28} cubic IMs, including the single unit monomial $E_4$ 
and three monomials $P_1,P_2,P_3$.

To find these chains, one should first solve the following equations:

\begin{equation}
\vec{k}^{i(ex)} \cdot (IM)_a=d(\vec{k}^{i(ex)}),
\end{equation}
where $a=1,2,3$. To solve these equations and to obtain {19} of the
chains, we should look for extended vector solutions
among the set of 4-dimensional extensions of the following RWVs: 

\begin{eqnarray}
\vec{k}_1=(1);\,\vec{k}_2=(1,1);\,
\vec{k}_3=(1,1,1),\,\vec{k}_3=(1,1,2),\,\vec{k}_3=(1,2,3).
\end{eqnarray}
The second way, that completes the structure of $K3$ spaces along this 
line, is to consider quartic monomials. This can be done in two steps, 
first expanding the unit monomial in terms of conics and then 
expanding the conics in terms of quartics:

\begin{eqnarray} 
&&E_4 \, \longrightarrow \, \{C_1,C_2: \frac{1}{2}(C_1+C_2)=E_4\} 
\nonumber\\
&&C_1  \, \longrightarrow \,\{P_1,P_2:\frac{1}{2} \, (P_1+P_2)=E_4\}.
\end{eqnarray}
We refer to this expansion as the Diophantine $D_4[1,1,2/4]$ expansion,
and call the expansion through cubics the $D_4[1,1,1/3]$ expansion. 

Twenty chains can be found this way by solving the following 
equations:

\begin{eqnarray}
\vec{k}^{i(ex)} \cdot P_{1}=\vec{k}^{i(ex)} \cdot P_{2}
=d(\vec{k}^{i(ex)}), \nonumber\\
\vec{k}^{i(ex)} \cdot C_2=d(\vec{k}^{i(ex)}), 
\end{eqnarray}
where we look again for solutions among the extended vectors.
The $D_4[1,1,1/3]$ and 
$D_4[1,1,2/4]$ expansions give {20} chains each, which combine to give 
the full {22} chains. The last type of Diophantine expansion 
is the double conic expansion, which is described by the
expression $D_4[1,1/2]+ D^\prime_4[1,1/2]$, i.e.:

\begin{eqnarray} 
&&E_4 \, \longrightarrow \, \{C_1,C_2: \frac{1}{2}(C_1+C_2)=E_4\} \nonumber\\
&&E_4 \, \longrightarrow \, \{C_3,C_4: \frac{1}{2}(C_3+C_4)=E_4\} \nonumber\\
\end{eqnarray}
Solving the corresponding equations leads to {14} such cases, which are 
contained among the 22 already found.

We thus see that the IM method yields the same 22 chains of $K3$ spaces
that were found previously via the normal expansion technique: indeed,
just two types of Diophantine expansion were sufficient. To understand
these expansions more deeply, look at Tables~\ref{Tabelda}, 
\ref{Tabeldb}, \ref{Tabeldc}. The monomials
$P_1,P_2,P_3$, corresponding to the $D_4[1,1,1/3]$ Diophantine 
expansion,
together with the unit monomial $E_4$, produce the full set of third-order
{IMs} for $K3$. The numbers {3} and {4} we term the degrees of the
corresponding {IMs}. The monomials $P_1,P_2,C_2$ together with the unit
monomial $E_4$ produce the set of fourth-degree {IM}s, and these Tables
show the different types of third- and fourth-degree {IM}s.
 
The third-degree {IM}s can be involved in the following four types of
triples of cubic and/or conic monomials: $(Cub)^3$,$(Cub)^2 (Con)$, $(Cub)
(Con)^2$ and $(Con)^3$. Similarly, one can characterize according to the
choice of $P_1$ and $P_2$ the following six types of fourth-degree {IM}s:
$(Qrt)^2$, $(Qrt) (Cub)$, $(Qrt) (Con)$, $(Cub)^2$, $(Cub) (Con)$ and
$(con)^2$. These tables show a very important correlation between the
types of {IM}s and the type of planar intersection, which is indicated in
these Tables by the special index of the number $N$. Thus, one can see
that the chains ${III} - \{4\}_\Delta$ and ${XII} - \{5\}_\Delta$ are
determined by the third-degree {IM}s, $(Cub)^3$ and $(Cub)^2 (Con)$,
respectively. Correspondingly, the chains ${VIII} - \{5\}_\Delta$ and $IX 
-
\{5\}_\Delta$ are determined by the fourth-degree {IM}s, $(Qrt)^2$, and
differ from each other by the types of conic, $C_2$. We note that the {7}
Weierstrass chains, $XV - \{7\}_\Delta$,....,$XXI - \{7\}_\Delta$ have the
third-degree and fourth-degree {IM} types $(Cub)^2 (Con)$ and $(Qrt)
(Cub)$, respectively.

In the next five Tables, we illustrate the relation between the types of
intersections and the types of {IM}s that determine the {22} $K3$ chains.
We divided the {22} chains into sets according the type of arity-2
intersection. Due to mirror $ \sigma \leftrightarrow \pi$ symmetry, these
intersections indicate immediately the type of fibre structure of the $K3$
space or its dual $K3^*$. In the holomorphic quotient approach, a fibre is
determined by the type of intersection in the mirror partner of the
$CY_d$ space~\cite{Cox}. In cases where $\sigma=\pi$, it directly
determines the fibre structure of the $CY_d$ space, and in the opposite
case it determines the fibre structure of the mirror Calabi-Yau manifold.
In the following Tables we mainly follow the notations of our previous
article~\cite{AENV1}. For example, the notion $\{10\}_{\Delta}$ means that
the intersection of arity-2 corresponds to the 10 point-monomials
producing the Batyrev reflexive triangle, the $\{9\}_{\Box}$ corresponds
to the 9 point-monomials producing the reflexive planar square figure,
etc.. From these Tables, using the equations $\vec{k}^{i(ex)} \cdot {IM}_a
=d_i$, one can easily convince oneself of the equivalence between the
Batyrev reflexive polyhedra description and the IM approach. For these
cases, the number of {IMs} is $a = 3$ and the number of solutions is $i =
2$.

%\newpage

{\scriptsize
\begin{table}[!ht]
\centering
\caption{\it 
The list of $K3$ {IM}s with the
Diophantine conditions:$\{3\}\rightarrow  1/3(P_1+P_2+P_3)=E_4$
and $\{4\}\rightarrow  1/4(P_1+P_2+2C_2)=E_4$, $1/2(C_1+C_2)= E_4$.
Here we present the chains with triangle $\{\sigma=10 ,\,4\}$
and $\{\sigma=9,\,5\}$ intersections~\cite{AENV1}.}
\label{Tabelda}
%\scriptsize
\vspace{.05in}
\begin{tabular}{||c|c||c||c|c|c||c|c||}
\hline
${ N} $&${ {\vec k}_i(eld)} $&$ Chain
$&$P_1               
$&$P_2                $&$ P_3/C_2
$&${IM}$&${IM}-type
$\\
\hline\hline \hline 
$I-\{10\}_{\Delta}$&$(1,1,1,1)[4]
$&$ (0,1,1,1)+(1,0,0,0)
$&$(1,3,0,0)               
$&$(1,0,3,0)          
$&$(1,0,0,3)
$&$\{3\}$&$(Cub)^3   
$ \\
$ $&$
$&$ 
$&$(3,0,0,1)               
$&$(0,2,1,1)          
$&$(0,1,2,1)
$&$\{3\}$&$(Cub)(Con)^2   
$ \\
$ $&$
$&$ 
$&$(1,2,0,1)               
$&$(0,1,2,1)          
$&$(2,0,1,1)
$&$\{3\}$&$(Con)^3    
$ \\ \hline
$ $&$
$&$ 
$&$(1,0,0,3)               
$&$(1,2,0,1)          
$&$(1,1,2,0)          
$&$\{4\}$&$(Cub)(Con)
$ \\\hline \hline
$II-\{10\}_{\Delta} $&$(1,1,2,2)[6]
$&$(0,1,1,1)+(1,0,1,1)
$&$(3,3,0,0)                
$&$(0,0,3,0)          
$&$(0,0,0,3)
$&$\{3\}$&$(Cub)^3
$ \\ 
$ $&$
$&$ 
$&$(3,3,0,0)               
$&$(0,0,2,1)          
$&$(0,0,1,2)
$&$\{3\}$&$(Cub)(Con)^2    
$ \\
$ $&$
$&$ 
$&$(0,0,0,3)               
$&$(2,2,1,0)          
$&$(1,1,2,0)
$&$\{3\}$&$ (Cub)(Con)^2 
$ \\
$ $&$
$&$ 
$&$(2,2,1,0)               
$&$(1,1,0,2)          
$&$(0,0,2,1)
$&$\{3\}$&$(Con)^3  
$ \\ \hline
$   $&$
$&$
$&$(0,0,0,3)                
$&$(2,2,0,1)          
$&$(1,1,2,0)          
$&$\{4\}$&$(Cub)(Con)
$ \\ 
$   $&$
$&$
$&$(3,3,0,0)                
$&$(1,1,2,0)          
$&$(0,0,1,2)          
$&$\{4\}$&$(Cub)(Con)
$ \\\hline\hline
$III-\{4\}_{\Delta} $&$(3,1,2,3)[9] 
$&$(0,1,1,1)+(3,0,1,2)
$&$(2,3,0,0)                         
$&$(1,0,3,0)          
$&$(0,0,0,3)
$&$\{3\}$&$(Cub)^3
$ \\
\hline\hline\hline
$IV-\{9\}_{\Delta}  $&$(1,1,1,2)[5] 
$&$(0,1,1,2)+(1,0,0,0)
$&$(1,3,0,1)                         
$&$(1,0,2,1)          
$&$(1,0,0,2)
$&$\{3\}$&$(Cub)(Con)^2
$ \\ \hline
$   $&$             
$&$
$&$(1,4,0,0)                         
$&$(1,0,4,0)          
$&$(1,0,0,2)          
$&$\{4\}$&$(Qrt)^2
$ \\ 
$   $&$             
$&$
$&$(1,3,1,0)                         
$&$(1,1,3,0)          
$&$                   
$&$\{4\}$&$(Cub)^2
$ \\
$   $&$             
$&$
$&$(1,4,0,0)                         
$&$(1,0,0,2)          
$&$(1,0,2,1)          
$&$\{4\}$&$(Qrt)(Con)
$ \\\hline \hline 
$V-\{9\}_{\Delta}$&$(1,1,1,3)[6] 
$&$(0,1,1,2)+(1,0,0,1)
$&$(2,3,1,0)                         
$&$(1,0,2,1)          
$&$(0,0,0,2)
$&$\{3\}$&$ (Cub)(Con)^2   
$ \\\hline
$   $&$              
$&$
$&$(2,4,0,0)                         
$&$(2,0,4,0)          
$&$(0,0,0,2)          
$&$\{4\}$&$(Qrt)^2
$ \\
$   $&$              
$&$
$&$(2,3,1,0)                         
$&$(2,1,3,0)          
$&$(0,0,0,2)                  
$&$\{4\}$&$(Cub)^2
$ \\
$   $&$              
$&$
$&$(2,4,0,0)                         
$&$(0,0,0,2)          
$&$(1,0,2,1)          
$&$\{4\}$&$(Qrt)(Con)
$ \\\hline \hline
$VI-\{9\}_{\Delta} $&$(1,1,2,4)[8]  
$&$(0,1,1,2)+(1,0,1,2)
$&$(3,3,1,0)                         
$&$(0,0,2,1)          
$&$(0,0,0,2)
$&$\{3\}$&$(Cub)(Con)^2
$ \\ 
$   $&$  
$&$
$&$(1,1,3,0)                         
$&$(2,2,0,1)          
$&$(0,0,0,2)
$&$\{3\}$&$(Cub)(Con)^2
$ \\ \hline
$   $&$              
$&$
$&$(4,4,0,0)                         
$&$(0,0,4,0)          
$&$(0,0,0,2)          
$&$\{4\}$&$(Qrt)^2
$ \\
$   $&$              
$&$
$&$(3,3,1,0)                         
$&$(1,1,3,0)          
$&$(0,0,0,2)                  
$&$\{4\}$&$(Cub)^2
$ \\  
$   $&$              
$&$
$&$(4,4,0,0)                         
$&$(0,0,0,2)          
$&$(0,0,2,1)          $&$\{4\}$&$(Qrt)(Con)
$ \\\hline \hline
$VII-\{9\}_{\Delta} $&$(1,1,1,1)[4] 
$&$(0,1,1,2)+(2,1,1,0)                         
$&$(0,3,1,0)                         
$&$(2,0,0,2)          
$&$(1,0,2,1)
 $&$\{3\}$&$(Cub)(Con)^2
$ \\ \hline
$   $&$              
$&$
$&$(0,4,0,0)           
$&$(0,0,4,0)          
$&$(2,0,0,2)          
$&$\{4\}$&$(Qrt)^2
$ \\
$   $&$              
$&$                        
$&$(0,3,1,0)           
$&$(0,1,3,0)          
$&$(2,0,0,2)                  
$&$\{4\}$&$(Cub)^2
$ \\
$   $&$              
$&$                        
$&$(0,4,0,0)           
$&$(2,0,0,2)          
$&$(1,0,2,1)          
$&$\{4\}$&$(Qrt)(Con)
$ \\\hline\hline
$VIII-\{5\}_{\Delta} $&$ (1,2,3,2)[8] 
$&$(0,1,1,2)+(1,1,2,0) 
$&$(4,0,0,2)                         
$&$(0,4,0,0)                   
$&$(0,0,2,1)          
$&$\{4\}$&$(Qrt)^2
$\\\hline\hline
$IX-\{5\}_{\Delta} $&$ (2,2,1,5)[10] 
$&$(0,1,1,2)+(2,1,0,3) 
$&$(3,0,4,0)                         
$&$(1,4,0,0)          
$&$(0,0,0,2)          
$&$\{4\}$&$(Qrt)^2
$\\\hline \hline \hline
\end{tabular}
\end{table}

%\newpage
\begin{table}[!ht]
\centering
\caption{\it The list of $K3$ {IM}s with the
Diophantine conditions:$\{3\} \rightarrow 1/3(P_1+P_2+P_3)=E_4$
and $\{4\} \rightarrow 1/4(P_1+P_2+2C_2)=E_4$, $1/2(C_1+C_2)= E_4$.
Here we present the chains with `quadratic-rhombus' intersections 
$\sigma$~\cite{AENV1}.}
\label{Tabeldb}
\scriptsize
\vspace{.05in}
\begin{tabular}{|c|c|c||c|c|c||c|c|}
\hline
${ N} $&${ {\vec k}_i(eld)} 
$&$ chain
$&$ P_1               
$&$ P_2               
$&$ P_3/C_2
$&$ {IM}$&$ {IM}-type
$\\
\hline\hline\hline
$X-\{9\}_{\Box}    $&$(1,1,1,1)[4]   
$&$(0,0,1,1)+(1,1,0,0) 
$&$(2,0,2,0)                         
$&$(1,1,0,2)         
$&$ (0,2,1,1)
$&$\{3\}$&$ (Con)^3
$\\ \hline
$   $&$   
$&$
$&$(2,0,2,0)                         
$&$(2,0,0,2)          
$&$(0,2,1,1)          
$&$\{4\}$&$ (Con)^2
$\\\hline
$    $&$   
$&$
$&$(2,0,2,0)                         
$&$(0,2,0,2)          
$&$                   
$&$\{2+2\}$&$[(Con)^2]^2 
$\\
$    $&$   
$&$
$&$(2,0,0,2)                         
$&$(0,2,2,0)           
$&$                    
$&$\{2+2\}$&$[(Con)^2]^2 
$\\\hline\hline
$XI-\{9\}_{\Box}    $&$(1,1,1,2)[5] 
$&$(0,0,1,1) + (1,1,0,1) 
$&$(3,0,2,0)                         
$&$(0,2,1,1)           
$&$(0,1,0,2)
$&$\{3\} $&$(Cub)(Con)^2
$ \\ 
$      $&$
$&$
$&$(2,1,2,0)                         
$&$(0,2,1,1)           
$&$(1,0,0,2)
$&$\{3\}$&$(Con)^3
$ \\ \hline
$   $&$              
$&$
$&$(3,0,2,0)                         
$&$(1,2,2,0)           
$&$(0,1,0,2)           
$&$\{4\}$&$(Cub)(Con)
$\\
$      $&$ 
$&$
$&$(3,0,2,0)                         
$&$(1,0,0,2)           
$&$(0,2,1,1)           
$&$\{4\}$&$ (Cub)(Con)
$\\\hline\hline
$XII-\{5\}_{\Box} $&$(1,1,1,1)[4]  
$&$(0,1,2,3)+(3,2,1,0) 
$&$(1,0,3,0)                         
$&$(0,3,0,1)           
$&$(2,0,0,2)
$&$\{3\}$&$ (Cub)^2(Con)
$\\\hline\hline \hline
$XIII-\{8\}_{\Box} $&$(1,1,2,3)[7] 
$&$(0,1,1,1)+(1,0,1,2)
$&$(3,2,1,0)                         
$&$(0,0,2,1)           
$&$(0,1,0,2)
$&$\{3\}$&$ (Cub)(Con)^2
$ \\ 
$      $&$
$&$
$&$(1,0,3,0)                         
$&$(2,2,0,1)           
$&$(0,1,0,2)
$&$\{3\}$&$ (Cub)(Con)^2
$ \\ \hline
$   $&$              
$&$
$&$(4,3,0,0)                         
$&$(0,1,0,2)           
$&$(0,0,2,1)           
$&$\{4\}$&$(Qrt)(Con)
$\\
$      $&$              
$&$
$&$(3,2,1,0)                         
$&$(1,0,3,0)            
$&$(0,1,0,2)            
$&$\{4\}$&$(Cub)^2
$\\\hline\hline
$XIV-\{6\}_{\Box}  $&$(1,1,1,2)[5]  
$&$(0,1,1,2)+(2,1,3,0) 
$&$(0,3,1,0)                         
$&$(3,0,0,2)            
$&$(0,0,2,1)
$&$\{3\}$&$ (Cub)^2(Con)
$ \\ \hline
$   $&$              
$&$
$&$(1,4,0,0)                         
$&$(3,0,0,2)            
$&$(0,0,2,1)            
$&$\{4\}$&$(Qrt)(Cub)
$\\\hline\hline\hline
$XXII-\{7\}_{\Box}$&$ (1,2,1,2)[6]  
$&$(0,1,1,2)+(1,1,0,0) 
$&$(1,1,3,0)                         
$&$(0,2,0,1)           
$&$(2,0,0,2)
$&$\{3\} $&$(Cub)(Con)^2
$ \\ \hline
$   $&$              
$&$
$&$(2,0,4,0)                         
$&$(2,0,0,2)            
$&$ (0,2,0,1)           
$&$\{4\}$&$(Qrt)(Con)
$\\
\hline\hline\hline
\end{tabular}
\end{table}
}

%\newpage

{\scriptsize
\begin{table}[!ht]
\centering
\caption{\it  
The list of $K3$ {IM}s with the
Diophantine conditions:$\{3\}\rightarrow  1/3(P_1+P_2+P_3)=E_4$
and $\{4\} \rightarrow 1/4(P_1+P_2+2C_2)=E_4$, $1/2(C_1+C_2)= E_4$.
Here we are present those with $\{7\}_\Delta$ chains~\cite{AENV1}.}
\label{Tabeldc}
\vspace{.05in}
\begin{tabular}{|c|c|c||c|c|c||c|c||}
\hline
${ N} $&${ {\vec k}_i(eld)} $&$ Chain
$&$ P_1               
$&$ P_2               
$&$ P_3/C_2
$&${SIM}$&${SIM}-type
$\\\hline\hline
$XV-\{7\}_{\Delta}   $&$ (1,1,2,3)[7]  
$&$(1,2,3,0)+ (0,0,0,1)
$&$(3,0,1,1)                         
$&$(0,3,0,1)          
$&$(0,0,2,1)
$&$\{3\}$&$ (Cub)^2(Con)
$\\ \hline
$     $&$               
$&$
$&$(4,1,0,1)                         
$&$(0,3,0,1)          
$&$(0,0,2,1)          
$&$\{4\}$&$(Qrt)(Cub)
$\\\hline
$     $&$               
$&$
$&$(6,0,0,1)                         
$&$                   
$&$
$&$  \{6\}     $&$ M_6
$\\\hline\hline
$XVI-\{7\}_{\Delta}  $&$ (1,1,2,4)[8]  
$&$(0,1,2,3)+(1,0,0,1) 
$&$(1,3,0,1)                                                 
$&$(2,0,3,0)          
$&$(0,0,0,2)
$&$\{3\}$&$ (Cub)^2(Con)
$\\ \hline
$     $&$               
$&$
$&$(2,4,1,0)                         
$&$(2,0,3,0)           
$&$(0,0,0,2)           
$&$\{4\}$&$(Qrt)(Cub)
$\\  \hline
$     $&$               
$&$ 
$&$(2,6,0,0)                         
$&$                    
$&$
$&$  \{6\}    $&$ M_6
$\\\hline\hline
$XVII-\{7\}_{\Delta}     $&$(1,1,3,4)[9]
$&$ (0,1,2,3)+(1,0,1,1)
$&$(2,3,0,1)                         
$&$(0,0,3,0)            
$&$(1,0,0,2)
$&$\{3\}$&$ (Cub)^2(Con)
$\\ \hline
$     $&$               
$&$
$&$(2,4,1,0)                         
$&$(0,0,3,0)            
$&$(1,0,0,2)            
$&$\{4\}$&$(Qrt)(Cub)
$\\\hline
$         $&$           
$&$ 
$&$(3,6,0,0)                         
$&$                     
$&$                     
$&$ \{6\}     $&$ M_6
$\\\hline\hline
$XVIII-\{7\}_{\Delta}   $&$(1,1,3,5)[10]  
$&$ (0,1,2,3)+(1,0,1,2) 
$&$(3,2,0,1)                         
$&$(0,1,3,0)            
$&$(0,0,0,2)
$&$ \{3\}$&$(Cub)^2(Con)
$\\ \hline
$     $&$               
$&$
$&$(4,3,1,0)                         
$&$(0,1,3,0)           
$&$(0,0,0,2)             
$&$\{4\}$&$(Qrt)(Cub)
$\\ \hline
$        $&$           
$&$  
$&$(6,4,0,0)                         
$&$                      
$&$                      
$&$\{6\}   $&$  M_6
$\\\hline\hline
$XIX-\{7\}_{\Delta}   $&$ (1,1,4,6)[12]  
$&$ (0,1,2,3)+(1,0,2,3)
$&$(3,3,0,1)                         
$&$(0,0,3,0)            
$&$(0,0,0,2)
$&$\{3\}$&$ (Cub)^2(Con)
$\\ \hline
$     $&$               
$&$
$&$(4,4,1,0)                         
$&$(0,0,3,0)            
$&$(0,0,0,2)            
$&$\{4\}$&$(Qrt)(Cub)
$\\\hline
$   $&$               
$&$  
$&$(6,6,0,0)                         
$&$                     
$&$
$&$ \{6\}     $&$M_6
$\\\hline\hline
$XX-\{7\}_{\Delta}      $&$ (1,1,1,3)[6]    
$&$(0,1,2,3)+(2,1,0,3)
$&$(0,3,0,1)                      
$&$(3,0,3,0)            
$&$(0,0,0,2)
$&$\{3\}$&$ (Cub)^2(Con)
$\\ \hline
$     $&$               
$&$
$&$(3,0,3,0)                         
$&$(1,4,1,0)            
$&$(0,0,0,2)            
$&$\{4\}$&$(Qrt)(Cub)
$\\\hline
$        $&$                 
$&$
$&$(0,6,0,0)                         
$&$                     
$&$                     
$&$  \{6\}    $&$ M_6
$\\\hline\hline
$XXI-\{7\}_{\Delta}     $&$ (3,2,4,3)[12]   
$&$(0,1,2,3)+(3,1,2,0)
$&$(1,3,0,1)                         
$&$(0,0,3,0)            
$&$(2,0,0,2)
$&$\{3\}$&$ (Cub)^2(Con)
$\\ \hline
$     $&$               
$&$
$&$(0,4,1,0)                         
$&$(0,0,3,0)           
$&$(2,0,0,2)            
$&$\{4\}$&$(Qrt)(Cub)
$\\\hline
$        $&$                 
$&$
$&$(0,6,0,0)                         
$&$                     
$&$                     
$&$ \{6\}   $&$ M_6
$\\\hline \hline
\end{tabular}
\end{table}
}
%\normalsize

We see from these Tables that the {IM}s determine completely the fibration
structures of the {22} $K3$ chains:

{\begin{eqnarray}
\{IM\}_4 &\mapsto&\biggl ( 1\cdot\{4\}_{\Delta} \biggl)+
\biggl({\bf 2\cdot\{10\}_{\Delta}}\biggl ) \nonumber\\
&+&\biggl (2 \cdot \{5\}_{\Delta}+ 1 \cdot \{5\}_{\Box} \biggl )\nonumber\\
&+&\biggl ({\bf4 \cdot\{9\}_{\Delta}}+ 2 \cdot \{9\}_{\Box} \biggl ) \nonumber\\
&+&\biggl({\bf 7\cdot\{ 7\}_{\Delta}} +1\cdot\{ 7\}_{\Box}\biggl )\nonumber\\
&+& \biggl(1 \cdot \{ 6\}_{\Box}\biggl )
+\biggl(1 \cdot \{ 8\}_{\Box}  \biggl )\nonumber\\
&\mapsto & \{22\}
\end{eqnarray}}
This expansion in terms of fibration structures is very helpful for
extending these $K3$ results to more general $CY_d$ spaces, via recurrence
relations. As we show later, each of the terms $\{10,4,...\}_{\Delta,
\Box, ...}$ in the expansion has its own recurrence relation, of which we
later derive several examples, indicated in bold script: 
${\bf 2\cdot\{10\}_{\Delta}}$, etc., providing complete results in
any number of dimensions for the numbers of $CY_d$ spaces with these particular
fibrations. A similar recurrence formula could be derived for any analogous
fibration.

We note that the examples with three cubic monomials, $(Cub)^3$, uniquely
determine the chains with intersections $\{10\}_{\Delta}$ and
$\{4\}_{\Delta}$, the {IM}s with two quartic monomials $(Qrt)^2$ and
conics $C_2$ determine the chains with $\{9\}_{\Delta}$ and
$\{5\}_{\Delta}$, etc.. The difference between $\{10\}_{\Delta}$
($\{9\}_{\Delta}$) and $\{4\}_{\Delta}$ ($\{5\}_{\Delta}$) is that the
first chains are also determined by other {IM}s, so one should look more
carefully at the structures of the {IM}s. We make a very important
observation in the $K3$ case, namely that each type of {IM} corresponds to
a different type of intersection.  For example, the {7} different
Weierstrass {IM}s, $\{3\} \sim(Cub)^2 (Con)$ or $\{4\}\sim (Qrt)(Con)$,
correspond exactly to the {7} different chains with a $\{7\}_{\Delta}$
intersection. Thus we can expect that knowing the structures of the {IM}s
and their recurrences in higher dimensions $n$, one can find all the
$CY_d$ spaces:  $d = n + 2$ which have some particular fibration. We
discuss this point further in the next sections.

\subsubsection{Applications to $CY_3$ Spaces and their Fibrations}

We now consider here all possible double Diophantine expansions of conic
monomials in 5 dimensions, and also expansions in 3 monomials of the unit
monomial. Once these triples have been obtained, the extended vectors are
obtained as follows. Given one triple of monomials in $n$ dimensions, we
look for all possible extended $k$ vectors with all components zero,
except for at most three, with the property that the scalar products with
the three monomials all give the sum of the $k$-vector components.

In practice, we form all possible $3 \times 3$ matrices obtained by
choosing all possible sets of 3 (out of the $n$)  components of the three
monomials, and solve the corresponding system of linear equations, where
the constant term is a column vector with all components equal to unity.
If the solutions for the 3 unknowns are all positive, and are such that,
when multiplied by the determinant of the matrix (to obtain integer
values) and summed, they give the determinant itself, then the 3 values
(divided if necessary by the greatest common divisor) are the components
of a $k$ vector. In case the determinant is null, one looks for solutions
among the set of 5 known good solutions ($(1)$, $(1,1)$, $(1,1,1)$,
$(1,1,2)$, $(1,2,3)$), and then sets all components to zero, except those
three.

Once we have the solutions, i.e., the set of $k$ vectors, we arrange them
in sets of $(n-2)$ to get the chains. Sets that turn out to have the same
component null are rejected. We also reject sets that have the following
quantity equal to zero: the determinant of the $(n-2) \times (n-2)$ matrix
obtained by multiplying the $(n-2) \times n$ matrix of $n-2$ $n$-dimensional
$\vec k$ vectors by its transpose, on the right.

The most time-consuming procedure is that of counting the distinct chains.
A permutation of the order of the $k$ vectors and/or a permutation of the
order of some components give rise to the same chain. The computation time
was reduced by some restrictive cuts on the checks.
   
In the 5-dimensional case, from the expansion of the unit monomial into 3
monomials, $D_5[1,1,1/3]$, we get 116 triples of monomials of maximal
degree 3. From the expansion on the second $D_r$ line of the conic
monomials in terms of two monomials, $D_5[112/4]$, we get 164 triples of
monomials with maximal degree 4. From the first set we get 231 distinct
3-vector chains, whereas from the second one we get 225 different chains.
Some chains are present in both sets, and the final number is 259, in
complete agreement with the 259 chains found previously by the normal
expansion and intersection technique.

We have already mentioned that, among the set of distinct chains, some can
be obtained as linear combinations with positive coefficients of other
chains. We have therefore looked for the minimal set of irreducible chains
that are nececessary to get all the others. In this way, we have confirmed 
that it contains 161 members, as derived previously using the RWV
expansion route. Among the {259} reducible chains ({161} irreducible chains)
there are {11}({7}) false vectors, corresponding to the minimal choices of the
coefficients, $m=n=l=1$. These false vectors are not reflexive because
they are not allowed at the level of arity {2}.  For example, the {IM}
composed of the cubic monomials $P_1=(0,0,0,3,0)$, $P_2=(0,3,2,0,1)$ and
$P_3=(3,0,1,0,2)$ gives the following arity-3 chain:
$\vec{k}^{1ex}=(0,0,1,1,1)$, $\vec{k}^{2ex}=(0,1,0,2,3)$,
$\vec{k}^{3ex}=(3,0,1,0,2)$. The choice $m = n = l = 1$ corresponds to the
vector $\vec {k}=(1,1,4,5,4)$, which has the following incorrect arity-2
expansion: $(1,1,4,5,4)=(1,1,0,1,0)+4 \cdot (0,0,1,1,1)$.  It follows from
this expansion that this vector is outside the arity-2 chain, $m \cdot
(1,1,0,1,0) + n \cdot (0,0,1,1,1)$, where the possible coefficients
corresponding to the arity-2 chain should satisfy the constraints $m
\leq 3$ and $n \leq 3$. The `good' intersection of these vectors
$\vec{k}^{ex}=(0,0,1,1,1)$ and $\vec{k}^{ex}=(1,1,0,1,0)$ gives a
reflexive polyhedron with {30} points and a mirror reflexive polyhedron
with {6} points. The corresponding arity-2 chain consists only of the
following {RWV}s:

\begin{eqnarray}
\vec{k}^{eld}&=& (1,1,1,2,1),\qquad m=1; n=1, \nonumber\\
\vec{k}^{2}&=& (2,2,1,3,1),\qquad m=2; n=1, \nonumber\\
\vec{k}^{3}&=& (3,3,1,4,1),\qquad m=3; n=1, \nonumber\\
\vec{k}^{3}&=& (3,3,2,5,2),\qquad m=3; n=2 .\nonumber\\
\end{eqnarray}
Note that the entire list of {RWV}s in the chain is determined by the 
structure of the {IM}.
The abovementioned arity-3 chain is normal: for example,
the weight vector 
\begin{eqnarray}
 (3,2,2,5,9)=(0,0,1,1,1)+ 2 \cdot (0,1,0,2,3)+ (3,0,1,0,2)
\end{eqnarray}
with $m=1,n=2,l=1$ is reflexive, as can be checked by Diophantine
expansion of one of the cubic monomials $P_2=(0,3,2,0,1)$
and the two monomials  $M_1=(0,6,0,0,1)$ and $M_2=(0,0,4,0,1)$
and fixed other monomials from the considered {IM}.

There are fixed types and numbers of IMs which determine the structures of
the full 259 (irreducible 161) chains, and they are similar to those we
already indicated for the $K3$ case, as seen, for example, in the
following Tables~\ref{Tabeld1} and ~\ref{Tabeld2}.

\begin{eqnarray}
\{IM\}_5&\mapsto&\biggl ( 9\cdot\{4\}_{\Delta} +
{\bf 4\cdot\{10\}_{\Delta}}\biggl )
 \nonumber\\
&+&\biggl (16 \cdot \{5\}_{\Delta}+ 5 \cdot \{5\}_{\Box}+
1 \cdot \{5\}_{\Box'} \biggl )\nonumber\\
&+&\biggl ({\bf 11\cdot\{9\}_{\Delta}}+ 5 \cdot \{9\}_{\Box}+ 
1 \cdot \{9\}_{\Box'}\biggl ) \nonumber\\
&+&\biggl({\bf 28\cdot\{ 7\}_{\Delta}} +7\cdot\{ 7\}_{\Box}+
1\cdot\{ 7\}_{Quint} \biggl )\nonumber\\
&+& \biggl(8 \cdot \{ 6\}_{\Box} +1 \cdot \{ 6\}_{Quint}\biggl )\nonumber\\
&+&\biggl(6 \cdot \{ 8\}_{\Box} +1 \cdot \{ 8\}_{Quint} \biggl )
\nonumber\\
&\mapsto& \{161\}
\end{eqnarray}

{\scriptsize
\begin{table}[!ht]
\centering
\caption{\it The $CY_3$ {IM}s with the
Diophantine conditions:$\{3\} \rightarrow 1/3(P_1+P_2+P_3)=E_5$
and $\{4\} \rightarrow 1/4(P_1+P_2+2C_2)=E_5$, $1/2(C_1+C_2)= E_5$.
Here we present the chains with triangle $\{10+4\}$
and $\{9+5\}$ intersections~\cite{AENV1}.}
\label{Tabeld1}
%\scriptsize
\vspace{.05in}
\begin{tabular}{||c|c||c||c|c|c||c|c||}
\hline
${ \sigma}                $&${ {\vec k}_i(eld)} 
$&$ Chain
$&$P_1               
$&$P_2                    
$&$ P_3/C_2
$&${IM}$&${IM}-type
$\\
\hline\hline \hline 
${\{10\}}_{\Delta}        $&$(1,1,1,1,1)[5]
$&$ (0,0,0,0,1)+
$&$(0,0,3,1,1)               
$&$(0,3,0,1,1)          
$&$(3,0,0,1,1)
$&$\{3\}$&$(Cub)^3   
$ \\
$ $&$
$&$ (0,0,0,1,0)+
$&$(3,0,0,1,1)               
$&$(0,2,1,1,1)          
$&$(0,1,2,1,1)
$&$\{3\}$&$(Cub)(Con)^2   
$ \\
$                         $&$
$&$(1,1,1,0,0)+
$&$(0,2,1,1,1)             
$&$(1,0,2,1,1)         
$&$(2,1,0,1,1)
$&$\{3\}$&$(Con)^3    
$ \\ 
$                         $&$
$&$ 
$&$(3,0,0,1,1)              
$&$(1,0,2,1,1)         
$&$(0,2,1,1,1)      
$&$\{4\}$&$(Cub)(Con)
$ \\ \hline \hline
${\{4\}}_{\Delta}          $&$(3,3,1,4,4)[15]
$&$(0,0,1,1,1)+
$&$(0,1,0,3,0)               
$&$(1,0,0,0,3)           
$&$(2,2,3,0,0)
$&$\{3\}$&$(Cub)^3   
$ \\
$                          $&$
$&$ (0,3,0,1,2)+
$&$-               
$&$-                    
$&$-                     
$&$-    $&$-  
$ \\
$                          $&$
$&$(3,0,0,2,1)+
$&$ -              
$&$ -                     
$&$ -                     
$&$ -  $&$ -
$ \\\hline \hline
${\{9\}}_{\Delta}         $&$(1,1,1,1,1)[5]
$&$(0,0,1,1,2)+
$&$(0,0,0,4,0)               
$&$(0,0,4,0,0)          
$&$(2,2,0,0,2)          
$&$\{4\}$&$(Qrt)^2   
$ \\
$ $&$
$&$ (0,2,1,1,0)+
$&$(0,0,1,3,0)               
$&$(0,0,3,1,0)         
$&$(2,2,0,0,2)        
$&$\{4\}$&$(Qrt)^2   
$ \\
$ $&$
$&$(2,0,1,1,0)+
$&$(0,0,1,3,0)              
$&$(1,1,2,0,1)         
$&$(2,2,0,0,2) 
$&$ \{3\}$&$(Cub)(Con)^2   
$ \\ 
$                         $&$
$&$ (1,0,0,0,1,2)                
$&$-         
$&$-
$&$-          
$&$-    $&$-
$ \\\hline \hline
${\{9\}}_{\Box^\prime}         $&$(1,1,1,2,2)[7]
$&$(0,0,0,1,1)+
$&$(2,0,0,0,2)               
$&$(2,0,1,1,1)          
$&$(0,2,0,0,2)          
$&$\{2\}^4$&$(Con)^4   
$ \\
$ $&$
$&$(0,0,1,0,1)+
$&$-               
$&$-         
$&$-         
$&$   $&$  
$ \\
$ $&$
$&$(1,1,0,0,0)+
$&$-             
$&$-         
$&$-  
$&$    $&$   
$ \\ \hline \hline
${\{5\}}_{\Delta}         $&$(2,1,1,3,5)[12]
$&$(0,0,1,1,2)+
$&$(2,4,4,0,0)               
$&$(0,0,0,4,0)          
$&$(1,0,0,0,2)          
$&$\{4\}$&$(Qrt)^2   
$ \\
$                         $&$
$&$(0,1,0,1,2)+
$&$-               
$&$-                      
$&$-                       
$&$-    $&$-   
$ \\
$                         $&$
$&$(2,0,0,1,1)
$&$ -             
$&$ -                      
$&$ -                     
$&$ -   $&$ -  
$ \\\hline \hline
${\{9\}}_{\Box}         $&$(1,1,2,1,5)[10]
$&$(0,0,0,1,1)+
$&$(4,4,0,2,0)               
$&$(0,0,4,2,0)          
$&$(0,0,0,0,2)          
$&$ \{4\}$&$(Quart)^2 
$ \\
$ $&$
$&$(0,1,1,0,2)+
$&$(3,3,1,2,0)               
$&$(1,1,3,2,0)-         
$&$ (0,0,0,0,2)         
$&$\{4\}$&$(Cub)^2   
$ \\
$ $&$
$&$(1,0,1,0,2)+
$&$(3,3,1,2,0)             
$&$(0,0,2,1,1)       
$&$(0,0,0,0,2)  
$&$ \{3\}$&$(Cub)(Con)^2 
$ \\\hline \hline  
${\{5\}}_{\Box}         $&$(3,3,4,5,3)[19]
$&$(0,0,1,2,3)+
$&$(1,0,0,3,0)               
$&$(0,1,3,0,1)          
$&$(2,2,0,0,2)          
$&$\{3\}^2$&$(Cub)^2(Con)   
$ \\
$ $&$
$&$(0,3,1,2,0)+
$&$-               
$&$-         
$&$-         
$&$   $&$  
$ \\
$ $&$
$&$(3,0,2,1,0)+
$&$-             
$&$-         
$&$-  
$&$    $&$   
$ \\ \hline \hline
${\{5\}}_{\Box^\prime}         $&$(2,1,2,1,2)[8]
$&$(0,0,0,1,1)+
$&$(0,0,2,0,2)               
$&$(1,0,2,2,0)          
$&$(1,2,0,0,2)          
$&$\{2\}^2$&$(Con)^2   
$ \\
$ $&$
$&$(0,1,1,0,0)+
$&$-               
$&$-         
$&$-         
$&$   $&$  
$ \\
$ $&$
$&$(2,0,1,0,1)+
$&$-             
$&$-         
$&$-  
$&$    $&$   
$ \\ \hline \hline
\end{tabular}
\end{table}
}
%\normalsize

{\scriptsize
\begin{table}[!ht]
\centering
\caption{\it The $CY_3$ {IM}s with the
Diophantine conditions:$\{3\} \rightarrow 1/3(P_1+P_2+P_3)=E_5$
and $\{4\} \rightarrow 1/4(P_1+P_2+2C_2)=E_5$, $1/2(C_1+C_2)= E_5$.
Here we present the chains with box and quintuple $\{8+6\}$ intersections 
and $\{7+7\}$ intersections~\cite{AENV1}.}
\label{Tabeld2}
%\scriptsize
\vspace{.05in}
\begin{tabular}{||c|c||c||c|c|c||c|c||}
\hline
${ \sigma}                $&${ {\vec k}_i(eld)} 
$&$ Chain
$&$P_1               
$&$P_2                    
$&$ P_3/C_2
$&${IM}$&${IM}-type
$\\
\hline\hline \hline 
${\{8\}}_{\Box}        $&$(1,1,1,3,5)[11]
$&$ (0,0,1,1,1)+
$&$(4,4,3,0,0)               
$&$(0,0,1,0,2)          
$&$(0,0,0,2,1)
$&$\{4\}$&$(Qrt)(Con)   
$ \\
$ $&$
$&$(0,1,0,1,2)+
$&$(3,3,2,1,0)               
$&$(1,1,0,3,0)          
$&$(0,0,1,0,2)
$&$\{4\}$&$(Cub)(Con) 
$ \\
$                         $&$
$&$(1,0,0,1,2)+
$&$(3,3,2,1,0)            
$&$(0,0,1,0,2)         
$&$(0,0,0,2,1)
$&$\{3\}$&$(Cub)(Con)^2    
$ \\ \hline \hline
${\{6\}}_{\Box}          $&$(2,2,2,1,5)[15]
$&$(0,0,0,1,1)+
$&$(1,4,0,2,0)               
$&$(3,0,2,2,0)           
$&$(0,0,1,0,2)
$&$\{4\}$&$(Qrt)(Cub)   
$ \\
$                          $&$
$&$(0,1,2,0,1)+
$&$(3,0,2,2,0)               
$&$(0,3,1,0,1)                    
$&$(0,0,0,1,2)                     
$&$\{3\}    $&$ (Cub)^2(Con)
$ \\
$                          $&$
$&$(2,1,0,0,3)+
$&$ -              
$&$ -                     
$&$ -                     
$&$ -  $&$ -
$ \\\hline \hline
${\{8\}}_{Quint}         $&$(1,1,2,1,1)[6]
$&$(0,0,0,1,1)+
$&$(3,2,0,2,0)               
$&$(0,1,1,0,2)          
$&$(0,0,2,1,1)          
$&$\{3\}$&$(Cub)(Con)^2   
$ \\
$ $&$
$&$ (0,1,1,0,0)+
$&$(2,2,0,1,1)               
$&$(1,0,2,2,0)         
$&$(0,1,1,0,2)        
$&$\{2\}^3$&$(Con)^3   
$ \\
$ $&$
$&$(1,0,1,0,1)
$&$             
$&$         
$&$
$&$   
$ \\\hline \hline
${\{6\}}_{Quint}         $&$(2,1,1,3,5)[12]
$&$(0,0,1,1,1)+
$&$(1,1,0,3,0)               
$&$(2,0,1,0,2)          
$&$(0,2,2,0,1)          
$&$\{3\}$&$(Cub)(Con)^2  
$ \\
$                         $&$
$&$(0,1,0,1,2)+
$&$(0,2,2,0,1)               
$&$(0,2,1,2,0)                      
$&$(2,0,1,0,2)             
$&$\{2\}^2    $&$(Con)^2   
$ \\
$                         $&$
$&$(1,0,2,1,0)
$&$ -             
$&$ -                      
$&$ -                     
$&$ -   $&$ -  
$ \\\hline \hline
${\{7\}}_{\Box}         $&$(2,2,2,1,3)[10]
$&$(0,0,0,1,1)+
$&$(0,4,0,2,0)               
$&$(2,0,2,2,0)          
$&$(1,0,1,0,2)          
$&$ \{4\}$&$(Quart)(Con) 
$ \\\hline
$ $&$
$&$(0,1,2,0,1)+
$&$(3,3,0,1,1)               
$&$(0,0,2,2,0)-         
$&$(0,0,1,0,2)         
$&$\{3\}$&$(Cub)(Con)^2   
$ \\
$ $&$
$&$(2,1,0,0,1)+
$&$             
$&$      
$&$ 
$&$
$ \\ \hline \hline
${\{7\}}_{Quint}         $&$(1,2,3,1,2)[19]
$&$(0,0,0,1,1)+
$&$(2,1,0,3,0)               
$&$(1,0,1,0,2)          
$&$(0,2,2,0,1)          
$&$\{3\}^2$&$(Cub)^2(Con)   
$ \\
$ $&$
$&$(0,1,1,0,1)+
$&$(0,3,2,1,0)               
$&$(2,0,0,2,1)         
$&$(1,0,1,0,2)         
$&$ \{3\}^2  $&$ (Cub)^2(Con) 
$ \\
$ $&$
$&$(1,1,2,0,0)+
$&$(0,2,2,0,1)             
$&$ (1,0,1,0,2)        
$&$(2,0,0,2,1)  
$&$\{2\}^2    $&$   (Con)^2
$ \\\hline\hline
\end{tabular}
\end{table}
}
%\normalsize

We stress that finding the IMs corresponding to the $E_r$ line is possible
in any dimension, because of simple recurrence relations for conic, cubic,
quartic and sextic monomials. As an illustration, we discuss the IMs
leading to Weierstrass fibrations, which have just the following two
structures: $\{3\} \rightarrow (Cub)^2 (Con)$ and
$\{4\} \rightarrow (Qrt) (Cub)$. 
There is an additional link between these two types of Weierstrass IMs and
the sextic monomials $M_6$, as seen in the Tables. As for the case of $K3$
spaces with Weierstrass fibres, here also one can put each IM into
correspondence with a sextic monomial $M_6$, with the following property:
$1/3(M_6 + 2 P_2)=C_1$, where $1/2(C_1+C_2)=E_4$. The last property
enables us to find an exact link between Weierstrass IMs and the types of
sextic monomials. This link is not one-to-one, but one to $i$, where $i =
p - 1$ depends on the number $p$ of zero components in the monomial $M_6$.
For instance, the monomial $M_6=(6,6,6,0,0)$ has two zero components and,
consequently $i=1$, whereas the monomial $M_6=(6,6,0,0,0)$ has three zero
components and consequently $i=2$, and the monomial $M_6=(6,0,0,0,0)$ has
four zero components and consequently $i=3$. In the case of the monomial
$M_6=(6,6,6,0,0)$, the condition $1/3(M_6 + 2 P_2)=C_1$ can be satisfied
in just one way, with $P_2=(3,3,3,0,1)$ and $C_1=(0,0,0,0,2)$. However, in
the case of the monomial $M_6=(6,6,0,0,0)$ there are already two ways of
satisfying the condition $1/3(M_6 + 2 P_2)=C_1$, namely $P_2=(3,3,0,0,1)$,
$C_1=(0,0,0,0,2)$ and $P_2'=(3,3,1,0,1)$, $C_1'=(0,0,2,0,2)$,
respectively. The three solutions for the condition $1/3(M_6 + 2 P_2)=C_1$
in the case of the monomial $M_6=(6,0,0,0,0)$ can be see in
Tables~\ref{Tabsix} and \ref{Tabelw1} to \ref{Tabelw5}. Using the fact 
that there exist these
multiple examples of Weierstrasss IMs for sextic monomials, we obtain the
following expression for the number of Weierstrass IMs, $\{3\}$ and
$\{4\}$:

\begin{equation}
N_{W}(n) = C_{n+3}^{n-3} = \frac{(n+3)!}{(6!)(n-3)!},
\end{equation}
valid for all dimensions.

{\scriptsize
\begin{table}[!ht]
\centering
\caption{\it The $CY_3$ {IM}s with the
Diophantine conditions:$\{3\}:1/3(P_1+P_2+P_3)=E_5$
and $\{4\}:1/4(P_1+P_2+2C_2)=E_5$, $1/2(C_1+C_2)= E_5$.
Here we start presenting the 28 Weierstrass sets of {IMs}~\cite{AENV1}. }
\label{Tabelw1}
\vspace{.05in}
\begin{tabular}{|c|c||c||c|c|c|c||c|c||}
\hline
${ N} $&${ {\vec k}_i(eld)} $&$ Chain
$&$P_1               
$&$P_2                  $&$ P_3
$&$C_2                  $&${IM}$&${IM}-type
$\\\hline\hline
$1   $&$ (1,1,1,6,9)[18]  
$&$(0,0,1,2,3,)+
$&$(0,0,0,3,0)                         
$&$(3,3,3,0,1)          $&$(0,0,0,0,2)
$&$-                    $&$\{3\}$&$ (Cub)^2(Con)
$\\
$     $&$ h_{21}={272}              
$&$(0,1,0,2,3)+
$&$(4,4,4,1,0)                         
$&$(0,0,0,3,0)          $&$-
$&$(0,0,0,0,2)          $&$\{4\}$&$(Qrt)(Cub)
$\\
$     $&$ h_{11}={2}              
$&$(1,0,0,2,3)\,
$&$(6,6,6,0,0)                         
$&$                    $&$
$&$                    $&$ \{6\} $&$ M_6
$\\\hline\hline
$2   $&$ (1,1,1,5,8)[16]  
$&$(0,0,1,1,2,)+
$&$(0,0,1,3,0)                         
$&$(3,3,2,0,1)         $&$(0,0,0,0,2)
$&$-                   $&$\{3\}$&$ (Cub)^2(Con)
$\\
$     $&$ h_{21}={231}              
$&$(0,1,0,2,3)+
$&$(4,4,3,1,0)                         
$&$(0,0,1,3,0)         $&$-
$&$(0,0,0,0,2)         $&$\{4\}$&$(Qrt)(Cub)
$\\
$     $&$  h_{11}={3}             
$&$(1,0,0,2,3)\,
$&$(6,6,4,0,0)                         
$&$                    $&$
$&$                    $&$ \{6\}   $&$ M_6
$\\\hline\hline
$3   $&$ (1,1,1,5,7)[15]  
$&$(0,0,1,1,1,)+
$&$(0,0,0,3,0)                         
$&$(3,3,2,0,1)         $&$(0,0,1,0,2)
$&$-                   $&$\{3\}$&$ (Cub)^2(Con)
$\\
$     $&$ h_{21}={208}              
$&$(0,1,0,2,3)+
$&$(4,4,2,1,0)                         
$&$(0,0,0,3,0)         $&$-
$&$(0,0,1,0,2)         $&$\{4\}$&$(Qrt)(Cub)
$\\
$     $&$   h_{11}={4}            
$&$(1,0,0,2,3)\,
$&$(6,6,3,0,0)                         
$&$                    $&$
$&$                    $&$ \{6\} $&$ M_6
$\\\hline\hline
$4   $&$ (1,1,1,4,7)[14]  
$&$(0,0,1,0,1,)+
$&$(0,0,2,3,0)                         
$&$(3,3,1,0,1)         $&$(0,0,0,0,2)
$&$-                   $&$\{3\}$&$ (Cub)^2(Con)
$\\ 
$     $&$ h_{21}={195}              
$&$(0,1,0,2,3)+
$&$(4,4,2,1,0)                         
$&$(0,0,2,3,0)         $&$-
$&$(0,0,0,0,2)         $&$\{4\}$&$(Qrt)(Cub)
$\\
$     $&$  h_{11}={3}             
$&$(1,0,0,2,3)\,
$&$(6,6,2,0,0)                         
$&$                    $&$
$&$                    $&$ \{6\}   $&$ M_6
$\\\hline\hline
$5   $&$ (1,1,1,4,6)[13]  
$&$(0,0,1,0,0)+
$&$(0,0,1,3,0)                         
$&$(3,3,1,0,1)         $&$(0,0,1,0,2)
$&$-                   $&$\{3\}$&$ (Cub)^2(Con)
$\\
$     $&$   h_{21}={173}             
$&$(0,1,0,2,3)+
$&$(4,4,1,1,0)                         
$&$(0,0,1,3,0)         $&$-
$&$(0,0,1,0,2)         $&$\{4\}$&$(Qrt)(Cub)
$\\
$     $&$  h_{11}={5}             
$&$(1,0,0,2,3)\,
$&$(6,6,1,0,0)                         
$&$                    $&$
$&$                    $&$ \{6\}   $&$ M_6
$\\\hline\hline
$6   $&$ (2,1,2,4,9)[18]  
$&$(0,0,1,2,3)+
$&$(3,0,0,3,0)                         
$&$(0,3,3,0,1)         $&$(0,0,0,0,2)
$&$-                   $&$\{3\}$&$ (Cub)^2(Con)
$\\ 
$     $&$   h_{21}={123}             
$&$(0,1,0,2,3)+
$&$(1,4,4,1,0)                         
$&$(3,0,0,3,0)         $&$-
$&$(0,0,0,0,2)         $&$\{4\}$&$(Qrt)(Cub)
$\\
$     $&$  h_{11}={3}              
$&$(2,0,1,0,3)\,
$&$(0,6,6,0,0)                         
$&$                    $&$
$&$                    $&$ \{6\}   $&$ M_6
$\\\hline\hline
$7   $&$ (3,2,1,6,6)[18]  
$&$(0,0,1,2,3)+
$&$(0,0,0,3,0)                         
$&$(1,3,3,0,1)         $&$(2,0,0,0,2)
$&$-                   $&$\{3\}$&$ (Cub)^2(Con)
$\\ 
$     $&$  h_{21}={79}             
$&$(0,1,0,2,3)+
$&$(0,4,4,1,0)                         
$&$(0,0,0,3,0)         $&$-
$&$(2,0,0,0,2)         $&$\{4\}$&$(Qrt)(Cub)
$\\
$     $&$   h_{11}={7}             
$&$(3,0,1,2,0)\,
$&$(0,6,6,0,0)                         
$&$                    $&$
$&$                    $&$ \{6\}   $&$ M_6
$\\\hline\hline
\end{tabular}
\end{table}
}
%\normalsize
%\newpage

{\scriptsize
\begin{table}[!ht]
\centering
\caption{\it
Continuation of the previous Table.}
\label{Tabelw2}
\vspace{.05in}
\begin{tabular}{|c|c||c||c|c|c|c||c|c||}
\hline
${ N} $&${ {\vec k}_i(eld)} $&$ Chain
$&$P_1               
$&$P_2                  $&$ P_3
$&$C_2                  $&${IM}$&${IM}-type
$\\\hline\hline
$8   $&$ (1,1,1,4,7)[16]  
$&$(0,0,1,1,2)+
$&$(0,1,1,3,0)                         
$&$(3,2,2,0,1)         $&$(0,0,0,0,2)
$&$-                   $&$\{3\}$&$ (Cub)^2(Con)
$\\ 
$     $&$  h_{21}={195}             
$&$(0,1,0,1,2)+
$&$(4,3,3,1,0)                         
$&$(0,1,1,3,0)         $&$-
$&$(0,0,0,0,2)         $&$\{4\}$&$(Qrt)(Cub)
$\\
$     $&$ h_{11}={3}              
$&$(1,0,0,2,3)\,
$&$(6,4,4,0,0)                         
$&$                    $&$
$&$                    $&$ \{6\}   $&$ M_6
$\\\hline\hline
$9   $&$ (1,1,1,4,6)[16]  
$&$(0,0,1,1,1)+
$&$(0,1,0,3,0)                         
$&$(3,2,2,0,1)         $&$(0,0,1,0,2)
$&$-                   $&$\{3\}$&$ (Cub)^2(Con)
$\\ 
$     $&$    h_{21}={173}             
$&$(0,1,0,1,2)+
$&$(4,3,2,1,0)                         
$&$(0,1,0,3,0)         $&$-
$&$(0,0,1,0,2)         $&$\{4\}$&$(Qrt)(Cub)
$\\
$     $&$  h_{11}={5}             
$&$(1,0,0,2,3)\,
$&$(6,4,3,0,0)                         
$&$                    $&$
$&$                    $&$ \{6\}   $&$ M_6
$\\\hline\hline
$10   $&$ (1,1,1,3,6)[16]  
$&$(0,0,1,0,1)+
$&$(0,1,2,3,0)                         
$&$(3,2,1,0,1)         $&$(0,0,0,0,2)
$&$-                   $&$\{3\}$&$ (Cub)^2(Con)
$\\ 
$     $&$  h_{21}={165}             
$&$(0,1,0,1,2)+
$&$(4,3,2,1,0)                         
$&$(0,1,2,3,0)         $&$-
$&$(0,0,0,0,2)         $&$\{4\}$&$(Qrt)(Cub)
$\\
$     $&$  h_{11}={3}             
$&$(1,0,0,2,3)\,
$&$(6,4,2,0,0)                         
$&$                    $&$
$&$                    $&$ \{6\}   $&$ M_6
$\\\hline\hline
$11   $&$ (1,1,1,3,5)[16]  
$&$(0,0,1,0,0)+
$&$(0,1,1,3,0)                         
$&$(3,2,1,0,1)         $&$(0,0,1,0,2)
$&$-                   $&$\{3\}$&$ (Cub)^2(Con)
$\\
$     $&$  h_{21}={144}             
$&$(0,1,0,1,2)+
$&$(4,3,1,1,0)                         
$&$(0,1,1,3,0)         $&$-
$&$(0,0,1,0,2)         $&$\{4\}$&$(Qrt)(Cub)
$\\
$     $&$   h_{11}={4}            
$&$(1,0,0,2,3)\,
$&$(6,4,1,0,0)                         
$&$                    $&$
$&$                    $&$ \{6\}   $&$ M_6
$\\\hline\hline
$12   $&$ (1,2,2,4,5)[14]  
$&$(0,0,1,1,2)+
$&$(0,0,1,3,0)                         
$&$(3,1,2,0,1)         $&$(0,2,0,0,2)
$&$-                   $&$\{3\}$&$ (Cub)^2(Con)
$\\ 
$     $&$ h_{21}={83}              
$&$(0,2,1,1,0)+
$&$(4,0,3,1,0)                         
$&$(0,0,1,3,0)         $&$-
$&$(0,2,0,0,2)         $&$\{4\}$&$(Qrt)(Cub)
$\\
$     $&$  h_{11}={5}             
$&$(1,0,0,2,3)\,
$&$(6,0,4,0,0)                         
$&$                    $&$
$&$                    $&$ \{6\}   $&$ M_6
$\\
$12'   $&$ (3,2,1,5,5)[16]  
$&$(0,0,1,1,2)+
$&$(0,0,1,3,0)                         
$&$(3,1,2,0,1)         $&$(0,2,0,0,2)
$&$-                   $&$\{3\}$&$ (Cub)^2(Con)
$\\ 
$     $&$   h_{21}={67}            
$&$(0,1,0,2,3)+
$&$(4,0,3,1,0)                         
$&$(0,0,1,3,0)         $&$-
$&$(0,2,0,0,2)         $&$\{4\}$&$(Qrt)(Cub)
$\\
$     $&$  h_{11}={9}             
$&$(3,1,0,2,0)\,
$&$(6,0,4,0,0)                         
$&$                    $&$
$&$                    $&$ \{6\}   $&$ M_6
$\\\hline\hline
$13   $&$ (1,2,4,2,5)[14]  
$&$(0,0,1,1,2)+
$&$(0,3,3,1,0)                         
$&$(3,0,0,2,1)         $&$(0,0,0,0,2)
$&$-                   $&$\{3\}$&$ (Cub)^2(Con)
$\\ 
$     $&$   h_{21}={83}            
$&$(0,2,1,1,0)+
$&$(4,1,1,3,0)                         
$&$(0,3,3,1,0)         $&$-
$&$(0,0,0,0,2)         $&$\{4\}$&$(Qrt)(Cub)
$\\
$     $&$  h_{11}={5}             
$&$(1,0,2,0,3)\,
$&$(6,0,0,4,0)                         
$&$                    $&$
$&$                    $&$ \{6\}   $&$ M_6
$\\
$13'  $&$ (2,2,1,3,8)[16]  
$&$(0,0,1,1,2)+
$&$(0,0,1,3,0)                         
$&$(3,1,2,0,1)         $&$(0,0,0,0,2)
$&$-                   $&$\{3\}$&$ (Cub)^2(Con)
$\\
$     $&$   h_{11}={103}            
$&$(0,1,0,2,3)+
$&$(4,0,3,1,0)                         
$&$(0,0,1,3,0)         $&$-
$&$(0,0,0,0,2)         $&$\{4\}$&$(Qrt)(Cub)
$\\
$     $&$  h_{11}={3}             
$&$(2,1,0,0,3)\,
$&$(6,0,0,4,0)                         
$&$                    $&$
$&$                    $&$ \{6\}   $&$ M_6
$\\\hline\hline
\end{tabular}
\end{table}
\normalsize

\begin{table}[!ht]
\centering
\caption{\it Continuation of the previous Table.}
\label{Tabelw3}
\scriptsize
\vspace{.05in}
\begin{tabular}{|c|c||c||c|c|c|c||c|c||}
\hline
${ N} $&${ {\vec k}_i(eld)} $&$ Chain
$&$P_1               
$&$P_2                  $&$ P_3
$&$C_2                  $&${IM}$&${IM}-type
$\\\hline\hline
$14   $&$ (1,1,1,4,5)[15]  
$&$(0,0,1,1,1,)+
$&$(0,0,0,3,0)                         
$&$(3,2,2,0,1)          $&$(0,0,0,0,2)
$&$-                    $&$\{3\}$&$ (Cub)^2(Con)
$\\
$     $&$  h_{11}={154}             
$&$(0,1,0,1,1)+
$&$(4,2,2,1,0)                         
$&$(0,0,0,3,0)          $&$-
$&$(0,0,0,0,2)          $&$\{4\}$&$(Qrt)(Cub)
$\\
$     $&$ h_{11}={4}              
$&$(1,0,0,2,3)\,
$&$(6,3,3,0,0)                         
$&$                    $&$
$&$                    $&$ \{6\}   $&$ M_6
$\\\hline\hline
$15   $&$ (1,1,1,3,5)[11]  
$&$(0,0,1,0,1,)+
$&$(0,0,2,3,0)                         
$&$(3,2,1,0,1)          $&$(0,1,0,0,2)
$&$-                    $&$\{3\}$&$ (Cub)^2(Con)
$\\ 
$     $&$  h_{21}={144}             
$&$(0,1,0,1,1)+
$&$(4,2,2,1,0)                         
$&$(0,0,2,3,0)          $&$-
$&$(0,1,0,0,2)          $&$\{4\}$&$(Qrt)(Cub)
$\\
$     $&$ h_{11}={4}              
$&$(1,0,0,2,3)\,
$&$(6,3,2,0,0)                         
$&$                    $&$
$&$                    $&$ \{6\}   $&$ M_6
$\\\hline\hline
$16   $&$ (1,1,1,3,4)[10]  
$&$(0,0,1,0,0,)+
$&$(0,0,1,3,0)                         
$&$(3,2,1,0,1)          $&$(0,1,1,0,2)
$&$-                    $&$\{3\}$&$ (Cub)^2(Con)
$\\ 
$     $&$ h_{21}={126}              
$&$(0,1,0,1,1)+
$&$(4,2,1,1,0)                         
$&$(0,0,1,3,0)          $&$-
$&$(0,1,1,0,2)          $&$\{4\}$&$(Qrt)(Cub)
$\\
$     $&$  h_{11}={4}             
$&$(1,0,0,2,3)\,
$&$(6,3,1,0,0)                         
$&$                    $&$
$&$                    $&$ \{6\}   $&$ M_6
$\\\hline\hline
$17   $&$ (3,2,1,5,4)[15]  
$&$(0,0,1,1,1)+
$&$(0,0,0,3,0)                         
$&$(1,3,2,0,1)          $&$(2,0,1,0,2)
$&$-                    $&$\{3\}$&$ (Cub)^2(Con)
$\\ 
$     $&$ h_{21}={66}              
$&$(0,1,0,2,3)+
$&$(0,4,2,1,0)                         
$&$(0,0,0,3,0)          $&$-
$&$(2,0,1,0,2)          $&$\{4\}$&$(Qrt)(Cub)
$\\
$     $&$  h_{11}={3}             
$&$(3,1,0,2,0)\,
$&$(0,6,3,0,0)                         
$&$                    $&$
$&$                    $&$ \{6\}   $&$ M_6
$\\
$17'   $&$ (1,1,4,2,4)[15]  
$&$(0,0,1,1,1)+
$&$(0,0,0,3,0)                         
$&$(1,3,2,0,1)          $&$(2,0,1,0,2)
$&$-                    $&$\{3\}$&$ (Cub)^2(Con)
$\\
$     $&$    h_{21}={101}           
$&$(0,1,0,1,1)+
$&$(0,4,2,1,0)                         
$&$(0,0,0,3,0)          $&$-
$&$(2,0,1,0,2)          $&$\{4\}$&$(Qrt)(Cub)
$\\
$     $&$ h_{11}={5}              
$&$(1,0,3,0,2)\,
$&$(0,6,3,0,0)                         
$&$                    $&$
$&$                    $&$ \{6\}   $&$ M_6
$\\\hline\hline
$18   $&$ (2,2,1,3,7)[15]  
$&$(0,0,1,1,1)+
$&$(3,0,0,3,0)                         
$&$(0,3,2,0,1)          $&$(0,0,1,0,2)
$&$-                    $&$\{3\}$&$ (Cub)^2(Con)
$\\ 
$     $&$ h_{21}={89}              
$&$(0,1,0,2,3)+
$&$(1,4,2,1,0)                         
$&$(3,0,0,3,0)          $&$-
$&$(0,0,1,0,2)          $&$\{4\}$&$(Qrt)(Cub)
$\\
$     $&$ h_{11}={5}              
$&$(2,1,0,0,3)\,
$&$(0,6,3,0,0)                         
$&$                    $&$
$&$                    $&$ \{6\}   $&$ M_6
$\\
$18'   $&$ (1,1,3,2,5)[12]  
$&$(0,0,1,1,1)+
$&$(3,0,0,3,0)                         
$&$(0,3,2,0,1)          $&$(0,0,1,0,2)
$&$-                    $&$\{3\}$&$ (Cub)^2(Con)
$\\ 
$     $&$    h_{21}={105}           
$&$(0,1,0,1,1)+
$&$(1,4,2,1,0)                         
$&$(3,0,0,3,0)          $&$-
$&$(0,0,1,0,2)          $&$\{4\}$&$(Qrt)(Cub)
$\\
$     $&$  h_{11}={3}             
$&$(1,0,2,0,3)\,
$&$(0,6,3,0,0)                         
$&$                    $&$
$&$                    $&$ \{6\}   $&$ M_6
$\\\hline\hline
\end{tabular}
\end{table}
}
%\normalsize

{\scriptsize
\begin{table}[!ht]
\centering
\caption{\it Continuation of the previous Table.}
\label{Tabelw4}
\vspace{.05in}
\begin{tabular}{|c|c||c||c|c|c|c||c|c||}
\hline
${ N} $&${ {\vec k}_i(eld)} $&$ Chain
$&$P_1               
$&$P_2                  $&$ P_3
$&$C_2                  $&${IM}$&${IM}-type
$\\\hline\hline
$19   $&$ (1,2,1,1,5)[10]  
$&$(0,0,0,1,1)+
$&$(0,3,2,2,0)                         
$&$(3,0,1,1,1)          $&$(0,0,0,0,2)
$&$-                    $&$\{3\}$&$ (Cub)^2(Con)
$\\ 
$     $&$  h_{21}={145}             
$&$(0,0,1,0,1)+
$&$(4,1,2,2,0)                         
$&$(0,3,2,2,0)          $&$-
$&$(0,0,0,0,2)          $&$\{4\}$&$(Qrt)(Cub)
$\\
$     $&$   h_{11}={1}            
$&$(1,2,0,0,3)\,
$&$(6,0,2,2,0)                         
$&$                    $&$
$&$                    $&$ \{6\}   $&$ M_6
$\\\hline\hline
$20   $&$ (1,2,1,1,4)[9]  
$&$(0,0,0,1,0)+
$&$(0,3,2,1,0)                         
$&$(3,0,1,1,1)          $&$(0,0,0,1,2)
$&$-                    $&$\{3\}$&$ (Cub)^2(Con)
$\\ 
$     $&$   h_{21}={123}              
$&$(0,0,1,0,1)+
$&$(4,1,2,1,0)                         
$&$(0,3,2,1,0)          $&$-
$&$(0,0,0,1,2)          $&$\{4\}$&$(Qrt)(Cub)
$\\
$     $&$  h_{11}={3}             
$&$(1,2,0,0,3)\,
$&$(6,0,2,1,0)                         
$&$                    $&$
$&$                    $&$ \{6\}   $&$ M_6
$\\\hline\hline
$21   $&$ (1,2,1,2,4)[9]  
$&$(0,0,0,1,1)+
$&$(0,3,0,2,0)                         
$&$(3,0,1,1,1)          $&$(0,0,2,0,2)
$&$-                    $&$\{3\}$&$ (Cub)^2(Con)
$\\ 
$     $&$    h_{21}={99}            
$&$(0,0,1,1,0)+
$&$(4,1,0,2,0)                         
$&$(0,3,0,2,0)          $&$-
$&$(0,0,2,0,2)          $&$\{4\}$&$(Qrt)(Cub)
$\\
$     $&$  h_{11}={3}             
$&$(1,2,0,0,3)\,
$&$(6,0,0,2,0)                         
$&$                    $&$
$&$                    $&$ \{6\}   $&$ M_6
$\\
$21'   $&$ (2,4,3,1,4)[14]  
$&$(0,0,0,1,1)+
$&$(0,3,0,2,0)                         
$&$(3,0,1,1,1)          $&$(0,0,2,0,2)
$&$-                    $&$\{3\}$&$ (Cub)^2(Con)
$\\ 
$     $&$   h_{21}={64}            
$&$(1,2,0,0,3)+
$&$(4,1,0,2,0)                         
$&$(0,3,0,2,0)          $&$-
$&$(0,0,2,0,2)          $&$\{4\}$&$(Qrt)(Cub)
$\\
$     $&$  h_{11}={6}             
$&$(1,2,3,0,0)\,
$&$(6,0,0,2,0)                         
$&$                    $&$
$&$                    $&$ \{6\}   $&$ M_6
$\\
$21''   $&$ (2,4,5,3,4)[18]  
$&$(1,2,3,0,0)+
$&$(0,3,0,2,0)                         
$&$(3,0,1,1,1)          $&$(0,0,2,0,2)
$&$-                    $&$\{3\}$&$ (Cub)^2(Con)
$\\ 
$     $&$   h_{21}={39}            
$&$(1,2,0,0,3)+
$&$(4,1,0,2,0)                         
$&$(0,3,0,2,0)          $&$-
$&$(0,0,2,0,2)          $&$\{4\}$&$(Qrt)(Cub)
$\\
$     $&$   h_{11}={7}            
$&$(0,0,2,3,1)\,
$&$(6,0,0,2,0)                         
$&$                    $&$
$&$                    $&$ \{6\}   $&$ M_6
$\\
$21'''   $&$ (1,2,2,1,2)[9]  
$&$(1,2,3,0,0)+
$&$(0,3,0,2,0)                         
$&$(3,0,1,1,1)          $&$(0,0,2,0,2)
$&$-                    $&$\{3\}$&$ (Cub)^2(Con)
$\\ 
$     $&$  h_{21}={86}             
$&$(1,2,0,0,3)+
$&$(4,1,0,2,0)                         
$&$(0,3,0,2,0)          $&$-
$&$(0,0,2,0,2)          $&$\{4\}$&$(Qrt)(Cub)
$\\
$     $&$  h_{11}={2}             
$&$(0,0,1,2,1)\,
$&$(6,0,0,2,0)                         
$&$                    $&$
$&$                    $&$ \{6\}   $&$ M_6
$\\\hline\hline
$22   $&$ (2,2,2,1,7)[9]  
$&$(0,0,0,1,1)+
$&$(3,0,3,2,0)                         
$&$(0,3,0,1,1)          $&$(0,0,0,0,2)
$&$-                    $&$\{3\}$&$ (Cub)^2(Con)
$\\ 
$     $&$  h_{21}={122}             
$&$(0,1,2,0,3)+
$&$(1,4,1,2,0)                         
$&$(3,0,3,2,0)          $&$-
$&$(0,0,0,0,2)          $&$\{4\}$&$(Qrt)(Cub)
$\\
$     $&$ h_{11}={2}              
$&$(2,1,0,0,3)\,
$&$(0,6,0,2,0)                         
$&$                    $&$
$&$                    $&$ \{6\}   $&$ M_6
$\\\hline\hline
\end{tabular}
\end{table}

\begin{table}[!ht]
\centering
\caption{\it Continuation of the previous Table.}
\label{Tabelw5}
\scriptsize
\vspace{.05in}
\begin{tabular}{|c|c||c||c|c|c|c||c|c||}
\hline
${ N} $&${ {\vec k}_i(eld)} $&$ Chain
$&$P_1               
$&$P_2                  $&$ P_3
$&$C_2                  $&${IM}$&${IM}-type
$\\\hline\hline
$23   $&$ (1,2,3,1,1)[8]  
$&$(0,0,0,0,1,)+
$&$(0,3,0,1,1)                         
$&$(3,0,1,1,1)          $&$(0,0,2,1,1)
$&$-                    $&$\{3\}$&$ (Cub)^2(Con)
$\\ 
$     $&$    h_{21}={106}             
$&$(0,0,0,1,0)+
$&$(4,1,0,1,1)                         
$&$(0,3,0,1,1)          $&$-
$&$(0,0,2,1,1)          $&$\{4\}$&$(Qrt)(Cub)
$\\
$     $&$   h_{11}={2}            
$&$(1,2,3,0,0)\,
$&$(6,0,0,1,1)                         
$&$                    $&$
$&$                    $&$ \{6\}   $&$ M_6
$\\\hline\hline
$24   $&$ (3,2,4,3,1)[13]  
$&$(0,0,0,0,1,)+
$&$(0,0,3,0,1)                         
$&$(1,3,0,1,1)          $&$(2,0,0,2,1)
$&$-                    $&$\{3\}$&$ (Cub)^2(Con)
$\\ 
$     $&$   h_{21}={62}            
$&$(0,1,2,3,0)+
$&$(0,4,1,0,1)                         
$&$(0,0,3,1,1)          $&$-
$&$(2,0,0,2,1)          $&$\{4\}$&$(Qrt)(Cub)
$\\
$     $&$    h_{11}={5}           
$&$(3,1,2,0,0)\,
$&$(0,6,0,0,1)                         
$&$                    $&$
$&$                    $&$ \{6\}   $&$ M_6
$\\\hline\hline
$25   $&$ (3,2,4,6,1)[16]  
$&$(0,0,0,0,1,)+
$&$(0,3,0,1,1)                         
$&$(3,0,3,0,1)          $&$(0,0,0,2,1)
$&$-                    $&$\{3\}$&$ (Cub)^2(Con)
$\\ 
$     $&$  h_{21}={73}             
$&$(0,1,2,3,0)+
$&$(1,4,1,0,1)                         
$&$(3,0,3,0,1)          $&$-
$&$(0,0,0,2,1)          $&$\{4\}$&$(Qrt)(Cub)
$\\
$     $&$   h_{11}={5}            
$&$(2,1,0,3,0)\,
$&$(0,6,0,0,1)                         
$&$                    $&$
$&$                    $&$ \{6\}   $&$ M_6
$\\\hline\hline
$26   $&$ (1,1,1,2,1)[6'']  
$&$(0,0,1,2,3)+
$&$(0,0,0,3,0)                         
$&$(1,1,3,0,1)          $&$(2,2,0,0,2)
$&$-                    $&$\{3\}$&$ (Cub)^2(Con)
$\\ 
$     $&$  h_{21}={103}             
$&$(0,3,1,2,0)+
$&$(0,0,4,1,0)                         
$&$(0,0,0,3,0)          $&$-
$&$(2,2,0,0,2)          $&$\{4\}$&$(Qrt)(Cub)
$\\
$     $&$  h_{11}={1}             
$&$(3,0,1,2,0)\,
$&$(0,0,6,0,0)                         
$&$                    $&$
$&$                    $&$ \{6\}   $&$ M_6
$\\ \hline\hline
$27   $&$ (3,2,3,4,6)[18]  
$&$(0,0,1,2,3)+
$&$(0,3,0,3,0)                         
$&$(1,0,3,0,1)          $&$(2,0,0,0,2)
$&$-                    $&$\{3\}$&$ (Cub)^2(Con)
$\\
$     $&$   h_{21}={53}             
$&$(0,2,1,0,3)+
$&$(0,1,4,1,0)                         
$&$(0,3,0,3,0)          $&$-
$&$(2,0,0,0,2)          $&$\{4\}$&$(Qrt)(Cub)
$\\
$     $&$  h_{11}={5}             
$&$(3,0,1,2,0)\,
$&$(0,0,6,0,0)                         
$&$                    $&$
$&$                    $&$ \{6\}   $&$ M_6
$\\ \hline\hline
$28   $&$ (2,2,3,2,9)[18]  
$&$(0,0,1,2,3)+
$&$(3,3,0,3,0)                         
$&$(0,0,3,0,1)          $&$(0,0,0,0,2)
$&$-                    $&$\{3\}$&$ (Cub)^2(Con)
$\\
$     $&$  h_{21}={112}             
$&$(0,2,1,0,3)+
$&$(1,1,4,1,0)                         
$&$(3,3,0,3,0)          $&$-
$&$(0,0,0,0,2)          $&$\{4\}$&$(Qrt)(Cub)
$\\
$     $&$ h_{11}={4}              
$&$(2,0,1,0,3)\,
$&$(0,0,6,0,0)                         
$&$                    $&$
$&$                    $&$ \{6\}   $&$ M_6
$\\ \hline\hline
\end{tabular}
\end{table}
}
%\normalsize

Similar recurrence formulae can (in principle) be found for all other
types of IMs in arbitrary dimensions $n$. We now
discuss the example of the IM $\{3_{10}\}$. This type of IM is determined
by 3 cubic monomials with one important property: the differences between
each pair of monomials included in this IM, the combinations $P_i-P_j$,
should have coefficients divisible by 3.  Therefore, the degrees of these
monomials should be constructed only from the numbers $3, 1$ and 0. In
order to find the recurrence formula, one should look at all possible
expansions of the integer numbers ${n}, {n-1},...,{3}$ in terms of 3
integers. For example, using the compositions of 
the numbers ${3}, {4}, {5}, {6}, .....$,
respectively, one can get  the following results for the number of such 
{IMs},  for different dimensions N(n):
\begin{eqnarray}
&&E_r-line\nonumber\\
{3}&=&{1}+{1}+{1}, \nonumber\\
{4}&=&{2}+{1}+{1}, \nonumber\\
&&N(4)=2 \nonumber\\
&&........................\nonumber\\
{5}&=&{3}+{1}+{1}, \qquad {2}+{2}+{1} \nonumber\\
&&N(5)=2N(4)+2=4\nonumber\\
{6}&=&{4}+{1}+{1}, \qquad {3}+{2}+{1}, \qquad {2}+{2}+{2} \nonumber\\
&&N(6)=N(4)+3=7\nonumber\\
&&........................\nonumber\\
{7}&=&{5}+{1}+{1}, \qquad {4}+{2}+{1} , \qquad {3}+{3}+{1} \nonumber\\
&=&\qquad {3}+{2}+{2} \nonumber\\
&&N(7)=N(6)+4=11,\nonumber\\
&&........................\nonumber\\
\end{eqnarray}

Take into account that for  the case $n=3$, one has only the 
following combination of monomials:
$(3,0,0,), (0,3,0)$, $(0,0,3)$, which corresponds in our description to the
expansion of the number 3 in three units. 
So, for the $K3$ case with $n=4$, we
already have one $\{IM\}=(3,0,0,1), (0,3,0,1), (0,0,3,1)$, which again
corresponds to the expansion of the number {3} in terms of three units.  
However, there also appears another possibility: $\{IM\}=(3,3,0,0),
(0,0,3,0), (0,0,0,3)$, which corresponds to the unique expansion of the
number 4 in terms of integer positive numbers. For the $CY_3$ case with
$n=5$, there are the following IMs:

\begin{eqnarray}
\{IM\}&=&(3,0,0,1,1), (0,3,0,1,1), (0,0,3,1,1);\qquad 
\rightarrow 3(3)=1+1+1 \nonumber\\
\{IM\}&=&(3,3,0,0,1), (0,0,3,0,1), (0,0,0,3,1);\qquad
\rightarrow 4(3)=2+1+1 \nonumber\\
N(4)&=&2
\end{eqnarray}
\begin{eqnarray}
\{IM\}&=&(3,3,3,0,0), (0,0,0,3,0), (0,0,0,0,3);\qquad 
\rightarrow 5(3)=3+1+1\nonumber\\
\{IM\}&=&(3,3,0,0,0), (0,0,3,3,0), (0,0,0,0,3);\qquad 
\rightarrow 5(3)=2+2+1 \nonumber\\
N(5)&=&N(4)+2=4,
\end{eqnarray}
where the last two IMs correspond to the two expansions of the number {5}.
This recurrence can easily be continued to higher dimensions $n > 5$.  To
get the total number of IMs $\{10\}_{\Delta}$ in dimension $n$, 
one should sum
over all the numbers characterizing the possible decompositions: ${n},
{n-1}, ...{3}$. For example, for $CY_d$: $d + 2 = n = 4, 5, 6, 7, ...$,
these numbers are {2}, {4}, {7} ,{11},... (see Figure \ref{basunit}).

Similarly, one can find a recurrence relation for
$\{IM\}=\{{9}\}_{\Delta}$, which is constructed from two quartic
monomials, one conic and $E_n$. In this case the difference of the two
quartic monomials should be divisible by four. Taking into account all
possible quartics, after some effort, one can find the following formula
for the number of these IMs:

\begin{eqnarray}
N_{9\Delta} = \frac{1}{3} \cdot (n-2)(n^2-4n+6).
\end{eqnarray}   
This expression gives the following numbers:  $1, 4, 11, 24, 45, 76, 119,
176, 249, ...$ for $n=3, 4, 5, 6, 7, 8, 9, 10, 11, ..$, respectively.

\subsubsection{Applications to $CY_4$ Spaces}

In the 6-dimensional case, we have, analogously to the previous case, two
sets of triples of monomials: one of 377 and the other of 494. From the first
set we get 5,216 different 4-vector chains and from the second one we get
4953 chains. The union set has 5527 members, consistent with the 5,607
that we got by the `expansion and intersection' method. As in the
5-dimensional case, the difference can be made up if we include the IMs
with double conic monomials.

\begin{figure}[th!]
   \begin{center}
   \mbox{
   \epsfig{figure=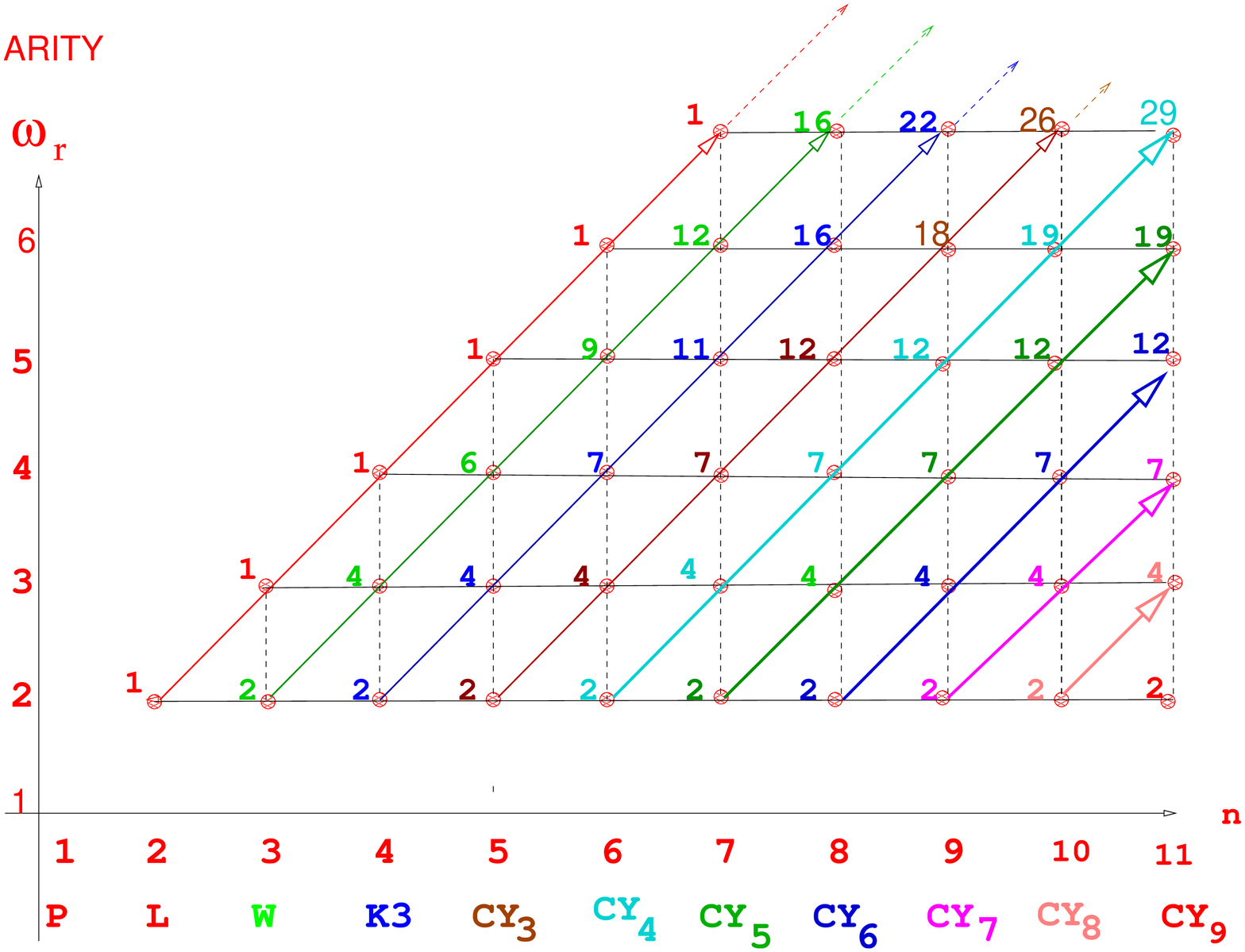,height=16cm,width=18cm}}
   \end{center}
   \caption{\it 
The numbers of recurrences of Calabi-Yau hypersurfaces  
with a $(1,...,1)_n$ fibre are calculable along all lines 
$n = r + p$ in the arity-dimension plot.}
\label{basunit}
\end{figure}

A further reduction in the number of chains has to be considered, from the
5,607 6-dimensional 4-vector chains to 2111 independent chains. We have
already mentioned that there are different types of IMs even among the
cubics $\{3\}$ and quartics $\{4\}$, and the number of different conics
grows monotonically with increasing dimension ${n}$. We have also already
remarked that there exists a recurrence formula for all types of IMs with
arbitrary dimension ${n}$, and have already discussed the reccurences of
the Weierstrass IMs $\{3_W\}$ and $\{4_W\}$.  The possible types of cubic
$\{3\}$, quartic $\{4\}$ and double conic IMs which describe the {2111}
irreducible $CY_3$ chains have different structures, corresponding to the
different types of intersections, that we can illustrate by the following
expression, see also Tables~\ref{Tabeld41}, \ref{Tabeld42} and 
\ref{Tabeld43}:

\begin{eqnarray}
\{IM\}_6&\mapsto&\biggl ( 37\cdot\{4\}_{\Delta} +
{\bf7\cdot\{10\}_{\Delta}}\biggl ) \nonumber\\
&+&\biggl (66 \cdot \{5\}_{\Delta}+ 27 \cdot \{5\}_{\Box}+
6 \cdot \{5\}_{\Box'} \biggl )\nonumber\\
&+&\biggl ({\bf24\cdot\{9\}_{\Delta}}+ 11 \cdot \{9\}_{\Box}+ 
5 \cdot \{9\}_{\Box'}\biggl ) \nonumber\\
&+&\biggl({\bf 84\cdot\{ 7\}_{\Delta}} +28\cdot\{ 7\}_{\Box}+
5\cdot\{ 7\}_{Quint}+1\cdot\{ 7\}_{Sixt} \biggl )\nonumber\\
&+& \biggl(36 \cdot \{ 6\}_{\Box} +5 \cdot \{ 6\}_{Quint}\biggl )\nonumber\\
&+&\biggl(21 \cdot \{ 8\}_{\Box} +5 \cdot \{ 8\}_{Quint} \biggl )\nonumber\\
&\mapsto& \{2111\}
\end{eqnarray}
The recurrence relation for Calabi-Yau spaces with elliptic fibres
$\{10\}_{\Delta}$ can be extended to the cases of $CY_d$ spaces with $K3$
fibres, described by $\vec{k}_4=(1,1,1,1)[4]$, whose algebraic equation
includes the 35-point monomial and its mirror with 5 points.
The ${IM}_4$ for this $K3$ space contains the
four quartic monomials $P_1,P_2,P_3,P_4$ obeying the Diophantine equation:
$(P_1+P_2+P_3+P_4)/4=E_4$. These monomials have in addition one very important
condition: $P_i - P_j$ should be divisible by {4} for each choice of {$i,
j = 1, 2, 3, 4, i \neq j$}. The types of different $n$-dimensional
$\{IM\}_4$, describing the $CY_d: n=d+2 \geq 4$ spaces with
$\{35\}_{\Delta}$ fibres are constructed only from the
numbers {4} and {0}. The number  {1} will play an additional role.
Therefore, similarly to the case of the third $E_r$ line,
the recurrence formulae for these {IMs} will be determined from the
expansions of positive integer numbers in terms of four positive integers,
i.e., (see Figure \ref{basunit}),
\begin{eqnarray}
&&K3-line \nonumber\\
n=4(4)&=&1+1+1+1 \nonumber\\
n=5(4)&=&2+1+1+1 \nonumber\\
&&N(5)=2 \nonumber\\
&&........................\nonumber\\
n=6(4)&=&3+1+1+1=2+2+1+1 \nonumber\\ 
&&N(6)=N(5)+2=4 \nonumber\\
&&........................\nonumber\\
n=7(4)&=&4+1+1+1=3+2+1+1=2+2+2+1 \nonumber\\
&&N(7)=N(6)+3=7 \nonumber\\
&&........................\nonumber\\
n=8(4)&=&5+1+1+1=4+2+1+1=3+3+1+1=\nonumber\\ 
&=&3+2+2+1=2+2+2+2 \nonumber\\ 
&&N(8)=N(7)+5=12 \nonumber\\
&&........................\nonumber\\
n=9(4)&=&6+1+1+1=5+2+1+1=4+3+1+1=\nonumber\\
&=& 4+2+2+1=3+3+2+1=3+2+2+2 \nonumber\\
&&N(9)=N(8)+6=18 \nonumber\\
&&........................\nonumber\\
n=10(4)&=&7+1+1+1=6+2+1+1=5+3+1+1=5+2+2+1= \nonumber \\
&=&4+4+1+1=4+3+2+1=3+3+3+1=3+3+2+2 \nonumber\\
&&N(10)=N(9)+8=26 \nonumber\\
&&........................\nonumber\\
n=11(4)&=&8+1+1+1=7+2+1+1=6+3+1+1=6+2+2+1=\nonumber\\
&=&5+4+1+1=5+3+2+1=5+2+2+2=4+4+2+1 \nonumber\\
&=&4+3+3+1=4+3+2+2=3+3+3+2 \nonumber\\ 
&&N(11)=N(10)+11=37 \nonumber\\
&&........................\nonumber\\
\end{eqnarray}
Thus, the corresponding numbers of these four $\{IMs\}$ with dimension $n$
are equal to sums of the possible expansions of the integers indicated in
this expression, i.e., for dimensions $ 5, 6, 7, 8, 9, 10, 11, ....$
they are equal to $2, 4, 7, 12, 18, 26, 37, ...$, respectively.

Similarly, this example can be extended to $CY_4$ ($CY_d$) spaces with 
a $CY_3$ ($CY_{d-1}$)
fibre described by the RWV $\vec{k}_5=(1,1,1,1,1)[5]$ 
($\vec{k}_n=(1,...,1)_n$). The results
for the types of the corresponding {IMs} can be obtained from the following 
expansions (see also Fig.~\ref{basunit}):
\begin{eqnarray}
&&CY_3-line \nonumber\\ 
n=5(5)&=&1+1+1+1+1 \nonumber\\ 
n=6(5)&=&2+1+1+1+1 \nonumber\\
&&N(6)=2\nonumber\\
&&........................\nonumber\\
n=7(5)&=&3+1+1+1+1=2+2+1+1+1\nonumber\\ 
&&N(7)=N(6)+2=4\nonumber\\
&&........................\nonumber\\
n=8(5)&=&4+1+1+1+1=3+2+1+1+1=2+2+2+1+1 \nonumber\\ 
&&N(8)=N(7)+3=7\nonumber\\ 
&&........................\nonumber\\
n=9(5)&=&5+1+1+1+1=4+2+1+1+1=3+3+1+1+1 \nonumber\\
&=&3+2+2+1+1=2+2+2+2+1 \nonumber\\ 
&&N(9)=N(8)+5=12\nonumber\\
&&........................\nonumber\\
n=10(5)&=&6+1+1+1+1=5+2+1+1+1=4+3+1+1+1\nonumber \\
&=&4+2+2+1+1=3+3+2+1+1=3+2+2+2+1=2+2+2+2+2 \nonumber \\
&&N(10)=N(9)+7=19\nonumber \\
&&........................\nonumber\\
n=11(5)&=&7+1+1+1+1=6+2+1+1+1=5+3+1+1+1\nonumber\\
&=&5+2+2+1+1=4+4+1+1+1=4+3+2+1+1=4+2+2+2+1\nonumber\\
&=&3+3+3+1+1=3+3+2+2+1=3+2+2+2+2 \nonumber\\
&&N(11)=N(10)+10=29\nonumber\\
&&........................\nonumber\\
\end{eqnarray}
The same approach can clearly be extended to establish the numbers of any 
other desired IMs.

{\scriptsize
\begin{table}[!ht]
\centering
\caption{\it The $CY_4$ {IM}s with the
Diophantine conditions:$\{3\}\rightarrow 1/3(P_1+P_2+P_3)=E_6$
and $\{4\}\rightarrow  1/4(P_1+P_2+2C_2)=E_6$, $1/2(C_1+C_2)= E_6$.
Here we present the chains with triangle $\{10+4\}$
and $\{9+5\}$ intersections.}
\label{Tabeld41}
%\scriptsize
\vspace{.05in}
\begin{tabular}{||c|c||c||c|c|c||c|c||}
\hline
${ \sigma}                $&${ {\vec k}_i(eld)} 
$&$ Chain
$&$P_1               
$&$P_2                    
$&$ P_3/C_2
$&${IM}$&${IM}-type
$\\
\hline\hline \hline 
${\{10\}}_{\Delta}        $&$(1,1,1,1,1,1)[6]
$&$ (0,0,1,0,1,1)+
$&$(0,1,1,1,0,3)               
$&$(0,1,1,1,3,0)          
$&$(3,1,1,1,0,0)
$&$\{3\}$&$(Cub)^3   
$ \\
$ $&$
$&$ (0,1,0,0,0,0)+
$&$(3,1,1,1,0,0)               
$&$(0,1,1,1,1,2)          
$&$(0,1,1,1,2,1)
$&$\{3\}$&$(Cub)(Con)^2   
$ \\
$                         $&$
$&$(0,0,0,1,0,0)+
$&$(2,1,1,1,0,1)               
$&$(1,1,1,1,2,0)          
$&$(0,1,1,1,1,2)
$&$\{3\}$&$(Con)^3    
$ \\ 
$                         $&$
$&$ (1,0,0,0,1,1)+
$&$(3,1,1,1,0,0)               
$&$(1,1,1,1,0,2)           
$&$(0,1,1,1,2,1)          
$&$\{4\}$&$(Cub)(Con)
$ \\ \hline \hline
${\{4\}}_{\Delta}          $&$(3,1,3,4,4,2)[17]
$&$(0,1,0,0,1,0)+
$&$(0,0,1,0,2,3)               
$&$(0,1,0,3,1,0)           
$&$(3,2,2,0,0,0)
$&$\{3\}$&$(Cub)^3   
$ \\ 
$                          $&$
$&$ (2,0,0,1,3,0)+
$&$-               
$&$-                    
$&$-                     
$&$-    $&$-  
$ \\
$                          $&$
$&$(0,0,3,2,0,1)+
$&$ -              
$&$ -                     
$&$ -                     
$&$ -  $&$ -
$ \\ 
$                          $&$
$&$ (0,0,0,1,0,0)                
$&$-         
$&$-
$&$-          
$&$-   $&$-
$ \\\hline \hline
${\{9\}}_{\Delta}         $&$(1,1,1,1,1,2)[7]
$&$(0,0,1,0,1,0)+
$&$(0,4,1,1,0,0)               
$&$(4,0,1,1,0,0)          
$&$(0,0,1,1,2,2)          
$&$\{4\}$&$(Qrt)^2   
$ \\
$ $&$
$&$ (0,1,0,0,0,0)+
$&$(1,3,1,1,0,0)               
$&$(3,1,1,1,0,0)         
$&$(0,0,1,1,2,2)         
$&$\{4\}$&$(Qrt)^2   
$ \\
$ $&$
$&$(0,0,0,1,0,0)+
$&$(3,1,1,1,0,0)              
$&$(0,2,1,1,1,1)         
$&$(0,0,1,1,2,2)  
$&$ \{3\}$&$(Cub)(Con)^2   
$ \\ 
$                         $&$
$&$ (1,0,0,0,1,2)                
$&$-         
$&$-
$&$-          
$&$-    $&$-
$ \\\hline \hline
${\{5\}}_{\Delta}         $&$(1,1,1,2,3,6)[14]
$&$(0,0,1,0,0,1)+
$&$(4,4,2,2,0,0)               
$&$(0,0,2,0,4,0)          
$&$(0,0,0,1,0,2)          
$&$\{4\}$&$(Qrt)^2   
$ \\
$                         $&$
$&$ (1,0,0,0,1,2)+
$&$-               
$&$-                      
$&$-                       
$&$-    $&$-   
$ \\
$                         $&$
$&$(0,1,0,0,1,2)+
$&$ -             
$&$ -                      
$&$ -                     
$&$ -   $&$ -  
$ \\ 
$                         $&$
$&$ (0,0,0,2,1,1)                
$&$ -                     
$&$ -                     
$&$-
$&$-    $&$-
$ \\\hline \hline
${\{9\}}_{\Box}         $&$(2,1,1,3,2,2)[11]
$&$(1,0,0,1,1,0)+
$&$(3,3,2,0,0,0)               
$&$(0,0,1,2,1,1)          
$&$(0,0,0,1,2,2)          
$&$\{3\}$&$(Cub)(Con)^2   
$ \\
$ $&$
$&$ (0,1,0,1,1,0)+
$&$(3,3,2,0,0,0)-               
$&$(1,1,2,0,2,2)-         
$&$(0,0,0,2,1,1)-         
$&$\{4\}$&$(Cub)(Con)   
$ \\
$ $&$
$&$(0,0,1,0,0,1)+
$&$(2,2,2,1,0,0)              
$&$(2,2,1,0,1,1)         
$&$(1,1,0,0,2,2)  
$&$ \{2\}^3$&$(Con)^3   
$ \\ 
$                         $&$
$&$ (1,0,0,1,0,1)                
$&$-         
$&$-
$&$-          
$&$-    $&$-
$ \\\hline \hline
${\{9\}}_{Quadr}         $&$(1,1,1,1,2,1)[7]
$&$(0,0,0,0,0,1)+
$&$(0,2,0,0,2,1)               
$&$(0,2,2,2,0,1)          
$&$(1,1,2,2,0,1)          
$&$\{2\}^4$&$(Con)^4   
$ \\
$ $&$
$&$ (0,0,0,1,1,0)+
$&$-               
$&$-         
$&$-         
$&$   $&$  
$ \\
$ $&$
$&$(0,0,1,0,1,0)+
$&$-             
$&$-         
$&$-  
$&$    $&$   
$ \\ 
$                         $&$
$&$ (1,1,0,0,0,0)                
$&$-         
$&$-
$&$-          
$&$-    $&$-
$ \\\hline \hline
${\{5\}}_{\Box}         $&$(3,2,1,3,4,6)[19]
$&$(0,0,1,0,0,0)+
$&$(3,3,1,1,0,0)               
$&$(0,0,1,0,3,1)          
$&$(0,0,1,2,0,2)          
$&$\{3\}^2$&$(Cub)^2(Con)   
$ \\
$ $&$
$&$ (1,0,0,3,2,0)+
$&$-               
$&$-         
$&$-         
$&$   $&$  
$ \\
$ $&$
$&$(2,0,0,0,1,3)+
$&$-             
$&$-         
$&$-  
$&$    $&$   
$ \\ 
$                         $&$
$&$ (0,2,0,0,1,3)                
$&$-         
$&$-
$&$-          
$&$-    $&$-
$ \\\hline \hline
${\{5\}}_{Quadr}         $&$(1,1,1,1,1,2)[7]
$&$(0,0,0,0,1,1)+
$&$(0,1,0,2,2,0)               
$&$(1,0,0,2,0,2)          
$&$(1,2,2,0,2,0)          
$&$\{2\}^2$&$(Con)^2   
$ \\
$ $&$
$&$ (0,0,1,1,0,0)+
$&$-               
$&$-         
$&$-         
$&$    $&$   
$ \\
$ $&$
$&$(0,2,0,1,0,1)+
$&$-             
$&$-         
$&$-  
$&$    $&$   
$ \\ 
$                         $&$
$&$ (2,0,0,1,1,0)                
$&$-         
$&$-
$&$-          
$&$-    $&$-
$ \\\hline \hline
\end{tabular}
\end{table}
}
%\normalsize

{\scriptsize
\begin{table}[!ht]
\centering
\caption{\it Continuation of the previous Table.}
\label{Tabeld42}
%\scriptsize
\vspace{.05in}
\begin{tabular}{||c|c||c||c|c|c||c|c||}
\hline
${ \sigma}                $&${ {\vec k}_i(eld)} 
$&$ Chain
$&$P_1               
$&$P_2                    
$&$ P_3/C_2
$&${IM}$&${IM}-type
$\\
\hline\hline  
${\{8\}}_{\Box}         $&$(1,1,1,1,2,5)[11]
$&$(0,0,1,0,0,1)+
$&$(4,3,2,2,0,0)               
$&$(0,1,0,0,0,2)          
$&$(0,0,1,1,2,1)          
$&$\{4\}$&$(Qrt)(Cub)   
$ \\
$ $&$
$&$ (0,0,0,1,0,1)+
$&$(3,2,2,2,1,0)               
$&$(0,1,0,0,0,2)
$&$(0,0,1,1,2,1)
$&$ \{3\}   $&$ (Cub)(Con)^2      
$ \\
$ $&$
$&$(1,0,0,0,1,2)+
$&$-             
$&$-         
$&$-  
$&$    $&$   
$ \\ 
$                         $&$
$&$ (0,1,0,0,1,1)                
$&$-         
$&$-
$&$-          
$&$-    $&$-
$ \\\hline \hline
${\{8\}}_{Quint}         $&$(2,1,1,3,2,1)[10]
$&$(0,0,1,1,0,0)+
$&$(3,2,2,0,0,0)               
$&$(0,1,0,2,1,1)          
$&$(0,0,1,1,2,2)          
$&$\{3\}$&$(Cub)(Con)2   
$ \\
$ $&$
$&$ (0,1,0,0,1,0)+
$&$(3,2,2,2,0,0)               
$&$(1,2,0,2,0,0)
$&$(0,0,1,0,2,2)
$&$ \{4\}   $&$ (Cub)(Con)      
$ \\
$ $&$
$&$(1,0,0,1,1,0)+
$&$(0,0,1,1,2,2)             
$&$(0,1,0,2,1,1)
$&$(1,0,2,0,2,2)  
$&$ \{2\}^3   $&$ (Con)^3  
$ \\ 
$                         $&$
$&$ (1,0,0,1,0,1)                
$&$-         
$&$-
$&$-          
$&$-    $&$-
$ \\\hline \hline
${\{6\}}_{\Box}         $&$(2,4,1,3,4,1)[15]
$&$(0,0,1,0,0,0)+
$&$(1,0,1,4,0,0)               
$&$(3,2,1,0,0,0)          
$&$(0,1,1,0,2,2)          
$&$\{4\}$&$(Qrt)(Cub)   
$ \\
$ $&$
$&$ (2,0,0,1,3,0)+
$&$(3,2,1,0,0,0)               
$&$(0,0,1,3,1,1)
$&$(0,1,1,0,2,2)
$&$ \{3\}   $&$ (Cub)^2(Con)      
$ \\
$ $&$
$&$(0,2,0,1,1,0)+
$&$-             
$&$-         
$&$-  
$&$    $&$   
$ \\ 
$                         $&$
$&$ (0,2,0,1,0,1)                
$&$-         
$&$-
$&$-          
$&$-    $&$-
$ \\\hline \hline
${\{6\}}_{Quint}         $&$(1,2,2,2,3,1)[11]
$&$(0,0,1,1,0,0)+
$&$(3,2,1,1,0,0)               
$&$(0,1,0,2,1,2)          
$&$(0,0,2,0,2,1)          
$&$\{3\}$&$(Cub)(Con)^2   
$ \\
$ $&$
$&$ (0,1,0,0,1,0)+
$&$ (0,0,2,0,2,1)              
$&$ (0,1,0,2,1,2)
$&$ (2,1,2,0,1,0)
$&$ \{2\}^2    $&$ (Con)^2      
$ \\
$ $&$
$&$(1,0,0,1,2,0)+
$&$-             
$&$-         
$&$-  
$&$    $&$   
$ \\ 
$                         $&$
$&$ (0,1,1,0,0,1)                
$&$-         
$&$-
$&$-          
$&$-    $&$-
$ \\\hline \hline
${\{7\}}_{\Box}         $&$(1,1,1,3,1,5)[12]
$&$(0,0,0,1,1,0)+
$&$(3,3,2,1,1,0)               
$&$(0,0,1,2,0,1)          
$&$(0,0,0,0,2,2)          
$&$\{3\}$&$(Cub)(Con)^2   
$ \\
$ $&$
$&$ (0,0,1,0,0,1)+
$&$ (4,4,2,0,2,0)              
$&$ (0,0,0,0,2,2)
$&$ (0,0,1,2,0,1)
$&$ \{4\}    $&$(Qrt) (Con)      
$ \\
$ $&$
$&$(1,0,0,1,0,2)+
$&$ -            
$&$-         
$&$-  
$&$    $&$   
$ \\ 
$                         $&$
$&$ (0,1,0,1,0,2)                
$&$-         
$&$-
$&$-          
$&$-    $&$-
$ \\\hline \hline
${\{7\}}_{Quint}         $&$(1,2,2,3,4,2)[14]
$&$(1,0,0,1,2,0)+
$&$(3,2,2,1,0,0)               
$&$(0,1,1,0,2,1)          
$&$(0,0,0,2,1,2)          
$&$\{3\}$&$(Cub)(Con)^2   
$ \\
$ $&$
$&$ (0,1,0,1,1,0)+
$&$ (3,2,2,1,0,0)              
$&$ (1,0,0,3,0,2)
$&$ (0,1,1,0,2,1)
$&$ \{4\}    $&$(Cub)^2      
$ \\
$ $&$
$&$(0,0,1,1,1,0)+
$&$-             
$&$-         
$&$-  
$&$    $&$   
$ \\ 
$                         $&$
$&$ (0,1,1,0,0,2)                
$&$-         
$&$-
$&$-          
$&$-    $&$-
$ \\\hline \hline
${\{7\}}_{Sixt}         $&$(1,2,1,2,1,2)[9]
$&$(0,0,0,0,1,1)+
$&$(0,2,1,1,2,0)               
$&$(0,2,2,0,1,1)          
$&$(1,1,0,2,2,0)          
$&$\{2\}^3$&$(Con)^3   
$ \\
$ $&$
$&$ (0,0,1,1,0,0)+
$&$ -            
$&$ -
$&$ -
$&$         $&$      
$ \\
$ $&$
$&$(1,1,0,0,0,0)+
$&$ -            
$&$-         
$&$-  
$&$    $&$   
$ \\ 
$                         $&$
$&$ (0,1,0,1,0,1)                
$&$-         
$&$-
$&$-          
$&$-    $&$-
$ \\\hline \hline
\end{tabular}
\end{table}
}
%\normalsize

{\scriptsize
\begin{table}[!ht]
\centering
\caption{\it  The $CY_4$ {IM}s with the
Diophantine conditions:$\{3\}\rightarrow 1/3(P_1+P_2+P_3)=E_6$
and $\{4\} \rightarrow 1/4(P_1+P_2+2C_2)=E_6$, $1/2(C_1+C_2)= E_6$,
$M_6+2P_1=3 C_1$. Here we present some of the 84 Weierstrass sets of 
{IM}s.}
\label{Tabeld43}
\vspace{.05in}
\begin{tabular}{|c||c||c|c|c|c||c||c|}
\hline
${ N} $&${ {\vec k}_i(eld)} 
$&$ \vec{k}^{1ex}       $&$ \vec{k}^{2ex}            
$&$ \vec{k}^{3ex}       $&$ \vec{k}^{4ex} 
$&$M_6                  
$\\\hline\hline
$1   $&$ (1,1,1,1,8,12)[24]  
$&$(1,0,0,0,2,3,)+      $&$(0,1,0,0,2,3)+                         
$&$(0,0,1,0,2,3)+       $&$(0,0,0,1,2,3)
$&$(6,6,6,6,0,0)                 
$\\\hline
$2   $&$ (1,1,1,1,7,11)[22]  
$&$(1,0,0,0,2,3)+       $&$(0,1,0,0,2,3)+                         
$&$(0,0,1,0,2,3)+       $&$(0,0,0,1,1,2)
$&$(6,6,6,4,0,0)                       
$\\\hline
$3   $&$ (1,1,1,1,7,10)[21]  
$&$(1,0,0,0,2,3)+      $&$(0,1,0,0,2,3)+                         
$&$(0,0,1,0,2,3)+      $&$(0,0,0,1,1,1)
$&$(6,6,6,3,0,0)                     
$\\\hline
$4   $&$ (1,1,1,1,6,10)[20]  
$&$(1,0,0,0,2,3)+      $&$(0,1,0,0,2,3)+                         
$&$(0,0,1,0,2,3)+      $&$(0,0,0,1,0,1)
$&$(6,6,6,2,0,0)                        
$\\\hline
$5   $&$ (1,1,1,1,6,9)[19]  
$&$(1,0,0,0,2,3)+       $&$(0,1,0,0,2,3)+                         
$&$(0,0,1,0,2,3)+       $&$(0,0,0,1,0,0)
$&$(6,6,6,1,0,0)                         
$\\\hline
$6   $&$ (2,1,1,6,8,6)[24]  
$&$(1,0,0,0,2,3)+      $&$(0,1,0,0,2,3)+                         
$&$(1,0,0,3,2,0)+      $&$(0,0,1,3,2,0)
$&$(6,6,6,0,0,0)                         
$\\
$6'  $&$ (2,1,1,9,8,3)[24]  
$&$(1,0,0,0,2,3)+      $&$(1,0,0,3,2,0)+                         
$&$(0,1,0,3,2,0)+      $&$(0,0,1,3,2,0)
$&$(6,6,6,0,0,0)                        
$\\\hline
$7   $&$ (2,1,1,4,4,12)[24]  
$&$(1,0,0,0,2,3)+      $&$(0,0,1,0,2,3)+                         
$&$(1,0,0,2,0,3)+      $&$(0,1,0,2,0,3)
$&$(6,6,6,0,0,0)                         
$\\
$7 ' $&$ (2,1,1,3,6,11)[24]  
$&$(1,0,0,0,2,3)+      $&$(0,1,0,0,2,3)+                         
$&$(0,0,1,0,2,3)+      $&$(1,0,0,3,0,2)
$&$(6,6,6,0,0,0)                         
$\\\hline\hline
$8   $&$ (1,1,1,1,8,12)[24]  
$&$(1,0,0,0,2,3,)+      $&$(0,1,0,0,2,3)+                         
$&$(0,0,1,0,2,3)+       $&$(0,0,0,1,2,3)
$&$(6,6,6,6,0,0)                      
$\\\hline
$9   $&$ (1,1,1,1,6,10)[20]  
$&$(1,0,0,0,2,3)+       $&$(0,1,0,0,2,3)+                         
$&$(0,0,1,0,1,2)+       $&$(0,0,0,1,1,2)
$&$(6,6,4,4,0,0)                         
$\\\hline
$10   $&$ (1,1,1,1,6,9)[19]  
$&$(1,0,0,0,2,3)+      $&$(0,1,0,0,2,3)+                         
$&$(0,0,1,0,1,2)+      $&$(0,0,0,1,1,1)
$&$(6,6,4,3,0,0)       )               
$\\\hline
$11   $&$ (1,1,1,1,5,9)[18]  
$&$(1,0,0,0,2,3)+      $&$(0,1,0,0,2,3)+                         
$&$(0,0,1,0,1,2)+      $&$(0,0,0,1,0,1)
$&$(6,6,4,2,0,0)                     
$\\\hline
$12   $&$ (1,1,1,1,5,8)[17]  
$&$(1,0,0,0,2,3)+      $&$(0,1,0,0,2,3)+                         
$&$(0,0,1,0,1,2)+      $&$(0,0,0,1,0,0)
$&$(6,6,4,1,0,0)                    
$\\\hline
$-                     $&$ - 
$&$ -                  $&$ -                      
$&$  -                 $&$ - 
$&$(6,...,...,0,0,0)                      
$\\\hline
$81   $&$ (4,3,3,3,3,8)[24]  
$&$(1,3,0,0,0,2)+      $&$(1,0,3,0,0,2)+                         
$&$(1,0,0,3,0,2)+      $&$(1,0,0,0,3,2)
$&$(6,0,0,0,0,0)                      
$\\\hline
$82   $&$ (4,4,4,6,3,3)[24]  
$&$(2,0,1,3,0,0)+      $&$(0,2,1,3,0,0)+                         
$&$(0,2,1,0,3,0)+      $&$(2,0,1,0,0,3)
$&$(0,0,6,0,0,0)                      
$\\
$82'   $&$ (6,2,4,6,3,3)[24]  
$&$(2,0,1,3,0,0)+      $&$(0,2,1,3,0,0)+                         
$&$(2,0,1,0,3,0)+      $&$(2,0,1,0,0,3)
$&$(0,0,6,0,0,0)                        
$\\\hline
$83   $&$ (4,2,2,4,9,3)[24]  
$&$(2,0,0,1,3,0)+      $&$(0,0,2,1,3,0)+                         
$&$(2,0,0,1,3,0)+      $&$(0,2,0,1,0,3)
$&$(0,0,0,6,0,0)                    
$\\
$83'   $&$ (4,2,2,4,9,3)[24]  
$&$(2,0,0,1,3,0)+      $&$(0,2,0,1,3,0)+                         
$&$(0,0,2,1,3,0)+      $&$(2,0,0,1,0,3)
$&$(0,0,0,6,0,0)                      
$\\\hline
$84  $&$ (1,1,1,1,2,6)[12]  
$&$(2,0,0,0,1,3)+      $&$(0,2,0,0,1,3)+                         
$&$(0,0,2,0,1,3)+      $&$(0,0,0,2,1,3)
$&$(0,0,0,0,6,0)                             
$\\\hline
\end{tabular}
\end{table}
}

%\normalsize

\section{Summary}

We have presented a Universal Calabi-Yau Algebra (UCYA) which provides
a two-parameter classification of $CY-d$ spaces in terms of arity and
dimension. This algebra is based on the following ingredients:

\begin{itemize}
\item{Universal composition rules} 
\item{Normal expansions and Diophantine decompositions}
\item{Mirror symmetry}
\end{itemize}
We have shown that this algebraic approach leads us to a natural
formalism for a unified description of complex geometry in all dimensions,
including $K3$ spaces and Calabi-Yau $d$-folds for any $d$.

Since the description of the UCYA is based on structures with two
integer parameters, the arity and dimension of the reflexive weight
vectors (RWVs), we have classified the structures of $CY_d$ spaces along the
diagonal $A_r,D_r,E_r,...$ lines in this plane. In this article we have
studied only the $d$-folds along the first three lines, presenting new
results for low $d$ and some recurrence formulae valid for all $d$.

As an alternative to the Batyrev reflexive polyhedron method, we have
proposed a new description of $CY_d$ spaces based on the structures of the
set of invariant monomials (IMs).
We have shown that the IM approach, which is based on Diophantine
decompositions, is a valuable alternative to the normal RWV expansion
approach. We have demonstrated this by comparing the results of both
approaches for the first three diagonal lines, $A_r,D_r$ and $E_r$, in the
arity-dimension plot for $CY_3$, $CY_4$ cases.

We have shown that recurrence relations for conic, cubic and quartic
monomials give us the formulae for the numbers of IMs in arbitrary
dimensions. This was illustrated in three cases, for $CY_d$ spaces with
$\{10\}_{\Delta}$, $\{9\}_{\Delta}$ and $\{7\}_{\Delta} $fibres. 
This confirms that, in the framework of the
UCYA, the Calabi-Yau `genome' can in principle be solved completely.

%\newpage
%\begin{figure}[th!]
%   \begin{center}
%   \mbox{
%   \epsfig{figure=gen3.eps,height=16cm,width=16cm}}
%   \end{center}
%   \caption{\it The genealogical tree-algebra for reflexive weighted vectors
%involving low dimensions up to $d=4$ (the age is equal 4).}
%\label{gen1}
%\end{figure}

\newpage
\begin{center}

{\bf Acknowledgements}\\

{~~}\\

\end{center}

G .V. thanks U. Aglietti, G. Altarelli, E. Alvarez, L. Alvarez-Gaum\'e, D.
Amati, G.~Costa, H. Dahmen, L. Fellin, G. Harigel, V. Kim, A. Kulikov, L.
Lipatov, A. Masiero, A.~Maslikov, V.~Mikhailov, S.  Petcov, V. Petrov, P.
Sorba, V. Savrin, E. Torrente, M. Virasoro, A. Zichichi, the CERN Theory
Division, ICTP Trieste, SISSA, INFN Sezione di Padova and LAPP-TH
Annecy-le Vieux for supporting his family after the terrible killing of
his daughter Maria in Protvino.

\end{document}